\documentclass[traditabstract]{aa}  
\usepackage{graphicx}
\usepackage{txfonts}
\usepackage{natbib}
\usepackage{hyperref}

\newcommand\patspeed{\mbox{km\,s$^{-1}$\,kpc$^{-1}$}}

\newcommand{\MSun}{\mbox{${M}_\odot$}}
\newcommand{\Msun}{\mbox{${M}_\odot$}}

\def\apgt{\ {\raise-.5ex\hbox{$\buildrel>\over\sim$}}\ }
\def\aplt{\ {\raise-.5ex\hbox{$\buildrel<\over\sim$}}\ }
\def\lteq{\ {\raise-.5ex\hbox{$\buildrel<\over-$}}\ }

\newcommand{\AMUSE}{{\tt AMUSE}}

\def\aap{\ {A\&A}\ }

\def\actaa{\ {Acta Astron.}\ }

\def\aj{\ {AJ}\ }

\def\apj{\ {ApJ}\ }
\def\apjl{\ {ApJL}\ }
\def\apjs{\ {ApJS}\ }

\def\araa{\ {ARA\&A}\ }

\def\bain{\ {Bul. Astron. Inst. Neth.}\ }

\def\icarus{\ {Icarus}\ }

\def\mnras{\ {MNRAS}\ }

\def\nat{\ {Nat}\ }

\def\na{\ {New Astron.}\ }

\def\pasj{\ {Publ. Astr. Soc. Japan}\ }

\def\physrep{\ {PhysRep}\ }

\def\sovast{\ {SvA}\ }
\def\ssr{\ {Space Sci. Rev.}\ }

\usepackage{color}

\begin{document}

\title{Oort cloud Ecology II:
       The chronology of the formation of the Oort cloud}
\titlerunning{Oort cloud Ecology II}
\authorrunning{Portegies Zwart}


\author{
Simon Portegies Zwart \inst{1},
Santiago Torres \inst{1,2},
Maxwell X. Cai \inst{1}
and
Anthony G. A. Brown \inst{1},
}



\institute{
$^1$ Leiden Observatory, Leiden University, PO Box 9513, 2300 RA,
     Leiden, The Netherlands\\
$^2$ Department of Physics and Astronomy, University of California, Los Angeles, CA 90095, USA 
}

\date{\today}

\abstract{ Jan Hendrik Oort hypothesized the existence of a distant
  cloud of cometary objects that orbit the Sun based on a spike in the
  reciprocal orbital separation at $1/a~\aplt~10^{-4}$\,au$^{-1}$. The
  Oort cloud is the source of long-period comets, but has not been
  observed directly, and its origin remains theoretical. Theories on
  its origin evoke a sequence of events that have been tested
  individually but never as a consistent chronology.

  We present a chronology of the formation and early evolution of the
  Oort cloud, and test the sequence of events by simulating the
  formation process in subsequent amalgamated steps. These simulations
  start with the Solar System being born with planets and asteroids in
  a stellar cluster orbiting the Galactic center. Upon ejection from
  its birth environment, we continue to follow the evolution of the
  Solar System while it navigates the Galaxy as an isolated planetary
  system.

  We conclude that the range in semi-major axis between $\sim 100$\,au
  and several $\sim 10^3$\,au still bears the signatures of the Sun
  being born in a $\apgt 1000$\,\Msun/pc$^3$ star cluster, and that
  most of the outer Oort cloud formed after the Solar System was
  ejected. The ejection of the Solar System, we argue, happened
  between $\sim 20$\,Myr and $50$\,Myr after its birth.

  Trailing and leading trails of asteroids and comets along the Sun's
  orbit in the Galactic potential are the by-product of the formation
  of the Oort cloud. These arms are composed of material that became
  unbound from the Solar System when the Oort cloud formed.
  
  Today, the bulk of the material in the Oort cloud ($\sim 70$\%)
  originates from the region in the circumstellar disk that was
  located between $\sim 15$\,au and $\sim 35$\,au, near the current
  location of the ice giants and the Centaur family of
  asteroids. According to our simulations, this population is
  eradicated if the ice-giant planets are born in orbital
  resonance. Planet migration or chaotic orbital reorganization
  occurring while the Solar System is still a cluster member is,
  according to our model, inconsistent with the presence of the Oort
  cloud. About half the inner Oort cloud, between $100$ and
  $10^4$\,au, and a quarter of the material in the outer Oort cloud,
  $\apgt 10^4$\,au, could be non-native to the Solar System but was
  captured from free-floating debris in the cluster or from the
  circumstellar disk of other stars in the birth
  cluster. Characterizing this population will help us to reconstruct
  the history of the Solar System.  }

\maketitle

\section{Introduction}

The formation of the Oort cloud \citep{1950BAN....11...91O} requires a
sequence of events on temporal and spatial scales that span more than
eight orders of magnitude, from the Solar System (on scales of years
and au) to the entire Galaxy (on scales of Gyr and kpc). As a
consequence, the formation of the Oort cloud was not a simple
happening; Occam's razor does not seem to apply here.

Individual events that led to the formation of the Oort cloud have in
the past been modeled separately to explain specific features
\citep[see e.g.,\,][]{1985prpl.conf.1100H,2004ASPC..323..371D,
2007AJ....134.1693H, 2008AJ....135.1161L,2008MNRAS.391.1350L,
2010A&A...509A..48P,2014M&PS...49....8R,2015SSRv..197..191D,
2018A&A...620A..45F}. However, the
chain of events  has never been tested as a causal sequence.
We present the results of computer simulations designed to
model the chronology of the formation of the Oort cloud in which
individual processes are included at their proper scales. Each phase
is simulated precisely and is connected to the subsequent phase of the
cascade. By doing so, we construct a consistent picture of the
formation and early evolution of the Oort cloud.

In our analysis, we assume that the Solar System formed, like most
stars \citep{2003ARA&A..41...57L,2010ARA&A..48...47A}, in a giant
molecular cloud in which gas contracts under its own gravity, and
stars form with disks around them
\citep{1990AJ.....99..924B,2007ARA&A..45..565M,2017MNRAS.472.4155G}. The
Solar System then probably formed in a cluster of stars which
interacted mutually before the Sun escaped the cluster
\citep{2009NewA...14..369P}. The circumstellar disk, a leftover from
the star-formation process, led to the coagulation of planets
\citep{1998Icar..131..171K,0004-637X-581-1-666,
  2006AJ....131.1837K,2008ARA&A..46..339W,2010AJ....139.1297L,
  2011ARA&A..49...67W,2020arXiv200705561E,2020arXiv200705562E,
  2020arXiv200705563S} and a large number of planetesimals
\citep{2007Natur.448.1022J,2017AREPS..45..359J,2018MNRAS.479.5136P,
  2021arXiv210208611J}\footnote{For lack of better terminology, we use
  asteroids to indicate planetesimals or comets. A glossary of terms
  is available in paper I \citep{2021A&A...647A.136P}, but see also
  \citep{2008ssbn.book...43G}.}. The presence of nearby stars in the
parent cluster will have affected the morphology of the gaseous disk
through tidal perturbations
\citep{1993MNRAS.261..190C,2005A&A...437..967P,2018MNRAS.478.2700W,2019MNRAS.483.4114C},
photo-evaporation
\citep{1998ApJ...499..758J,2004ApJ...611..360A,2007MNRAS.376.1350C},
and stellar winds \citep{2015ApJ...811..146O}.

The evolution of the circumstellar disk has profound consequences for
the newly formed planetary orbits
\citep{1998ApJ...508L.171L,2014A&A...565A.130B,
  2015A&A...577A.115V,2016ApJ...828...48V,2021MNRAS.501.1782C}, although it is currently difficult to quantify these effects. The
formation of the Oort cloud, however, seems somewhat problematic in
this picture \citep[but see][for a counter
argument]{2010Sci...329..187L,2019AJ....157..181V}.  We argue that the
majority of the Oort cloud formed after the Sun escaped the cluster
because perturbations of nearby stars would easily ionize an earlier
Oort cloud \citep[see also][]{2015AJ....150...26H}.

In the present paper, we present a study of the formation of the Oort cloud from a numerical
perspective. The chain of events that lead to the Oort cloud can now
be simulated in its entirety, although not fully self-consistently. We revisit this problem because we feel slightly overwhelmed by
the enormity of the  literature and the changing paradigm over
the years. We hope to contribute to a clearer picture of the processes
that appear to be relevant to the formation and evolution of the Oort
cloud.

We perform a numerical investigation and show that the Oort cloud was
not formed in any straightforward manner, but resulted from a
complex interplay during the
Sun's infancy between the Sun itself and the neighboring stars, free-floating debris, the Galactic tidal field, and
planetary scattering. Each of these components turn out to be
important, although their relative contributions are sometimes hard to
quantify.  The lack of a simple formation mechanism, and a conspiracy
of processes  instead, make the Oort cloud unique.  A complicated
story requires a complicated numerical approach that covers many time
and size scales. Apart from stellar evolution in the early
star-cluster phase, the physics is relatively simple. We deal with the
fundamentally chaotic nature of the underlying physics
\citep[see\,][]{1964ApJ...140..250M,1993ApJ...415..715G} by repeating
calculations with a different random seed. In this discussion, we
limit ourselves mainly to Newtonian physics \citep{Newton:1687}. We
address each component of this calculation separately but knit the
results together to from a homogeneous narrative on the formation and
evolution of the Oort cloud as part of the Solar System.

The timescale of Oort-cloud formation is probably closely connected to
the birth environment of the Solar System
\citep{2012Icar..217....1B}. If born in a star cluster, as argued by
\cite{2009ApJ...696L..13P,2010ARA&A..48...47A,2020RSOS....701271P,
  2020ApJ...897...60P}, with a characteristic size of $\sim 1$\,pc and
with $\sim 2500$ siblings \citep{2019A&A...622A..69P}, asteroids in
wide and highly elliptic orbits are vulnerable to being stripped from
the Solar System by the cluster potential or by passing stars
\citep[see also][]{2017A&A...603A.112N}.  \cite{2020ApJ...897...60P}
on the other hand, derive an even higher density cluster with a
density of up to $\sim 10^5$\,\MSun/pc$^3$.  If formed too early in
the lifetime of the Solar System, the outer parts of the Oort would be
lost due to stripping when a small change in relative velocity
$\delta v \equiv dv/v \apgt {\cal O}(10^{-4})$ is induced upon the
asteroids (see Fig.\,\ref{fig:schematic_overview}). So long as the Sun
is a cluster member, asteroids with an eccentricity of $e \apgt 0.98$
with a semi-major axis of $a \apgt 2400$\,au ($a(1-e) \apgt 50$\,au)
are easily lost.


In the following section, we discuss our current
understanding of the outer Solar System. We then describe
the numerical setup and its ingredients in
Sect.\,\ref{Sect:Methods}. In Sect.\,\ref{Sect:Discussion} we
discuss the simulation results, and describe a holistic view of the
evolution of the Sun and proto-Oort cloud in the Galaxy in
Sect.\,\ref{Sect:SSinGalaxy}. The consequences of resonant planetary
orbits are discussed in Sect.\,\ref{Sect:resonances}. These
arguments are supported by simulations presented in
Table\,\ref{Tab:Simulations_G}. To further explore the consequences of
alternative models on the formation of the Solar System, we include
calculations adopting planetary orbits in resonance. We conclude in
Sect.\,\ref{Sect:Conclusions}.

\section{Current understanding of the outer Solar System}\label{Sect:state}

We summarize our current understanding of the remote parts of the
Solar System, starting with the Kuiper-belt region, and subsequently
move on to the outer parts.

Four regimes in the outer parts of the Solar System are important for
this discussion, including: the dynamical class of trans-Neptunian
objects (TNOs), the parking zone\footnote{The parking zone is the
  region where asteroids (or dwarf planets) in orbit around the Sun
  are not affected by the giant planets, and are also hardly affected by
  the Galactic tidal field or an occasional encounter with a field
  star \citep{2015MNRAS.451.4663P,2020DDA....5120005V}. It was also called
  the {\em inert zone} by \cite{2020CeMDA.132...12S} and the {\em
    forbidden region} by \cite{2019MNRAS.490.2495C}}, the Hills cloud,
and the Oort cloud \citep[for a review
see][]{2019GSL.....6...12M}. The inner three regions could have been
populated and affected by encountering stars before the Sun escaped
the cluster. In that case, the presence and orbital distribution of
asteroids in the parking zone and Hills cloud could provide interesting
constraints on the dynamical evolution of the Sun in its birth cluster
before it was ejected \citep[see also][]{2020ApJ...901...92M}.

\subsection{The trans-Neptunian region and Kuiper belt}

The phase-space structure in the Edgeworth-Kuiper belt is complex
\citep{1943JBAA...53..181E,1951astr.conf..357K}. Close to the
perturbing influence of Neptune (the red curve in
figure\,\ref{fig:schematic_overview} marks the outer boundary of the
conveyor belt\footnote{Area in semi-major axis and eccentricity where
  an asteroid crosses the orbit of one or more of the major planets,
  causing their orbits to drift to higher eccentricity and larger
  semi-major axis while preserving pericenter distance.}; see the
dotted line in
figure\,\ref{fig:ae_simulations_scattered_and_sednitos}) we find
multiple families of Kuiper-belt objects. The most striking might be
the scattered disk
\citep{2001AJ....121.2804B,2002ARA&A..40...63L,2000ApJ...529L.103T,2020RNAAS...4..212N},
but there are other exotic orbital characteristics observed in this
region, including the warp \citep[at an inclination of
$i = 1^\circ.8^{+0^\circ.70}_{-0^\circ.4}$;][]{2017AJ....154...62V},
the mix of resonant families \citep{2003AJ....126..430C}, the broad
distribution in eccentricity and inclination of the Plutinos
\citep{1993Natur.365..819M,2001AJ....121.2804B}, the orbital topology
of the classical belt population \citep{2005AJ....129.1117E}, the
outer edge at the 1:2 mean-motion resonance with Neptune
\citep{2004Icar..170..492G}, and the extent of the scattered disk
\citep{2001AJ....122.1051G}. Apart from perhaps the population of
almost circular objects in the classical Kuiper belt
\citep{2000Natur.407..979T}, many of these characteristics can be
explained at least qualitatively with some incarnation of the Nice
model \citep[][see \S\,\ref{Sect:NiceModel} for more on the Nice
model]{2005Natur.435..466G,2005Natur.435..462M,
  2005Natur.435..459T,2008Icar..196..258L}. Some of these features in
the Kuiper belt originated $\apgt 50$\,Myr after the planets formed
\citep{2020RNAAS...4..212N}.

Here we demonstrate that it is possible to reconstruct part of the  history of the
Solar System from the kinematics and phase-space distribution
of the orbiting bodies \citep[see also][]{2020ApJ...901...92M}.
Although we consider the Edgeworth-Kuiper belt to be crucial to our
understanding of the formation and early evolution of the Solar System,
the present paper focuses on the Oort cloud and its formation.

\subsection{Sedna, other trans-Neptunian objects in the parking zone, and the Hills cloud}

To date, several asteroids have been observed beyond the Kuiper cliff
(at $\sim 48$\,au), between a few hundred and a few thousand astronomical units (au) with
a pericenter distance of $\apgt 50$\,au
\citep{2019AJ....157..139S,2019ApJS..244...19A}. These include the
dwarf planet 90377~Sedna \citep{2004ApJ...617..645B}, 2012 VP113
\citep{2014MPEC....F...40S}, and a dozen others. These objects can be divided
into two clusters, one population with their argument of pericenter of
$\sim 310^\circ$ \citep{2014Natur.507..471T}, and another at
the opposite side \citep{2019AJ....157..139S}. This small number of
known detached trans-Neptunian objects can, in part, be attributed to
observational selection effects because they are far
\citep{2019ApJS..244...19A}, red \citep{2009Icar..201..284B}, and
tend to have a relatively small albedo \citep{2006ApJ...639.1238R}.

Also from the outer regions of the Solar System, the area in phase
space where Sedna and family are found is hard to reach, meaning that
the asteroids in this region cannot have come from either direction,
the inner Solar System or the periphery.  This unreachable area in
parameter space is called the parking zone
\citep{2015MNRAS.453.3157J}, and is identified as such in
figure\,\ref{fig:schematic_overview}. Asteroids in this region keep
their orbital parameters for a very long time, possibly even as long
as the  age of the Solar System.  We note that, if the asteroids in the parking
zone cannot have come from the inner Solar System, and they do not
have an origin from the periphery, they either formed in situ or were
injected there by other means.  The several observed dwarf planets and
asteroids (such as Sedna) in this region preserve evidence for the
dynamical history of the Solar System
\citep{2008Icar..197..221K,2015MNRAS.453.3157J} and can be used to
constrain its origin.

These arguments led to the hypothesis that Sedna and family were
captured from another star in the parent cluster
\citep{2015MNRAS.453.3157J}. In this case, Sedna and family would have
been captured from the circumstellar debris disk around another star.
The population of asteroids in the parking zone may then form the
leftover evidence for such a captured population
\citep{2002MNRAS.333..835N,2004AJ....128.2564M,2011epsc.conf..633S,2021PSJ.....2...53N}.

Such a capture then must have happened more than $4$\,Gyr ago, and in
that time frame the influence of the ice-giant planets would have
caused the orbits of Sedna and family to change, in particular their
arguments of pericenter should have randomized.  But their orbits are
such that they pass the pericenter at the same side of the Sun
\citep{2014Natur.507..471T}.

Based on the alignment of the arguments of pericenter of the
Sedna family of objects, \cite{2016AJ....151...22B} and
\cite{2019PhR...805....1B} argued in favor of the existence of a
planet in the inner Oort cloud.  Such a planet in a wide orbit or
possibly even a stellar companion to the Sun could induce orbital
characteristics similar to those observed in Sedna and other detached
trans-Neptunian objects \citep{2020RNAAS...4..212N}, or even
affect the orbits of asteroids in the parking zone
\citep{2016MNRAS.457L..89M,2020AJ....160...50Z}.

The main motivation for an extra planet disappeared, as the alignment
of the arguments of pericenter turn out to be an observational
selection effect \citep{2019AJ....157..139S,2021arXiv210205601N}.
Although a binary companion is currently probably absent
\citep{2010MNRAS.407L..99M,1984Natur.311..636H,1984Natur.311..638H},
the Sun might have had one in the past \citep{2020ApJ...899L..24S}, particularly because such wide stellar companions are rather common among
other stars \citep{2013Natur.493..381K}.

Further out, we find the hypothetical Hills cloud
\citep{1981AJ.....86.1730H} between the outer edge of the parking zone
($\apgt 1\,000$\,au) and the inner edge of the Oort cloud
($\aplt 20\,000$\,au).  The origin of the objects in the Hills cloud
is unclear, but the population could be related to trans-Neptunian
objects
\citep{1981Icar...47..470F,2015SSRv..197..191D}. \cite{1981AJ.....86.1730H}
estimated the total mass to exceed that of the Oort cloud by as much as a
factor of 100 \citep[see however][who argue in favor for comparable
populations]{2000DPS....32.3602D}. The majority of trans-Neptunian
objects have prograde orbits \citep{2020tnss.book...61K} and isotropic
distributions in mean anomaly, longitude of the ascending node, and
the argument of perihelion \citep{2020PSJ.....1...28B}.  These support
an inner Solar System origin because asteroids scattered from the
circumstellar disk tend to have prograde orbits
\citep{2020ApJ...901...92M}, whereas the orbits of captured asteroids
may well be retrograde, depending on the encounter that introduced
them into the Solar System \citep[see
also][]{2018MNRAS.473.5432H,2021PSJ.....2...53N}. The dynamical
history of the Solar System in its birth cluster therefore plays an
important role in the formation and orbital topology of asteroids in
the parking zone and the Hills sphere \citep{2020RSOS....701271P}. Because of a lack of perturbing influences, signatures of a captured
population in the parking zone remain noticeable for much longer than
in other parts of the Solar System.

\subsection{The Oort cloud}

Discussions over the {\"O}pik-Oort cloud began in 1932 by {\"O}pik
\citep{1932PAAAS..67..169O} who discussed the origin of nearby
parabolic orbits in the Solar System, and in 1950 by Oort
\citep{1950BAN....11...91O} with the discovery of a spike in the
reciprocal orbital separation $1/a~\aplt~10^{-4}$\,au$^{-1}$ of
observed comets. \cite{1950BAN....11...91O} argued that the
long-period comets originated from a region between 25\,000\,au and
200\,000\,au from the Sun. Today, these estimates have not changed considerably \citep{2019MNRAS.490.2495C}.

The outer limit is considered to coincide with the Hill radius
\citep{1913AJ.....27..171H} of the Sun in the Galactic potential
\citep{1965SvA.....8..787C}. The  origin and precise location of the inner edge are less clear
\citep{1981AJ.....86.1730H,2008MNRAS.391.1350L} and are still debated
\citep{2015MNRAS.446.3788B}. What
defines the transition region between the Hills cloud and the Oort
cloud also remains unclear.

The mass of the Oort cloud is estimated to range from
$1.9$\,M$_{\oplus}$ \citep{1996ASPC..107..265W} to $38$\,M$_{\oplus}$
\citep{1983A&A...118...90W}. These estimates seem somewhat on the high
side when compared to those based on numerical simulations, which
arrive at $0.75 \pm 0.25$\,M$_{\oplus}$ \citep{2008A&A...492..251B} to
$1.0\pm0.4$M$_{\oplus}$ \citep{2000Icar..145..580F}.

With a typical comet-mass of a few times $10^{12}$ to $10^{14}$\,kg
\citep{1987ESASP.278..471R,2009MNRAS.393..192S,2011MNRAS.416..767S}
the Oort cloud contains ${\cal O}(10^{12})$ comet-sized objects.
Interestingly, this estimate is comparable to Oort's original estimate
of $\sim 10^{11}$ objects \citep{1950BAN....11...91O}, and to the
density of interstellar asteroids
\citep{2017AJ....153..133E,2018MNRAS.479L..17P,2019NatAs...3..594O,2020ApJ...903..114P}.

\subsection{The Formation and early evolution of the outer Solar System}

Estimates of the formation timescale of the Oort cloud range from
instantaneously after the formation of the Sun
\citep{1997Icar..129..106F}, synchronously with Jupiter's formation
\citep{1988Icar...75..146S,2000Icar..145..580F,2004come.book..153D}, to
after Jupiter formed and potentially migrated
\citep{2019MNRAS.485.5511S}, to slow growth over several 100\,Myr
\citep{2008Icar..197..221K,2017A&A...603A.112N}, and even to gigayear timescales
\citep{1987AJ.....94.1330D}.

Two main scenarios are popular for the formation of the Oort cloud:
One stipulates that the cloud formed mainly by ejecting inner Solar System
material through planet--disk interactions
\citep{2000DPS....32.3602D, 2004ASPC..323..371D}, and the other concerns formation  through two distinct processes: Local
disk asteroids are ejected into an inner region at $3000$–-$20\,000$
au, while the outer region at $\apgt 20\,000$ au is mainly accreted
from free-floating debris in the parent star cluster
\citep{1990CeMDA..49..265Z,1992CeMDA..54...37V,
2006Icar..184...59B,BRASSER2007413,2008A&A...492..251B,2010Sci...329..187L}. Planets could also be captured in this way \citep{2012ApJ...750...83P},
possibly explaining the origin of a hypothetical exoplanet in the
outer parts of the Solar System \citep{2016MNRAS.460L.109M}.

The relatively low mass in some of the numerically derived estimates
stem from the low efficiency, which ranges from $\sim 1.1$ \%
\citep{2008MNRAS.391.1350L,2009A&A...507.1041M}, $\sim 2$\,\%
\citep{2019MNRAS.490.2495C} to $\sim 4$\,\%
\citep{2010A&A...516A..72B}, at which point disk material is launched
into a bound Oort cloud \citep{2010A&A...509A..48P}; the majority of
objects are expected to escape the Solar System or hit the Sun. The
giant planets could therefore be insufficiently efficient to explain
the currently anticipated mass of the Oort cloud
\citep{2006Icar..184...59B,2008Icar..197..221K,2012Icar..217....1B}.

If the entire Oort cloud originates from the depletion of asteroids
between Uranus and Neptune \citep[see also][]{2013Icar..222...20F},
this region must have been populated by objects with a total mass of $100$ to $3800$\,M$_{\rm
Earth}$ \citep{2014Icar..231..110F,2014Icar..231...99F}. In a disk
with a density profile of $\rho \propto r^{-1.5}$ between 0.1\,au and
45\,au, about half the mass is between the orbits of the ice giants. To
supply the Oort cloud with sufficient material, the proto-planetary
disk must then have had a mass of $200$ to $7600$\,M$_{\oplus}$ (or
equivalently $\sim 0.001$ to $0.02$\,\MSun) in comet-sized asteroids,
which is consistent with other estimates \citep{2009ApJ...698..606C}.

Given the low efficiency of the ice-giants to produce the Oort cloud,
more than 90\% is likely to have come from elsewhere; such as the
circumstellar disks of other stars \citep{2010Sci...329..187L},
free-floating debris in the parent cluster
\citep{1990CeMDA..49..265Z,2000A&A...361..369N}, the capture of
interstellar objects
\citep{1982ApJ...255..307V,2020MNRAS.493L..59H,2021arXiv210406845P},
or accretion from the circumstellar disks of other stars
\citep{2000Icar..145..580F,2010Sci...329..187L}.

The orbits of captured asteroids could be rather distinct from those
produced by scattered asteroids from the proto-planetary disk
\citep{2019MNRAS.490...21H,2020MNRAS.492..268H}. In the Oort cloud,
both populations will be affected by the Galactic tidal field
\citep{1987AJ.....94.1330D,2006CeMDA..95..299F} and probably mix on a
timescale of a few hundred million years. Orbital inclinations isotropize on
this timescale, and eccentricities thermalize
\citep{2006CeMDA..96..341F,2015AJ....150...26H}.

Once fully developed, the Oort cloud erodes by injecting comets into
the inner Solar System through Kaib-Quinn jumpers
\citep{2008Icar..197..221K} or by a more gradual process
\citep[sometimes referred to as {\em
  creepers};][]{2014Icar..231..110F,2018A&A...620A..45F}, but also because of encounters with molecular clouds \citep{2011DDA....42.0903D} and
tidal interaction with the Galaxy
\citep{1986Icar...65...13H,2001AJ....121.2253L,2011MNRAS.411..947G,2019A&A...629A.139T}.
These external perturbations may be the main reason why comets are launched
from the Oort cloud into the inner Solar System. This process was
suggested to lead to periodic showers of comets
\citep{1984Natur.308..715D,2011MNRAS.411..947G} causing mass
extinctions every $\sim 29.7$ Myr
\citep{1984PNAS...81..801R,2020AsBio..20.1097R}.  Such a wide
companion was never found \citep{2014ApJ...781....4L}, and re-analysis
of the data shows that there is no statistically significant evidence
for a periodicity in these mass extinctions
\citep{1987Natur.330..248P,2000A&A...353..409J,2009IJAsB...8..213B}.


\subsection{The Nice model}\label{Sect:NiceModel}

Many of the model simulations above depend on some sort of chaotic
reorganization in the early Solar System. This could have happened
during or shortly after the formation of Jupiter
\citep{2006ChJAA...6..588L}
up to about half a gigayear after the planets formed 
\citep{2009EM&P..105..257N,2009EM&P..105..263L}. This chaotic
reorganization was introduced to explain the enhanced flux of
asteroids throughout the young Solar System
\citep{2005Natur.435..462M,2005Natur.435..459T,2005Natur.435..466G}, which could have lead to a peak in the flux of lunar impactors
\citep[late-heavy bombardment is a commonly used
term;][]{1965Icar....4..157H,1966Icar....5..406H,1984Natur.308..718A,
2001SSRv...96....9S,2001SSRv...96...55N,2002JGRE..107.5022R}.

Follow-up analyses indicated that the impact period could be between
4.2 and 3.5\,Gyr ago
\citep{2014P&SS...98..254F,2017OLEB...47..261Z,2017AREPS..45..619B},
rather than being a peak, rendering the need for a disturbance in the force in
the early Solar System unnecessary. Re-analysis of the lunar impact
statistics by \cite{2010LPICo1538.5226L} indicated that a high flux
of impactors is not needed to explain the lunar cratering record. This
study was further supported by the re-analysis of the 267 Apollo
samples on $^{40}$Ar/$^{39}$Ar isotope ratios, where it was demonstrated that the
material on the moon arrived in a continuously decreasing flux rather
than a peak \citep{Boehnke10802,2018LPICo2107.2033B}.

The Nice model is successful in explaining several aspects of the
Solar System, including the orbital topology of the giant planets
\citep{2007AJ....134.1790M}, the obliquity of Uranus
\citep{2019E&PSL.506..407W}, Trojan asteroids
\citep{2015aste.book..203E}, irregular moons
\citep{2010AJ....139..994B}, some morphologies of the Kuiper belt
\citep{2020tnss.book...25M}, and several other curious orbital
choreographies \citep{2008Icar..196..258L,2019MNRAS.485.5511S}.
Although, the Nice model seems to fail in explaining the global
structure of the Oort cloud (Fouchard et al.\, 2018, but see also
Shannon et al.\, 2019, for competing arguments based on the rocky
comet
C/2014~S3~(PANSTARRS))\nocite{2018A&A...620A..45F,2019MNRAS.485.5511S},
it might be hard to find a single alternative model that can explain
so many phenomena in the Solar System.  One aspect of the Nice model
that may be important for explaining the phenomena of the Solar System
is a period in which the orbits tend to change chaotically
\citep{2008ApJ...676..728T}.  Such a chaotic phase can be initiated by
planetary resonance ---as in the Nice model--- induced by a passing
star or by the last phase of evaporation of the circumstellar disk.

Several independently developed explanations exist, including
models based on internal dynamical processes
\citep{1995AJ....110.3073D,2000ApJ...528..351I}, encounters with other
stars
\citep{2000ApJ...528..351I,2005Icar..177..246K,2019A&A...629A.139T}, or
a small molecular cloudlet in a wide orbit around the Sun
\citep{2020A&A...642L..20E}. However, there is no global model that
explains the Solar System in unison.

This early phase in the evolution of the  Solar System appears far from
being understood from first principles, and more research on this
topic would help to understand the apparent stability and inertness of the current Solar System.

\section{Methods, models, and simulations}\label{Sect:Methods}

This section explains the numerical methods, the initial conditions,
and simulations that we used to qualify and quantify the formation and early
evolution of the Oort cloud. Our calculations are not self-consistent
because the results were not produced in a single simulation, but the
results are melded together to form a consistent understanding.

\begin{table*}
\begin{tabular}{p{2.0cm}p{14.0cm}}
\multicolumn{2}{l}{\bf Scattered and captured asteroids (see Sect.\,\ref{Sect:Isolated_disk})}\\
Designation & Simulation A. \\
Simulations & Two simulations of the Solar System with planets, and scattered and captured asteroids in the Milky way potential.
\\ \hline
Star & Single 1\,\MSun\, star
in the smooth potential of the Milky Way galaxy.
\\ 
Planets & Four planets per star in a circular disk, 
with Jupiter, Saturn, Uranus, and Neptune in
circular orbits with semi-major axes of $5.1$, $9.5$, $19.2$,
and $30.1$\,au, respectively. 
\\ 
Asteroids & One simulation with 2000 asteroids (from planetary system \# 157 of
sect.\,\ref{Sect:Scattered_asteroids} \cite[see also][]{Torres2020}.
One simulation with 1888 asteroids
(from the most probable encounter, which, according to 
\cite{2015MNRAS.453.3157J}, resulted in a population of Sedna-like objects.
\\ 
Numerics & {\tt Huayno} \citep{2012NewA...17..711P} coupled with the Galactic
model (see Table\,\ref{Tab:galaxy_paraeters}) via
bridge \citep{ZWART2020105240} in \AMUSE\, \citep{2018araa.book.....P}.
\\ 
Computer & run on Cartesius, and ALICE using GPU.
\\ 
Duration & The simulations started at an age of $\sim 100$\,Myr. By that time, the Solar System
has left the parent cluster, and continues for $1$\,Gyr.\\
\hline
\end{tabular}
\caption{Simulations of the Solar System with captured and scattered asteroids in the Galactic potential.}
\label{Tab:Simulations_D}
\end{table*} 

\begin{table*}
\begin{tabular}{p{2.0cm}p{14.0cm}}
\multicolumn{2}{l}{\bf Star cluster simulation (see Sect.\,\ref{Sect:primordial_star_cluster})}\\
Designation & Simulation B. \\
Simulations & Two-hundred simulations with 200 asteroids per star, and 24
simulations with 2000 asteroids per star in a cluster where stars interact with the asteroids.
\\ \hline
Stars & Two-thousand stars with a \cite{2001MNRAS.322..231K} mass function
between 0.08 and 100\,\MSun in R=1pc virialized Plummer
sphere \citep{1911MNRAS..71..460P}. No primordial binaries, no residual gas.
\\ 
Planets & Four planets per star with circular orbits.
Each star has the same planetary systems in a plane, but each planetary system has
a different random orientation in space.
In one model (extended), the planets Jupiter, Saturn, Uranus, and Neptune have initial semi-major axes $5.1$, $9.5$, $19.2$,
and $30.1$\,au, respectively. In the other model (compact), Jupiter, Saturn,
Neptune, and Uranus have initial semi-major axes of
$5.5$, $8.1$, $11.5$, and $14.2$\,au. 
\\ 
Asteroids & In a disk between 40\,au and 1000\,au in the extended
model and between 16\,au and 400\,au in the compact
model.
\\ 
Numerics & {\tt NBODY6++GPU} \citep{2015MNRAS.450.4070W} and {\tt REBOUND}
using IAS15 integrator \citep{2015MNRAS.446.1424R} coupled via the LonelyPlanets scheme
\citep{2019MNRAS.489.4311C} in \AMUSE\, \citep{2018araa.book.....P}.
\\ 
Computer & Run on a GPU-equipped Little-Green-Machine-II (LGM-II) computer.
\\ 
Duration & The simulations start with all stars on the zero-age main sequence
with four planets in the adopted initial orbits.
The calculation is continued for 100\,Myr.
\\ \hline
\end{tabular}
\caption{Simulations of the Solar System in its birth star
cluster using the {\em LonelyPlanets} approach (see
sect.\,\ref{Sect:LonelyPlanets}). The resulting orbital distribution
of asteroids from the Sun-representative star is used in this work
to further explore the evolution of the distribution of asteroids in
the Oort cloud. These simulations do not lead to a reliable
representation of abducted asteroids because the lonely planet
approach does not allow exchange interactions.
}
\label{Tab:Simulations_C}
\end{table*} 

\begin{figure*}
\center
\includegraphics[width=1.0\textwidth]{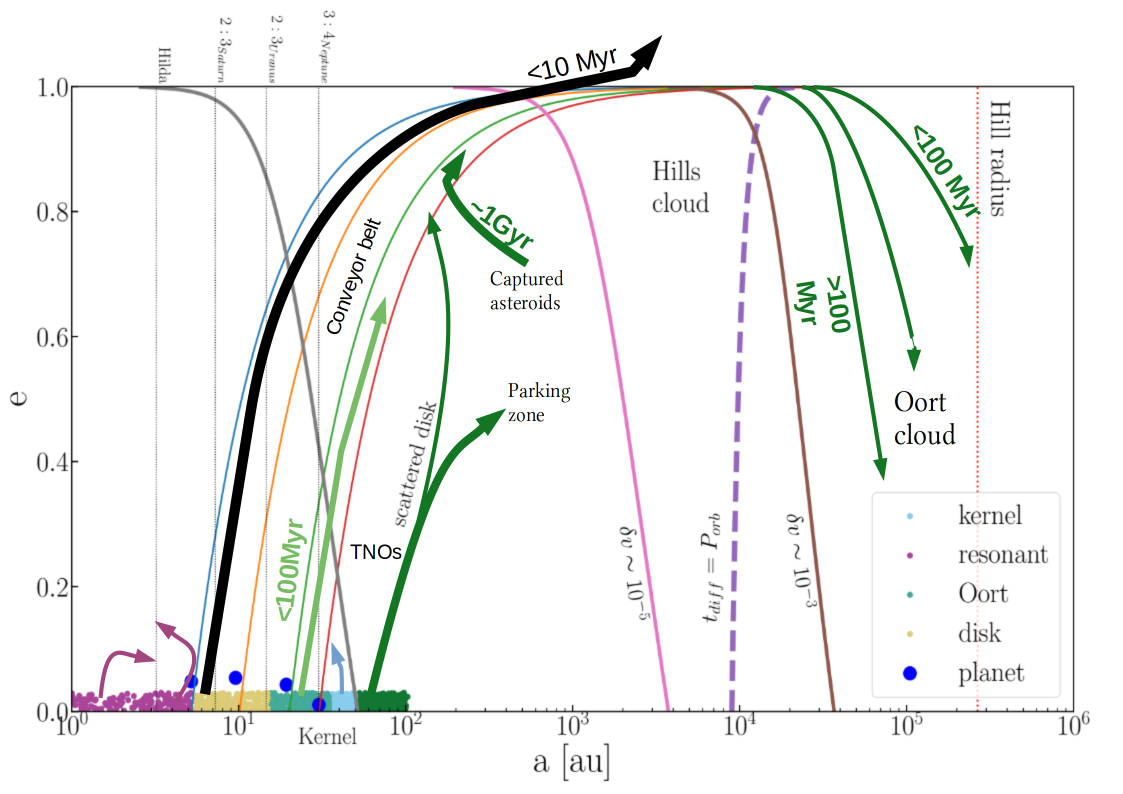}
\caption{Orbital migration of asteroids and how they end-up in the
  Oort cloud. Several of the thin curves are also plotted in
  figure\,\ref{fig:OCformation}. The four major planets are indicated
  with their current semi-major axis and eccentricity as blue dots.
  The initial circumstellar disk is presented in five colors,
  depending on the final destination of the asteroids populating the
  different disk sections. The arrows indicate the movement of
  asteroids originating from the disk or captured from other
  stars. The timescales presented near the arrows give an estimate of
  the timescale of migration. The colors indicated in the legend give
  the original inner disk (magenta), which mainly migrates away from
  resonant orbits \citep{2017Sci...357.1026D} (four important
  resonances are indicated with thin vertical dotted lines). Ogre
  indicates those asteroids that are ejected from the Solar System on
  a relatively short timescale ($\aplt 10$\,Myr). Light blue indicates
  the kernel and the resonant Kuiper belt. The light green and dark
  green curves show the migration patterns of the  asteroids that eventually
  reach the Oort cloud. These objects migrate from a semi-major axis
  of a few tens of au to beyond $10^4$\,au through a narrow neck of high
  eccentricity. The asteroids in the dark green area require external
  support from a stellar encounter to be able to migrate to the Oort
  cloud. This migration is more visible in
figure\,\ref{fig:OCformation}, where the eccentricity (y-axis) is in
  logarithmic units. We note that the closer an asteroid is to the Hill
  radius (vertical red dotted line to the right), the quicker its
  orbit will be circularized by the Galactic tidal field (see also
figure\,\ref{fig:ae_randomwalk}, where this is illustrated). The gray,
  pink, and brown curves indicate where the relative velocity kick
  imparted at apocenter by the Galactic tidal field to an asteroid is
  $\delta v = 10^{-8}$, $10^{-5}$, and $10^{-3}$ of the orbital
  velocity, respectively. The purple dashed curve indicates the orbital
  separation and eccentricity for which the tidal eccentricity damping
  timescale is equal to the orbital period
  \citep[see][]{1987AJ.....94.1330D}.
\label{fig:schematic_overview}
}
\end{figure*}

\begin{figure*}
\center
\includegraphics[width=1.0\textwidth]{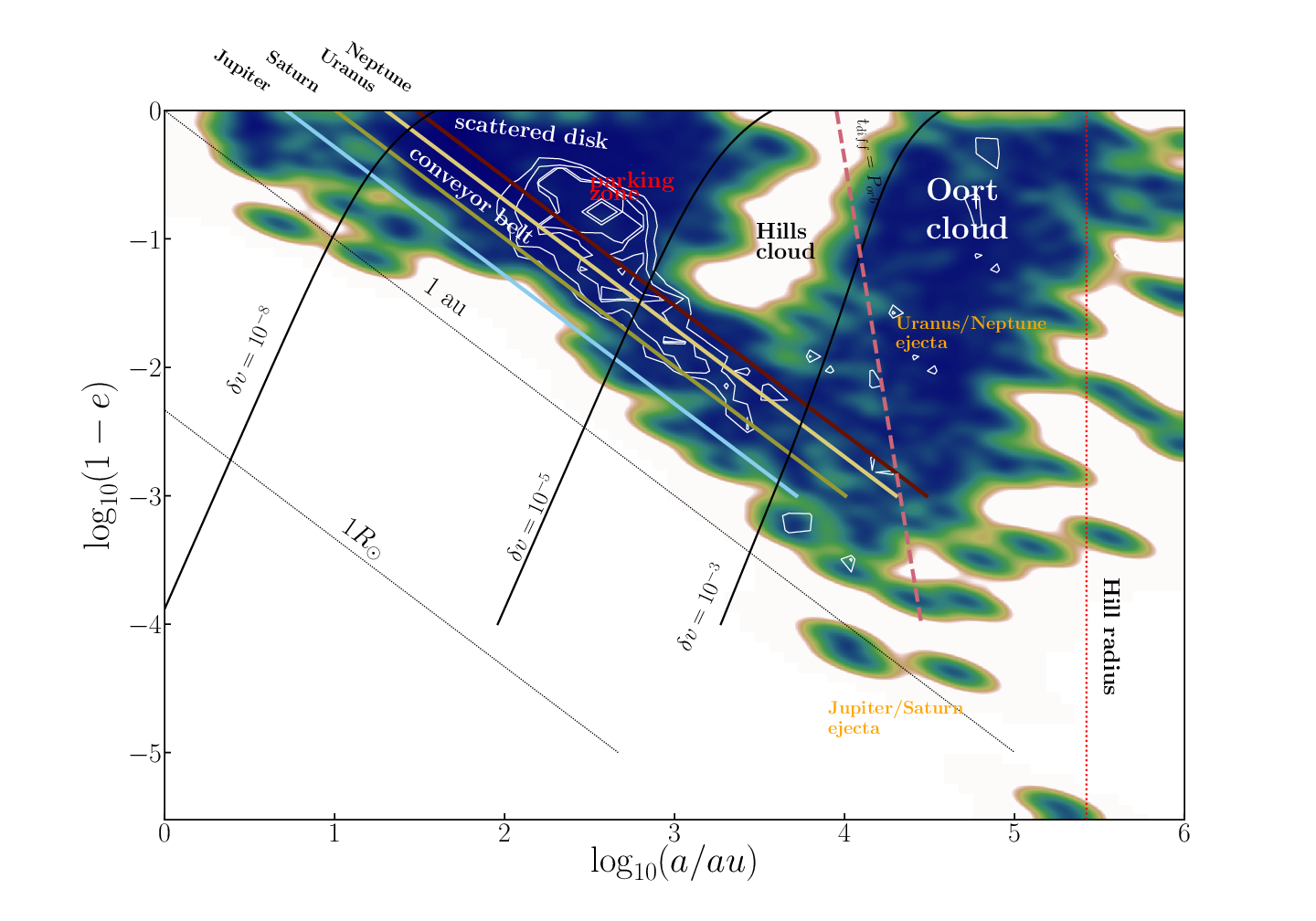}
\caption{Phase-space distribution of asteroids around the Sun
$\sim 100$\,Myr after escape from its parent cluster.
The shaded region presents a kernel-density estimation of the
simulation results, representing various families of objects. We
adopted a non-parametric Gaussian kernel density estimator with a
symmetric bandwidth of $0.02$ \citep{GaussianKDE1992S}. At this
moment, part of the Oort cloud is already in place (to the right), but
the formation process is still ongoing.
The colored diagonal curves (from top left to bottom right) indicate
orbits that cross those of the giant planets 
(see also figure\,\ref{fig:schematic_overview}). The dashed burgundy colored curve
to the right indicates where the orbital period, $P_{\rm orb}$, is equal to
the eccentricity damping-diffusion timescale by the Galactic tidal
field \citep[$t_{\rm diff}$, Eq.5 of][]{1987AJ.....94.1330D}.
The solid black curves indicate the Galaxy's perturbing influence in
terms of the velocity-kick imparted to an object. Relative velocity
perturbations of $\delta v = 10^{-3}$ (right), $10^{-5}$, and $10^{-8}$
(left) are indicated. The little area to the right of Neptune's
influence (between the red diagonal curve and the black curve
indicating a perturbation $\delta v = 10^{-8}$) corresponds to the
Kuiper-belt kernel distribution. The parking zone is indicated in
red. The Oort cloud is to the right of the rightmost solid black curve
(labeled as $\delta v = 10^{-3}$) and the burgundy colored dashed
curve.  The Hills cloud is to the left of this dividing line
(indicated in black). In our simulations, the Hills cloud at low
eccentricity ($e \aplt 0.95$) is mostly empty, but at higher
eccentricity (near the bottom of the figure) its population is
substantial, in particular along the conveyor belt.
The captured and scattered asteroids are indicated in white contours.
Locally, at the extreme, both populations have comparable phase-space
density.
At the age of 100\,Myr, a considerable fraction of the native disk
population has already reached the Oort cloud, or is on its way there
through the conveyor belt. Some captured asteroids are currently
migrating along the conveyor belt and a few have already reached
the Oort cloud. However, the majority of the scattered and captured asteroids are in the parking zone between $\sim 100$\,au and $\sim
1000$\,au, where they will stay for the duration of the simulation.
Jupiter and Saturn eject asteroids along the conveyor belt into
escaping orbits (indicated in orange). Uranus and Neptune eject
asteroids on a timescale considerably longer ($\apgt 100$\,Myr) than that for
Jupiter and Saturn ($\aplt 10$\,Myr), allowing these asteroids to be
circularized by the Galactic tidal field (also in orange). This is
also visible in the lower kernel density along the Jupiter--Saturn
conveyor belt in comparison with the Uranus--Neptune conveyor belt.
Eventually, the latter asteroids become members of the Oort cloud.
The red dotted curve indicates the Hill radius of the Sun in orbit
around the Galactic center, here at about 0.65\,pc.
The thin dotted diagonal curves indicate pericenter distances of 1\,au
and 1\,R$_\odot$. Comets from the Oort cloud may enter the inner Solar System (to the far bottom right and below the 1\,au curve).
\label{fig:OCformation}
}
\end{figure*}

We start with a discussion on the software framework, the Astrophysics
Multipurpose Software Environment (\AMUSE; see
Sect. \,\ref{Sect:AMUSE}), which was used for the simulations
presented here. We discuss each of the following processes:
\begin{itemize}
\item[{\bf A}] The evolution of the circumstellar disk of an isolated
Solar System (see Sect.\,\ref{Sect:Isolated_disk} and
Table\,\ref{Tab:Simulations_D}).

The calculation includes the four giant planets orbiting the Sun
together with a disk of mass-less particles. The entire setting is
integrated in the potential of the Galaxy, starting with the
approximate position in the Galaxy where the Sun was $1$\,Gyr ago, and
lasts for 1 Gyr.

\item[{\bf B}] The evolution of the Solar System in its birth cluster
(Sect.\,\ref{Sect:primordial_star_cluster} and
Table\,\ref{Tab:Simulations_C}).

The arguments here are based on two quite distinct calculations: first
a simulation of 2000 stars in a virialized cluster with a half-mass
radius of $\sim 1$\,pc. We study the evolution of debris disks under
the influence of the encounters with neighboring stars. In a second
calculation, we study the exchange efficiency and orbital
characteristics of the circumstellar disks that were perturbed in
encounters between two stars. These two simulations are used to
understand the orbital parameters of various populations of asteroids:
\begin{enumerate}
\item The captured asteroids and their migration towards the Oort
cloud (see Sect.\,\ref{Sect:Scattered_asteroids}),
\item and the scattered-disk asteroids along the conveyor belt (see
Sect.\,\ref{Sect:Captured_asteroids}).
\end{enumerate}

\item[{\bf C}] Transport of asteroids along the conveyor belt and
  eccentricity damping by the tidal field once the Oort cloud is
  reached (see Sect.\,\ref{Sect:Eventual_evaporation} and
  Table\,\ref{Tab:Simulations_E}).

The formation of the Oort cloud is supported by simulations of the
population of asteroids in the conveyor-belt region. These
calculations include the four giant planets (with various orbital
configurations) and the Galaxy's tidal field. Both are necessary to
study the transport of asteroids from the planetary region to the Oort
cloud. The eccentricity and semi-major axis of the asteroids increases
when evolving along the conveyor belt. This process is driven by the
major planets. Once the pericenter of the orbits of the asteroid detaches
from the major planets, the Galactic tidal field dampens the
eccentricity of the orbits.

\end{itemize}

figure\,\ref{fig:schematic_overview} presents a schematic overview
of the evolution of the Oort cloud. Asteroids near Jupiter and
Saturn are vulnerable to being ejected on a relatively short timescale
($\aplt 10$\,Myr), whereas asteroids originally formed between Uranus
and Neptune, and the scattered and captured populations (green
arrows), tend to reach the Oort cloud on a much longer timescale of
$\sim 100$\,Myr. All asteroids that eventually reach the Oort cloud
pass through the narrow neck at an eccentricity of $\apgt 0.998$ at a
semi-major axis of around $10^4$\,au, after which they are subjected to
eccentricity damping by interacting with the Galactic tidal field
(green arrows to the right in figure\,\ref{fig:schematic_overview}). figure\,\ref{fig:OCformation} presents a superposition of
simulation data $\sim 100$\,Myr after the Solar System escaped its
birth star cluster. This data originates from multiple calculations
that simulate various aspects of the formation of the Oort cloud.

\subsection{The Astrophysics Multipurpose Software Environment}
\label{Sect:AMUSE}

The calculations in this study are performed using the Astrophysics
Multipurpose Software Environment
\citep{2018araa.book.....P}. \AMUSE\, is a modular
language-independent framework for homogeneously interconnecting a
wide variety of astrophysical simulation codes. It is built on public
community codes that solve gravitational dynamics, hydrodynamics,
stellar evolution, and radiative transport using scripts that do not
require recompilation. The framework adopts {\em Noah's Arc}
philosophy, meaning that it has incorporated at least two codes that
solve the same physics \citep{2009NewA...14..369P}.

Most calculations are carried out using a combination of symplectic,
direct, N-body test-particle integration and the tidal field of the
Galaxy. For the former two, we use {\tt Huayno}, which adopts the
recursive Hamiltonian splitting strategy \citep[much like bridge;
  see][]{2007PASJ...59.1095F,ZWART2020105240} to generate a symplectic
integrator that conserves momentum to machine precision \citep[down to
  $10^{-14}$ in normalized units;][]{2012NewA...17..711P}. Additional
calculations were performed using the ABIE code \citep{ABIE2018}, {\tt
  NBODY6++GPU} \citep{2015MNRAS.450.4070W}, and {\tt REBOUND}
\citep{2012A&A...537A.128R} using the IAS15 scheme
\citep{2015MNRAS.446.1424R}. These methods are combined using the
\AMUSE\, framework in the LonelyPlanets approach \citep[][see also
  Sect.\,\ref{Sect:LonelyPlanets}]{2019MNRAS.489.4311C}.

Where appropriate, we include stellar mass loss in our calculations
using the {\tt SeBa} binary population synthesis package
\citep{1996A&A...309..179P,2016ComAC...3....6T}. For the Galaxy, we
take a slowly varying potential of the bar, bulge, spiral arms, disk,
and halo into account. The parameters for the Galaxy model are listed
in Table\,\ref{Tab:galaxy_paraeters} \citep[see
also][]{2015MNRAS.446..823M,2016MNRAS.457.1062M,2017MNRAS.464.2290M}.

\begin{table*}
\begin{tabular}{p{2.0cm}p{14.cm}}
  \multicolumn{2}{l}{\bf Simulation of the migration of asteroids from the conveyor belt to the Oort cloud (see Sect.\,\ref{Sect:Eventual_evaporation}).}\\
  Designation & Simulation C. \\
  simulations & Nine simulations with various initial realizations, but geared toward populating the Oort cloud from the conveyor belt.\\
  \hline
  Planets & Four planets to each star in a circular disk, 
            with Jupiter, Saturn, Uranus, and Neptune in
            orbits with semi-major axes of $5.1$, $9.5$, $19.2$, and
            $30.1$\,au, respectively. 
  \\ 
  Asteroids & $10^4$ asteroids per run distributed in the ecliptic plane
              in the conveyor belt on eccentric orbits of between 0.1 and 0.9 and with a
              pericenter distance of between 5.1\,au and 30.1\,au
              \citep[see also][]{1987AJ.....94.1330D}. Four simulations with
              the ecliptic plane in the Galactic plane and five simulations with an
              inclination of $60^\circ$ to the Galactic plane.
  \\ 
  Numerics & {\tt Huayno} \citep{2012NewA...17..711P} coupled with the Galactic
             model (see Table\,\ref{Tab:galaxy_paraeters}) via
             bridge \citep{ZWART2020105240} in \AMUSE\, \citep{2018araa.book.....P}. Seven
             simulations were conducted with standard precision of
             the {\tt Huayno} integrator, and two simulations at twice the
             precision (by reducing the time-step parameter $\eta$ from 0.02 to 0.01).
  \\ 
  Computer & Run on ALICE using a GPU.
  \\ 
  Duration & Start at an age of $\sim 100$\,Myr (once the Solar System
has             left its birth cluster, and continued for 1\,Gyr ($\sim 1$\,Gyr
             after the Sun leaves the cluster; see sect.\,\ref{Sect:SunEscapersCluster}).
  \\ \hline
\end{tabular}
\caption{Simulation of the Solar System with asteroids in the conveyor
  belt. These simulations are performed with a populated conveyor
  belt. The initial conditions for the asteroids are therefore not the
  result of earlier calculations. These conditions were selected to
  mimic the phase in which asteroids are launched onto the conveyor
  belt. As this process may last for up to some 100\,Myr, we
  shorten this by generating initial conditions when the
  asteroids are already in the conveyor-belt region. These simulations
  are mainly used to study the   eccentricity-damping process of the Galactic tidal field and the phase-space distribution of
  asteroids in the Oort cloud.}
\label{Tab:Simulations_E}
\end{table*}

\begin{table}
\caption{\bf Model parameters of the Milky Way.}
\label{Tab:galaxy_paraeters}
\begin{tabular}{lll} \hline
\hline
\multicolumn{3}{c}{ \rule{0pt}{3ex}\textit{Axisymmetric component}} \\ 
\rule{0pt}{4ex}Mass of the bulge ($M_\mathrm{b}$) & $1.4\times 10^{10}$ M$_{\odot}$ &\\ 
Scale length bulge ($b_\mathrm{1}$) & $0.3873$ kpc &\\
Disk mass ($M_\mathrm{d}$) & $8.56\times10^{10}$ M$_{\odot}$ &\\
Scale length 1 disk ($a_\mathrm{2}$) & $5.31$ kpc & 1) \\
Scale length 2 disk ($b_\mathrm{2}$) & $0.25$ kpc &\\
Halo mass ($M_\mathrm{h}$) & $1.07\times 10^{11} $ M$_{\odot}$ &\\
Scale length halo ($a_\mathrm{3}$) & 12 kpc &\\
\hline
\multicolumn{3}{c}{ \rule{0pt}{3ex}\textit{Central Bar}} \\
\rule{0pt}{4ex}Pattern speed ($\Omega_\mathrm{bar}$) & $40$ \patspeed & 2)\\ 
Mass ($M_\mathrm{bar}$) & $9.8\times10^9$ M$_{\odot}$& 4) \\ 
Semi-major axis ($a$) & $3.1$ kpc & 5)\\
Axis ratio ($b/a$) & $0.37$& 5) \\
Vertical axis ($c$) & 1 kpc & 6)\\
Present-day orientation & $20^\circ$ & 3)\\ 
\hline
\multicolumn{2}{c}{ \rule{0pt}{3ex}\textit{ Spiral arms}} \\ 
\rule{0pt}{4ex}Pattern speed ($\Omega_\mathrm{sp}$) & $20$ \patspeed & 2)\\
Number of spiral arms ($m$) & $2$ & 7)\\
Amplitude ($A_\mathrm{sp}$) & $3.9\times10^7$ M$_\odot$~kpc$^{-3}$ & 4) \\
Pitch angle ($i$) & $ 15.5^\circ$ & 4)\\
Scale length ($R_\mathrm{{\Sigma}}$) & $2.6$ kpc & 7)\\
Scale height ($H$) & $0.3$ kpc & 7)\\
Present-day orientation & $20 ^\circ$ & 5) \\
\hline
\end{tabular}\\
\textbf{References:} 1)
\cite{1973asqu.book.....A}; 2) \cite{2011MSAIS..18..185G}; \\
3) \cite{2011MNRAS.418.1176R}; 4) \cite{2012A&A...541A..64J}; \\
5) \cite{2015MNRAS.446..823M}; 6) \cite{2014A&A...569A..69M}; \\
7) \cite{2000A&A...358L..13D}; 8) \cite{2008ApJ...673..864J}
\end{table} 

The galactic tidal-field code is coupled to the various N-body codes
using the augmented non-linear response propagation pattern
\citep[called {\em rotating bridge} in][]{2017MNRAS.464.2290M}. The
interaction time-step between the Galactic tidal field and the Solar System was $100$\,yr. This time step is sufficient because the period of
the asteroids in such wide orbits exceeds 0.1\,Myr.

\begin{figure*}
\includegraphics[width=\textwidth]{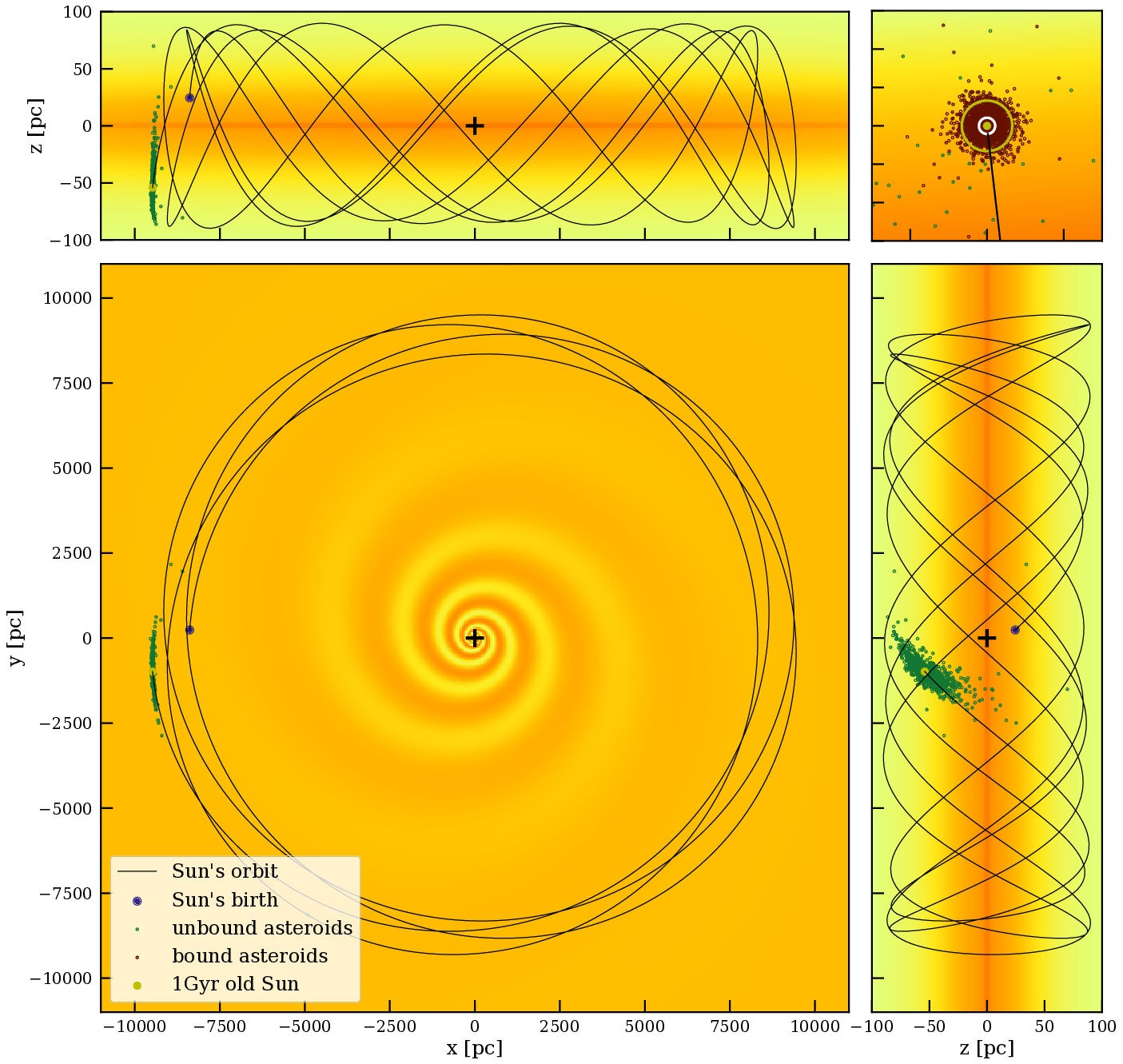}
\caption{Galactic distribution of asteroids 1\,Gyr after the Sun
escapes its birth cluster (still $\sim 3.6$\,Gyr in the past). The
colored background represents the adopted Galaxy potential: bottom
left, top left, and bottom right views show the various Cartesian
coordinates (the bar is not shown). The red and green dots show the
bound and unbound asteroids. The starting position of the Sun is also
indicated. The top-right corner panel shows a magnified view of 6 pc by
6 pc around the Sun. The outer circle shows the Hill radius at
$\sim 0.65$\,pc, and the inner circle represents the inner edge of the
Oort cloud at about $\sim 0.21$\,pc (at $\delta v = 10^{-3}$).}
\label{fig:Galactic_orbit}
\end{figure*}

\subsection{Evolution of the circumstellar disk of an isolated Solar System}
\label{Sect:Isolated_disk}

The simulations presented in this section are designed to help us to understand the
consequences of evolution of the newly born Solar System with a disk
of planets and asteroids as part of the Galactic tidal field. The
Solar System is initialized in the Galactic potential at $-8.4$\,kpc
along the x-axis and $17$\,pc above the Galactic plane. The initial
velocity of the Solar System was $v_\odot = (11.352, 233.105,
7.41)$\,km/s, which follows an almost circular orbit around the
Galactic center (see figure\,\ref{fig:Galactic_orbit}). From this
position and velocity we calculate the position of the  Sun in the Galaxy
backwards for $1$\,Gyr, which is where our calculations start. The
location of the Solar System 1\,Gyr ago was $(-4.451, 7.796,
53.99)$\,kpc, with a velocity of $(204.1, 101.3, 5.742)$\,km/s.

We initialized the Solar System with the four giant planets, Jupiter,
Saturn, Uranus, and Neptune, as they appear in their orbit on 1
January 2020 (2020-01-01T00:00:00.000 UTC), according to the JPL
Horizon database.\footnote{see
  \url{https://ssd.jpl.nasa.gov/?horizons_doc}} (We note here
that the choice of the initial epoch should not considerably effect either
the qualitative or the quantitative statistical results. However, for the
orbits of individual asteroids, the precise initial realization is
important;  see Sect.\,\ref{Sect:chaos} for a brief
discussion). The ecliptic was initialized with an angle of $60^\circ$
to the Galactic plane. The initial conditions for these simulations
are presented in Table\,\ref{Tab:Simulations_D}. A wider range of
initial realizations was tried, including changing the orbital phases,
and separations of the planets and the parameters for the debris disk
(see Sect.\,\ref{Sect:Discussion}). Varying the orbital phases does
not result in qualitative differences, but the results are relatively sensitive to changes in the initial orbital
separations of the  planets. After initializing the planets, we place a disk of test
particles in the ecliptic plane around the Sun. The disk was generated
using the routine {\tt ProtoPlanetaryDisk} in the \AMUSE\, framework
with a Toomre-Q parameter of $25$ \citep{1964ApJ...139.1217T}. Most
simulations are performed with disks of between 16\,au and 35\,au, but
other ranges were explored with a minimum of 3\,au and an upper limit
of 1000\,au.

\begin{figure*}
\center
\includegraphics[width=1.0\textwidth]{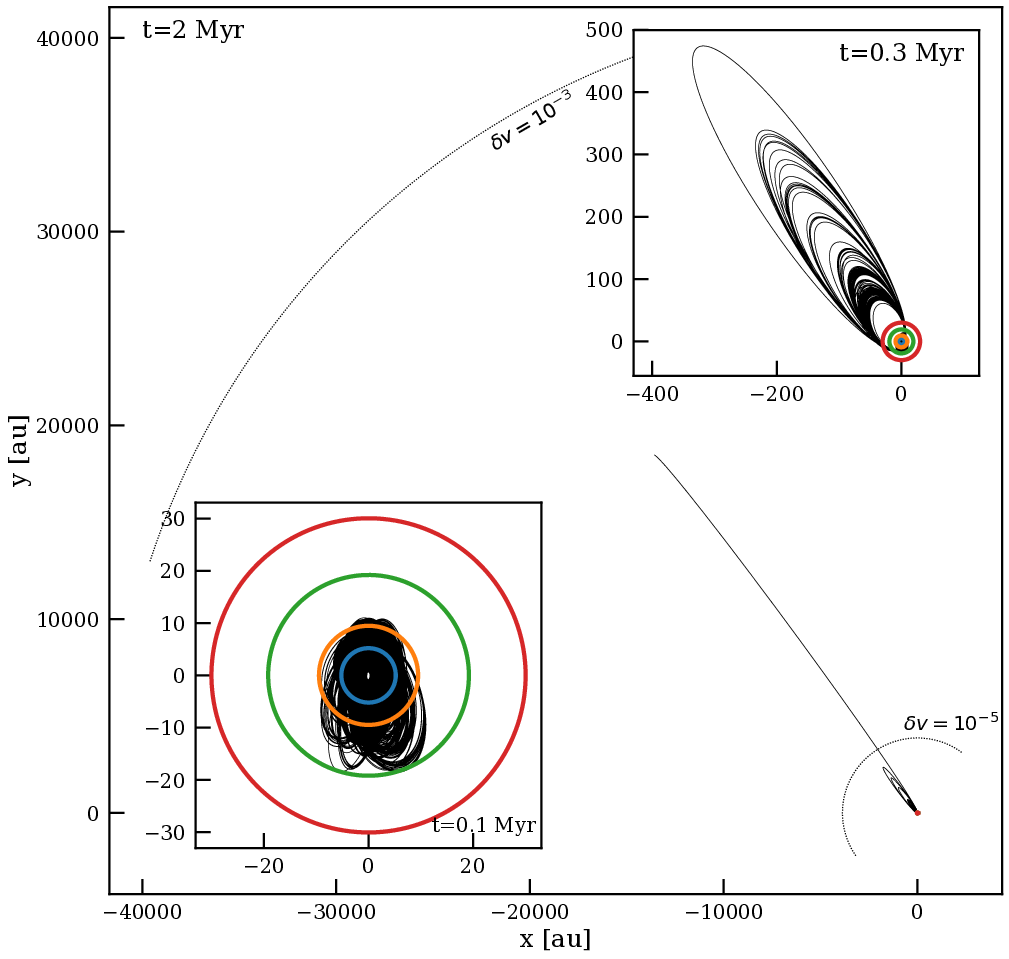}
\caption{First 2\,Myr of the orbital evolution of an asteroid born in a
circular orbit in 4:5 mean-motion resonance with Jupiter. Repeated
interactions with the giant planets (colored circles: blue, orange,
green, and red for Jupiter, Saturn, Uranus, and Neptune, respectively)
launch the asteroid along the conveyor belt into the Oort cloud until it
reaches the Hill sphere at a semi-major axis of $\apgt 14\,000$\,au
with eccentricity $\apgt 0.995$. The inset to the lower left gives
the first 0.1\,Myr of evolution, and the inset to the top right
gives the first 0.3\,Myr. Each frame is on a different scale.
\label{fig:asteroid_ejection}
}
\end{figure*}

In the simulations, Jupiter starts to populate the resonance regions
(such as the asteroids of the Hilda-family; in
figure\,\ref{fig:schematic_overview} the purple arrows to the bottom
left illustrates how they migrate) and launches asteroids into the
conveyor belt (light green and black curves) directly from the start
of the simulation. This process happens very early after the formation
of the Solar System which may then still be a cluster member. After
$1$\,Myr, $43.4$\% of the asteroids born between $3$ and $15$\,au are
still in resonant orbits, and $39.4$\,\% have escaped the Solar System
\citep[see also][]{2018DPS....5020001P,2019A&A...623A.169P}. Asteroids
initially on wider orbits are not affected momentously within this
short time-frame. figure\,\ref{fig:asteroid_ejection} illustrates this
process by plotting the orbit of a test particle integrated with the
Solar System and the tidal field of the Galaxy.

On a longer timescale, based on the simulations listed in
Table\,\ref{Tab:Simulations_E} and Table\,\ref{Tab:Simulations_G}, the
ice-giant planets transport asteroids along the conveyor belt into the
Oort cloud, where they slowly circularize and isotropize. After
$\sim 100$\,Myr, the orbital distribution of asteroids is similar to that of the present epoch; the stable resonant regions near the giant
planets are populated, and the non-resonant orbits as well as the
conveyor belt are depleted. Only $1.4$\,\% of the asteroids born
between $15$ and $50$\,au are still in the disk after 100\,Myr and
$59.2$\% have become unbound from the Solar System. The remainder of
the asteroids born between $1$5 and $50$\,au have migrated to the Oort
cloud at this point in time.

The giant planets  only scatter the asteroids in inclination by a few
degrees \cite[see also][]{2020arXiv200609657D}. By the time the
asteroids reach the inner Oort cloud, their inclinations are still not
much larger than a few tens of degrees around the ecliptic plane. Even
asteroids that are scattered along the conveyor belt to the Oort cloud
preserve a relatively low inclination with respect to the ecliptic
plane \cite[see also ][]{1987AJ.....94.1330D,2019MNRAS.490.2495C}. The
inclination distribution only isotropizes once the Galactic tidal
field starts to dominate the orbital evolution of the  asteroids \citep[see
also][]{2018A&A...620A..45F}.  Although the majority of the asteroids
escape the Solar System while being kicked out by the planets, the
number that remain bound is sufficient to explain the richness of the
predicted population of objects in the Oort cloud.


Nevertheless, the schematic view presented above is rather idealized
because we assume that the Solar System was born as a single star
orbiting in the Galactic potential. In the following section, we relax this assumption and study the consequences of the Sun being born
in a young stellar cluster. It turns out that being born in a
clustered environment has profound consequences for the efficiency of
the formation of the Oort cloud.

\subsection{The evolution of the Solar System in its birth cluster}
\label{Sect:primordial_star_cluster}

For the simulations presented in this section, the Solar System is
initialized as a member of a star cluster \citep[see
also][]{Torres2020,2020MNRAS.497.1807S,2020MNRAS.493.5062V}. Two
series of simulations were performed: one in which the Sun has four
giant planets in a compact configuration, and the other that shows a
more extended configuration (see Table\,\ref{Tab:Simulations_C}).
Simulations are performed with $2000$ asteroids in circular orbits
of between 16\,au and 35\,au for the compact, and another series with
asteroids in circular orbits
of between 40 and 1000\,au for the extended configuration
\citep[see also][]{Torres2020}. An additional set of 32 simulations
(not listed in Table\,\ref{Tab:Simulations_C}) was performed in which
the Solar System, with a disk of 1000 asteroids, experiences a single
close encounter at a distance of 225\,au or 400\,au with a relative
velocity of 1\,km/s. Here we varied the impact angle of the
encountering star from $0^\circ$ (in the ecliptic plane) to $30^\circ$, $60^\circ$, and $90^\circ$ (perpendicular to the ecliptic).

In these models, we ignore the inner disk (within $40$\,au for the
extended models and within $16$\,au for the compact models). This is
motivated by the fragility of the inner Solar System. Any encounter
that would perturb the inner region would probably leave the Solar System
unrecognizable today. Therefore, there seems to be no particular
reason to include the inner disk in the calculations. Each simulation
was performed up to 100\,Myr. This timescale is smaller than the
cluster lifetime, but is sufficient to support the conclusions of this
paper.

The cluster in those simulations is built using {\tt NBODY6++GPU}
\citep{2015MNRAS.450.4070W}, while we use {\tt REBOUND}
\citep{2012A&A...537A.128R} to integrate the planetesimals
\citep[using the IAS15 integration scheme
  from][]{2015MNRAS.446.1424R}.  The simulations are carried out using
the LonelyPlanets approach designed in
\cite{2017MNRAS.470.4337C,2018MNRAS.474.5114C,2019MNRAS.489.4311C}.

\subsubsection{Simulating planetary systems in a dense star cluster:
the LonelyPlanets approach}\label{Sect:LonelyPlanets}

In the LonelyPlanets module in \AMUSE\,, we first evolve a star cluster
without planets or asteroids for $100$\,Myr using NBODY6++GPU
\citep{2015MNRAS.450.4070W}. This calculation includes the $N$-body
dynamics of the stars, stellar evolution, and the interaction with the
Galactic tidal field (as described above).  Clusters are born
instantaneously without residual gas and with stars from a mass
function on the zero-age main sequence distributed in a virialized
Plummer sphere.  We store masses, positions, and velocities on the five
nearest neighbors for each star during the calculation at 1000-year
time intervals.  Interactions with a single nearest star provided
satisfactory statistics for the strongest encounters
\citep{2020AJ....160..126G}, but in our opinion lagged in weaker
perturbations. We therefore include the nearest five stars.

In the second pass through the data, each stored encounter is treated
as a scattering experiment lasting 1000 years.  After each scattering
experiment with five perturbing single stars and one star with planets,
we continue with the new set of five perturbing stars while carrying the
planetary system over from one scattering experiment to the next,
$1000$ years later.  This works as follows. The masses, positions, and
velocities of the five perturbing stars are recovered from file, and the
target star with planets and test particles as asteroids is integrated
together with the perturbers (see
Table\,\ref{Tab:Simulations_C}). After $1000$ years of integration, we
recover the next set of five perturbers from file and integrate them as a
new experiment together with the target star, planets, and asteroids.
We note that these five nearest neighbors may be different stars from one
snapshot to the next ($1000$ years later).  The presence of planets in
the scatter experiment leads to small perturbations of the encountering
stars, which affects their orbits.  Therefore, even if a subset
of perturbing stars is identical from one snapshot to the next, the
positions and velocities of these perturbers does not necessarily have
to be identical at the end of a scatter experiment and the beginning
of the next snapshot because the encounter database was built without
planets. Our approach is therefore inconsistent.

\begin{figure}
\includegraphics[clip, trim=0.0cm 0.0cm 0.0cm 1.3cm,width=\columnwidth]{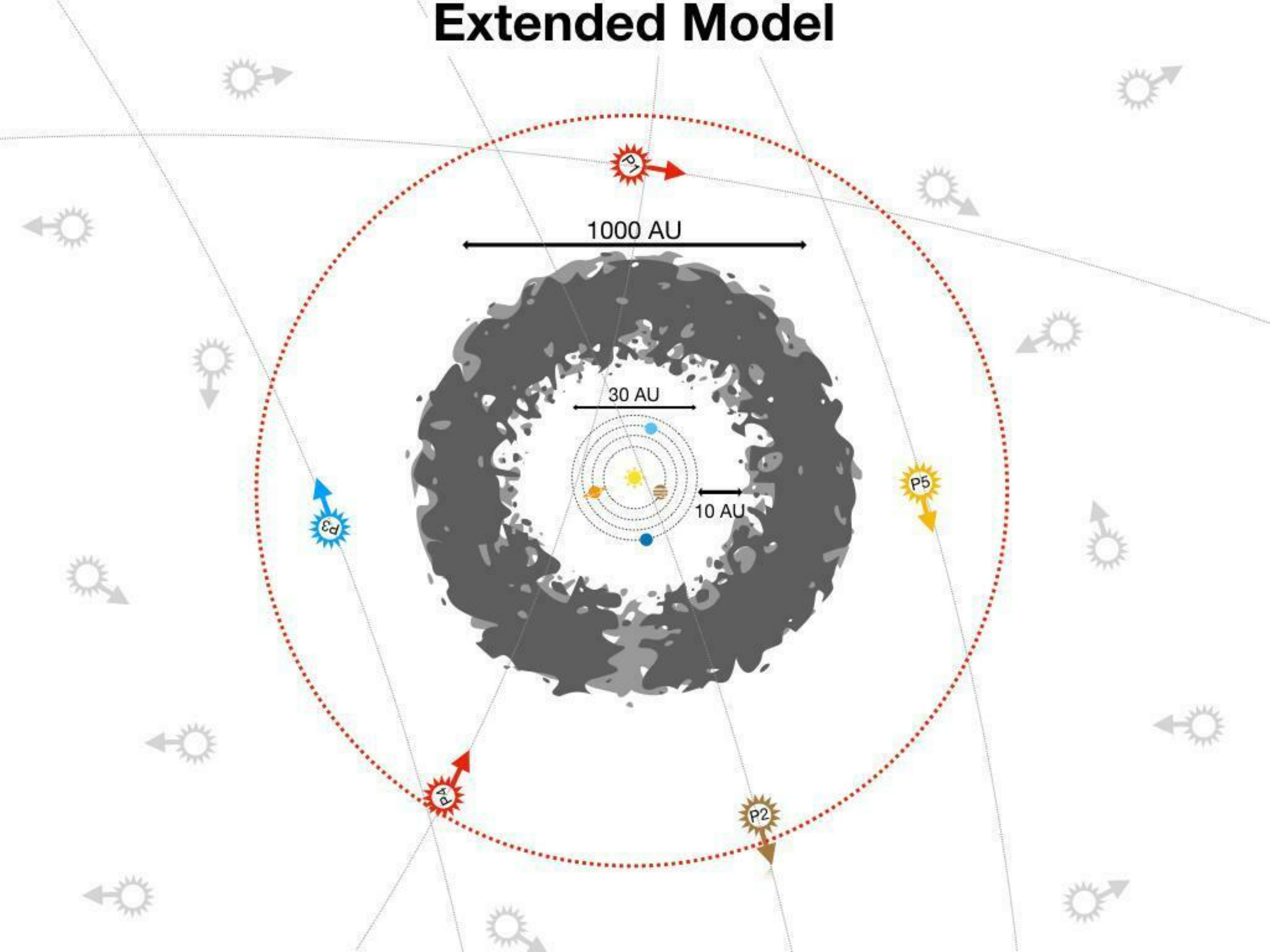}
\caption{Illustration of the extended Solar System model (not to
scale). The gray shaded region represents the debris disk with an
outer radius of 1000\,au. The dotted red circle indicates the boundary of the
neighboring sphere, beyond which we ignore the influence of
perturbations by passing stars. Gray symbols represent the stars in
the cluster that are ignored. The colored star symbols (P1, P2, P3,
P4, and P5) represent the five nearest stars that perturb the Solar System at a given time. The arrows indicate the direction of motion
of the perturbers \citep[figure from][]{Torres2020}.
\label{fig:illustration_Torrest_Fig4.1}
}
\end{figure}

The encounter between the five perturbing stars and the one star with
planets and asteroids is calculated using {\tt REBOUND}
\citep{2012A&A...537A.128R}. This process is repeated for each
encounter for the duration of the star-cluster simulation
($100$\,Myr).  figure\,\ref{fig:illustration_Torrest_Fig4.1} 
presents an illustration of this method with the six interacting stars
identified.

In the following two sections, we discuss two populations: the
captured and the scattered populations of asteroids. The amount of
mass (or the number of asteroids) that is transferred is similar to
the number of asteroids at the periphery of the disk, which is unbound or
scattered into highly eccentric and inclined orbits
\citep{2015MNRAS.453.3157J}.

\subsubsection{The influence of the cluster on the circumstellar disk
\label{Sect:Scattered_asteroids}}

When a star is a young cluster member, mutual encounters affect the
circumstellar disks \citep{2015A&A...577A.115V,2016ApJ...828...48V}.
However, most simulation studies address the effects of one star on
the disk of another star resulting in the deformation of the disks
\citep{2014A&A...565A.130B,2016MNRAS.460.3505R,2016MNRAS.455.3086X,2017A&A...599A..91B}.
A star in a cluster is exposed to multiple encounters, each having a
subsequent effect on the disk's morphology and mass.  Only a few
studies take these multiple interactions into account
\citep{2020AcA....70...53J,Torres2020}.

Moreover, mutual interactions also lead to the transport of material from
one star to another \citep[see also][]{1995MNRAS.274...85K}. We
address both processes separately. Here we describe the scattering of
the native disk due to stellar encounters. In
sect.\,\ref{Sect:Captured_asteroids}, we discuss asteroids that are abducted
from the  disk of another star.

In the calculations with LonelyPlanets, we follow the dynamical
evolution of the debris disk and planets around each star individually
(see previous paragraph). The disks are only resolved in the second
pass through the data. As a result, disks do not interact mutually,
but they are affected by multiple encounters with other stars.

Each star experiences multiple interactions with other stars. We focus
on one particular star in this simulation: the 1\,\MSun\, star \# 157
from Torres et al.\, (2020, their figure~2)\nocite{Torres2020}. We
picked this star because the orbital topology of its scattered
population resembles the Kuiper belt of the Solar System. The initial inner
edge of the disk in this calculation is $40$\,au. This seems rather
large, but a strong perturbation of the more inner region would have
had considerable repercussions for the entire planetary system,
contrary to what is observed in the Solar System.

In figure\,\ref{fig:ae_simulations_scattered_and_sednitos}, we present
the orbital distribution of the scattered disk population (green
dots). The small black lines from each dot point in the direction of
the evolution of its orbit over the next 300\,Myr. The population near
the inner part of the disk ($a\aplt 50$\,au) is affected by the outer
planets, causing an increase in their eccentricity. The population
between $50$ and $500$\,au is hardly affected by either the planets
or the other stars, and is hardly affected over the lifetime of the
Solar System, until the Sun evolves beyond the main sequence and sheds
its outer layers \citep[this process was studied extensively
by][]{2011MNRAS.417.2104V,2012MNRAS.422.1648V,2014MNRAS.437.1127V,2020AJ....160..232Z}.

Once the star and surviving disk leave the parent cluster,
$\sim 4.2$\,\% of its native outer-disk asteroids have been scattered
through interactions with other stars directly into the conveyor belt
region.  The majority of these asteroids move on a timescale of
$\sim 100$\,Myr along the conveyor belt until they reach the Oort
cloud (once they cross the curve $t_{\rm diff} = P_{\rm orb}$, or the
curve of $\delta v = 10^{-3}$). This process is illustrated in
figure\,\ref{fig:ae_simulations_scattered_and_sednitos} with the long
lines that start at the asteroids and point to their orbital parameters
300\,Myr later in time (in the direction of the Oort cloud).  We note
that in figure\,\ref{fig:ae_simulations_scattered_and_sednitos} the
curve for $\delta v = 10^{-5}$ is given, which roughly corresponds to
the inner edge of the Hills cloud. The difference between the
$\delta v = 10^{-5}$ and $\delta v = 10^{-3}$ is also illustrated in
Figs.\,\ref{fig:schematic_overview} and\,\ref{fig:asteroid_ejection}.

After $\sim 100$\,Myr, some asteroids of $\apgt 2.5$\,Neptune-Hill radii
from the orbit of  Neptune are still in the process of ascending the
conveyor belt (see the thin dotted curve to the right of the red curve
in figure\,\ref{fig:ae_simulations_scattered_and_sednitos}). This
small population of asteroids within the conveyor belt does not
migrate to the Oort cloud. \cite{2020arXiv200609657D} found a similar
population of lingering asteroids, and the migration process in their
calculations takes $\sim 68$\,Myr, which is comparable to the
$\sim 100$\,Myr in the calculations presented here.  Also,
\cite{2004Icar..172..372F} studied the migration rate of scattered
trans-Neptunian objects to the Oort cloud and argue that eventually,
after 5\,Gyr, half the scattered disk settles in the Oort cloud.

\subsubsection{The captured asteroids}
\label{Sect:Captured_asteroids}

Apart from scattering of the circumstellar disk, stellar encounters also
lead to the capture of asteroids. The majority are deposited in the
parking zone, possibly in highly inclined orbits
\citep{2016MNRAS.457.4218J}. Those that enter the conveyor belt
migrate towards the Oort cloud or are ejected. The fraction of
captured asteroids that enter the conveyor belt depends on the details
of the encounter, and the disk
parameters of the  encountered star. Although the numbers vary strongly depending on the parameters of the 
encounter, the typical fraction of captured asteroids is
comparable to the fraction of asteroids lost, at least in an
equal-mass encounter \citep[see also][]{2017A&A...599A..91B}.
Encounters in which the encountering star meets the prograde orbiting
debris disk are most effective in producing unbound asteroids
\citep{2021arXiv210406845P}.  The population of captured asteroids may
therefore have a high proportion of retrograde orbits compared to those
ejected from the inner Solar System.

We included a population of captured asteroids in our
analysis. However, the calculations using the LonelyPlanets approach
are not suitable for acquiring a census of their orbital distributions
\citep{2018MNRAS.474.5114C} because once an asteroid becomes unbound
from its parent star it is lost from the simulation. Instead of taking
the results of system \#~157 from \cite{Torres2020}, as we did for the
scattered disk, we adopt the results of \cite{2015MNRAS.453.3157J} for
the captured asteroids.  These latter authors performed $N$-body
calculations to study the mutual stellar encounter that could explain
the orbit of the dwarf planet Sedna. They argue that Sedna was
captured from the circumstellar disk of another star. This must have
happened while both stars were members of the same parent cluster.
The captured population from \cite{2015MNRAS.453.3157J} reproduces
Sedna's orbit as a captured object.

At present, Sedna is orbiting in the parking zone well outside the
conveyor belt and is not expected to reach the Oort cloud.  From a
Galactic point of view, Sedna is relatively close to the Solar
System. Asteroids within $\aplt 1000$\,au of the parent star are
hardly affected by the tidal field of the Galaxy.  The parking zone is
therefore composed of a roughly equal number of scattered and captured
asteroids. The orbits of asteroids in the parking zone are no expected
to affected by either the planets or by the Galactic field. Frozen in
time, this population may bear information about the mechanism that
brought it there.

\begin{figure}
\includegraphics[width=\columnwidth]{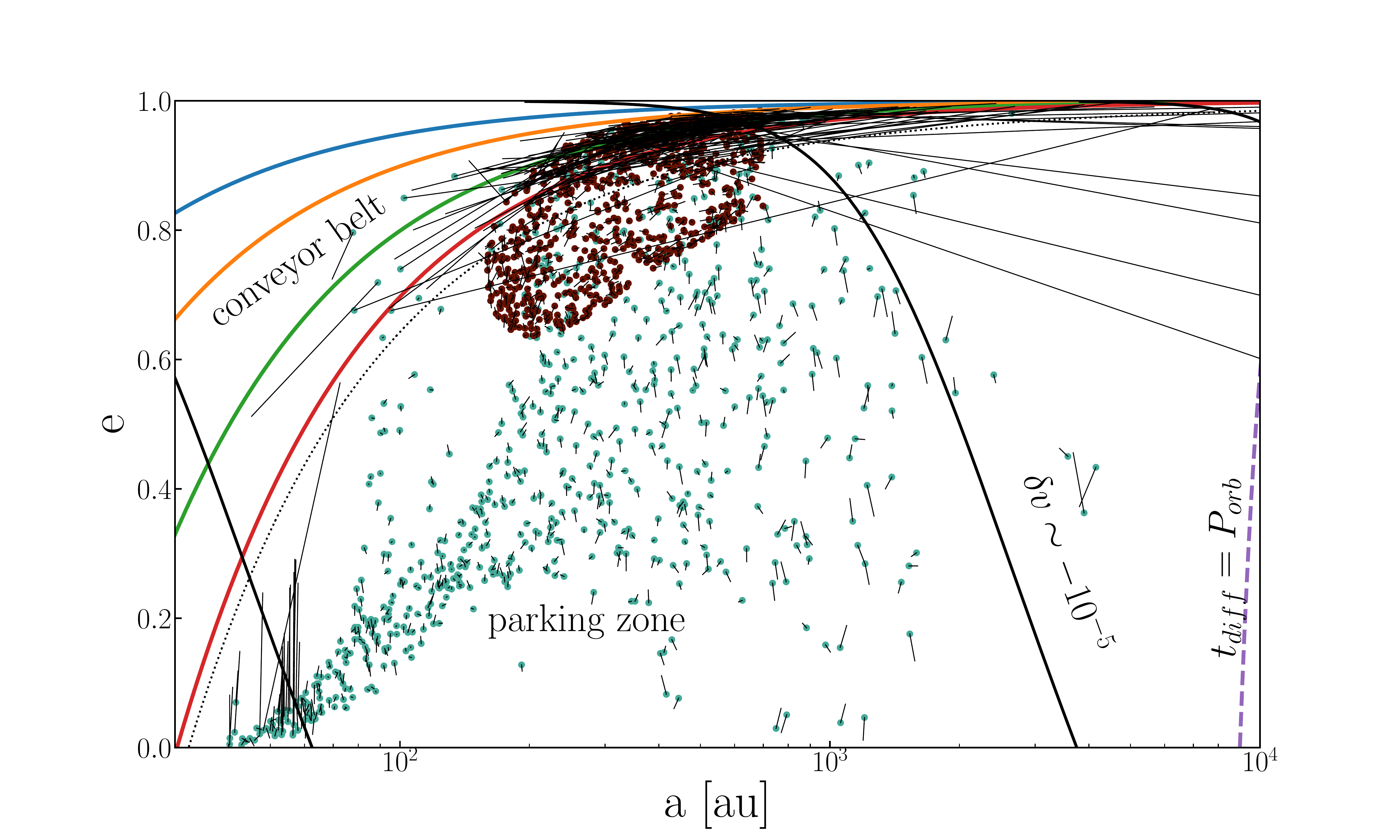}
\caption{Orbital distribution of the scattered (green dots) and
  captured (brown) asteroids. These conditions originate
  from the scatter and capture calculations,
  but form the initial conditions for subsequent calculations.
  The former are scattered from the circumstellar debris disk by a close encounter with another star
  in the parent cluster. The latter population is captured from another star
  with which the Sun interacted in its parent cluster. The
  simulations for both populations are described in
  Table\,\ref{Tab:Simulations_D}. The small thin black lines from each
dot-point point in the direction in which the orbit migrates over
300\,Myr (we adopted 300\,Myr here for the presentation, although we
realize that adopting 100\,Myr would have been a more systematic
choice). These data were acquired by integrating the Solar System in
the tidal field of the Galaxy. The asteroids that orbit the outer
and inner parking zones are hardly affected by the Galactic tidal
field. Those near the conveyor belt are strongly affected by the
giant planets and some eventually reach the Oort cloud. The thin
black dotted curve is parallel to the 2.5 Hill-radii to the right of Neptune's
pericenter influence (solid red curve), and indicates the extent to which
the planet still affects the orbits of the  asteroid. Simulation
parameters are listed in Table\,\ref{Tab:Simulations_D}.
\label{fig:ae_simulations_scattered_and_sednitos}
}
\end{figure}

We introduced the captured population into the Solar System and
integrated it as test particles together with the giant planets and
the background potential of the Galaxy over a timescale of
300\,Myr\footnote{The timescale of 300\,Myr was selected
  empirically. The migration timescale for highly inclined asteroids
  turns out to be about 100\,Myr (see
  Sect.\,\ref{Sect:Scattered_asteroids}). We therefore had to
  integrate for at least this time-span.  The calculations, however,
  are sufficiently expensive to not continue them for too long.}.  The
simulations are summarized in Table\,\ref{Tab:Simulations_D}.  The
resulting orbital parameters are presented in
figure\,\ref{fig:ae_simulations_scattered_and_sednitos}.

During integration, the captured asteroids that were introduced in the
conveyor belt region have been driven by the giant planets into the
Oort cloud region. The timescale on which they reach the Oort cloud
depends on their relative inclination to the ecliptic, and on the
pericenter distance. Highly inclined asteroids, for example, may have
their closest approach to the Sun far from the perturbing influence of
the giant planets. Asteroids orbiting in the ecliptic plane are more
strongly affected by the giant planets, and are scattered into the
conveyor belt on a shorter timescale.

\subsubsection{When did the Sun escape the parent cluster?}\label{Sect:SunEscapersCluster}

Here, we assume that stars form instantaneously in random positions
from a point-symmetric \cite{1911MNRAS..71..460P} potential in virial
equilibrium. Planets also form instantaneously in almost circular
orbits in a randomly oriented ecliptic plane.

These are strong (and, in our view, rather preposterous) assumptions, and
some of our results are affected by these choices. However,
relaxing any of these assumptions will open up parameter space and
lead to a dramatic increase in the number of calculations required to
acquire reliable uncertainties on the simulation results.  One improvement would be to take the hydrodynamical collapse of the
molecular cloud into account, including the processes of star
formation, stellar feedback, disk evolution, and planet formation.
However, these calculations are somewhat elaborate
\citep{2020MNRAS.499.5873C,2021arXiv210410892F}.

Even though the processes discussed in this paper are fundamental, the
timescales and efficiencies of the various processes are affected by
the initial conditions of the simulations
\citep{2006ApJ...641..504A}. The timescale over which, and the order
in which, the giant planets form are important for the presented model
and the efficiency at which asteroids are injected into the Oort
cloud. The conclusion that asteroids were ejected early on by the
giant planets while the Sun was a cluster member remains largely
unaffected. The Oort cloud will be prevented from forming so long as
the Sun is a member of a star cluster
\citep{2015AJ....150...26H}. Only after the Sun has left the birth
cluster is it possible to keep asteroids bound in orbits as wide as
the current Oort cloud up to the Hill radius in the Galactic
potential.

It is not trivial to constrain the moment at which the Sun escaped its
birth cluster.  In the first $\aplt 10$\,Myr, the cluster is expected
to have a relatively high density $\apgt 10^3$\,\MSun/pc$^{-3}$
\citep{2012MNRAS.423.1985L,2016MNRAS.457.1062M,2020ApJ...897...60P,2020ApJ...903..114P}. After
that, the cluster expansion is driven by stellar winds, radiation
feedback, and supernovae until it eventually dissolves in the Galaxy's
tidal field. This latter phase in which the cluster dynamics is
dominated by two-body relaxation and tidal stripping lasts for
approximately 100\,Myr
\citep{2006ApJ...641..504A,2020MNRAS.497.1807S}. While the cluster
expands with time, the local density gradually decreases, and the
encounter rate drops
\citep{2007MNRAS.380.1589B,2017MNRAS.470.4337C}. These time scales are
all short compared to the formation timescale for Jupiter of
$\aplt 4$\,Myr
\citep{2017PNAS..114.6712K,2010Icar..209..616M,2021Icar..35514087D}.

The Solar System may have survived this first dense phase rather
unharmed, leaving the cluster at a later stage, possibly around the
cluster's half-life timescale. According to our calculations
\citep[but also see][]{2008Icar..197..221K}, it seems more difficult to grow a
rich Oort cloud if the Sun stays in the parent cluster for
$\apgt 100$\,Myr, because this corresponds to the timescale on which
the ice-giant planets eject their local asteroids into the conveyor
belt. We therefore argue that the Solar System might have been
ejected from the parent cluster within $\sim 20$ or $50$\,Myr after
birth.

The effect of a nearby encounter on the cold Kuiper belt may not have
had a long-lasting impact, as this population might have regrown over
time
\citep{2005MNRAS.360..401A,2014MNRAS.444.2808P,2020ApJ...901...92M}. Other
signatures of the difference between a few strong encounters and
extended exposure to relatively weak perturbations are hard to
quantify. We did not explore the long-term survival of the Solar System in its birth cluster and its possible consequences for the  outer edge of the
planetary disk. It would be interesting to study this
aspect, but these calculations are elaborate, and the parameter space is
extended.

\subsection{Eccentricity damping of the asteroids in the Oort cloud}
\label{Sect:Eventual_evaporation}

\begin{figure}
\includegraphics[width=\columnwidth]{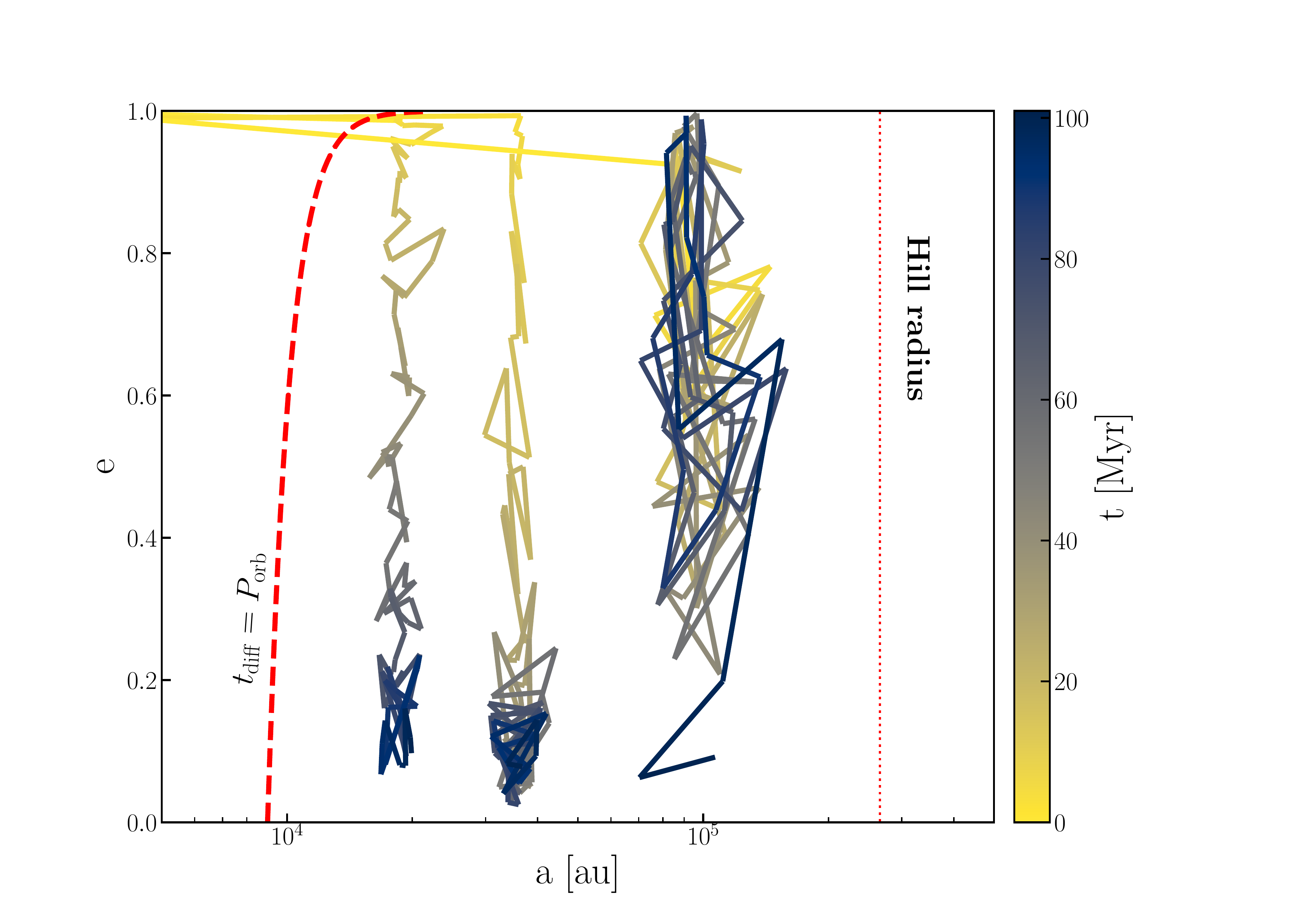}
\caption{Orbital evolution of three asteroids for 100\,Myr in 
semi-major axis and eccentricity. Each starts from the conveyor belt
(to the top left) until it circularizes. These asteroids are
launched into the conveyor belt by a giant planet. This process is
illustrated in figure\,\ref{fig:asteroid_ejection}. Once the Galactic
tidal field starts to dominate the orbital evolution of the
asteroids (to the right of the red dashed curve), it detaches them from
the influence of the  inner planets. The Galactic tidal field subsequently
drives the eccentricity evolution of the asteroid until it
circularizes. The timescale of the circularization process is
illustrated with the color bar. Asteroids further in the Oort cloud
are more strongly affected by the Galactic tidal field, and therefore
have a shorter circularization timescale. Not all asteroids
circularize to the low eccentricities of the three examples
presented here. The eventual eccentricity distribution of the Oort
cloud objects is presented in
figure\,\ref{fig:cumulative_eccentricity_distribution}. The
vertical dotted curve to the right indicates the Hill radius of the
Solar System in the Galactic potential. 
\label{fig:ae_randomwalk}
}
\end{figure}

\begin{figure}
\center
\includegraphics[width=1.0\columnwidth]{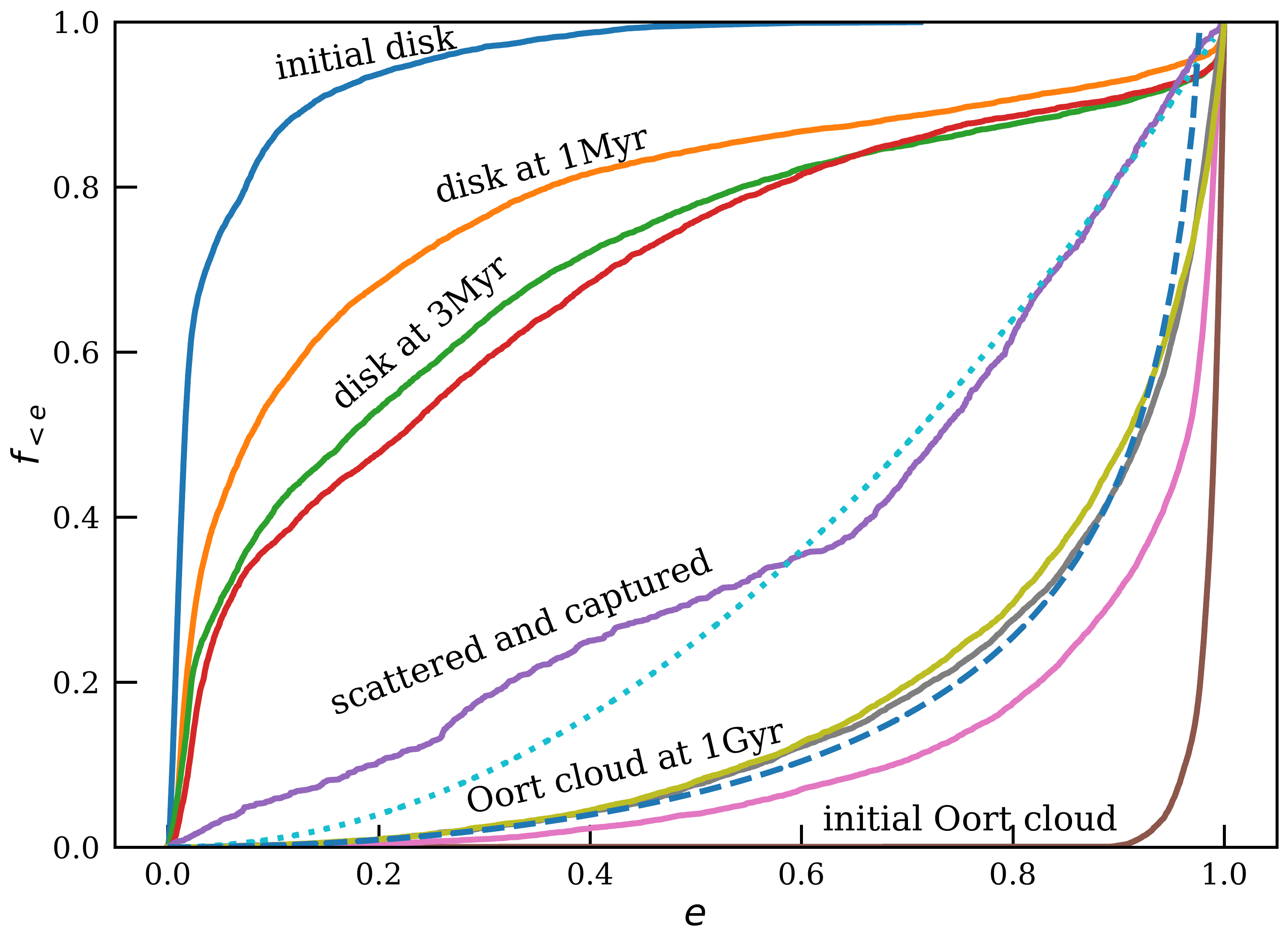}
\caption{Cumulative distribution of the eccentricity at various stages
of the evolution of the eventual Oort-cloud population of asteroids.
The dotted and dashed curves are presented to guide the eye. The
dotted curve gives the probability density function for the thermal
distribution in eccentricity ($f(e) \propto e^2$). The dashed curve
gives an eccentricity probability density function $f(e) \propto
(1-e^2)^{-2/9}$, which matches the eventual Oort cloud
distribution at 1\,Gyr.
The top four curves give the eccentricity distribution of the
initial disk around the Sun (see Table\,\ref{Tab:Simulations_F}), at
the ages of $1$\,Myr, $3$\,Myr, and $10$\,Myr (red curve).
The purple curve through the middle of the figure near the dotted
curve gives the distribution of the two populations of scattered
disk objects and the captured asteroids (see also
figure\,\ref{fig:ae_simulations_scattered_and_sednitos} and
Table\,\ref{Tab:Simulations_D}). Both the scattered and captured
populations are comparable in number.
The bottom four curves give the distribution of the Oort cloud
asteroids, from bottom to top immediately after the asteroids have entered the Oort
cloud (indicated with ``initial Oort cloud''), followed by the
distribution at ages of 100\,Myr, 500\,Myr, and 1\,Gyr (as
indicated, see also Table\,\ref{Tab:Simulations_E}). 
This figure contains data from simulations for which the parameters
are listed in Table\,\ref{Tab:Simulations_F} (top curves),
\ref{Tab:Simulations_D} (middle purple curve), and
\ref{Tab:Simulations_E} (bottom set of curves).
\label{fig:cumulative_eccentricity_distribution}
}
\end{figure}

While the gas-giant and ice-giant planets launch asteroids further
into the Oort cloud, the Galactic tidal field gradually becomes a
stronger perturber. Eventually, when asteroids cross the purple-dashed
curve in figure\,\ref{fig:ae_simulations_scattered_and_sednitos} (far
to the right, but more visible in Figs.~\ref{fig:schematic_overview}
and \ref{fig:OCformation}) their orbits become strongly affected by
the Galactic tidal field through von Zeipel-Lidov-Kozai resonance
\citep{1910AN....183..345V,1962PSS..9..719L,1962AJ.....67..591K};
\citep[see][for a historical overview on the
terminology]{2019MEEP....7....1I}.  This leads to damping of the
eccentricity \citep[as is also demonstrated
in][]{2019AJ....157..181V}. In our simulations, this proceeds through
a random walk in semi-major axis and eccentricity but directed towards
lower eccentricities. In figure\,\ref{fig:ae_randomwalk} we illustrate
this process with three asteroids from one of our simulations. These
particles start in the narrow neck between 1000\,au and 3000\,au at an
eccentricity $\apgt 0.998$ while they are kicked by one of the giant
planets along the conveyor belt into the Oort cloud region. The tidal
field of the Galaxy subsequently reduces the orbital eccentricity of
these asteroids.
This eccentricity-damping process proceeds as a
random walk in eccentricity and orbital separation. The influence of
the Galactic tidal field near the apocenter of the asteroid may cause
the orbital eccentricity to increase as well as decrease, whereas the
semi-major axis is less strongly affected. The global trend is a
reduction in eccentricity.
Asteroids further in the Oort cloud are more strongly affected by the
Galactic tidal field, and therefore have a shorter circularization
timescale. Not all asteroids circularize to the low eccentricities of
the three examples presented figure\,\ref{fig:ae_randomwalk}.
Eventually, after a few hundred million
years, we find empirically that the cumulative probability density
function for the eccentricity approaches
$f(e) \propto (1-e^2)^{-2/9}$.
Figure\,\ref{fig:cumulative_eccentricity_distribution}
presents the evolution of the eccentricity distribution of the Oort
cloud (four lower curves). Table\,\ref{Tab:Simulations_E} presents the
conditions for the simulations that support this section.

Asteroids in the disk start with almost circular orbits (uppermost
blue curve in
figure\,\ref{fig:cumulative_eccentricity_distribution}). In time, this
distribution becomes more skewed to higher eccentricity, mainly due to
the injection of objects into the conveyor belt (orange curve). The
scattered and captured asteroid populations have, upon capture, a
steeper eccentricity distribution (purple curve), which approaches the
thermal distribution (dotted curve).

A fraction of the disk particles together with some of the scattered
disk and captured asteroids migrate further along the conveyor belt
until their eccentricity distribution resembles the lowest (red) curve
in figure\,\ref{fig:cumulative_eccentricity_distribution}. This latter
distribution continues to evolve over time until it resembles the purple
curve (indicated with the Oort cloud at 1\,Gyr). Further evolution of
the Oort cloud is slow, on a timescale of $\apgt 1$\,Gyr, and
is dominated by evaporation.  By counting the number of objects that
re-enter the inner Solar System, we derive a rate of $\sim
2$--$6 \times 10^{-12}$ comets per asteroid in the Oort cloud per year
that find their way back into the inner Solar System within 10\,au
(see Sect.\,\ref{Sect:retrogade_orbits}). This leads to one or two
comet arrivals in the inner Solar System per year, which is consistent
with earlier estimates
\citep{1986EM&P...36..263B,1986Icar...65...13H,2011MNRAS.411..947G,2020CeMDA.132...43F}.

\subsection{Summary of the chain of simulations}\label{Sect:Methods_Summary}

Now that we have briefly discussed each of the simulations, we can
construct a more holistic view of the formation and early evolution of
the Oort cloud. Here, we summarize this model and in the following section
we discuss the consequences.  

Asteroids in the conveyor belt cross the orbit of the giant
planets. Each time this happens, they receive a small kick causing
their orbits to drift with constant pericenter distance to higher
eccentricity and larger semi-major axis \citep[illustrated in
figure\,\ref{fig:asteroid_ejection}, but see
also][]{1987AJ.....94.1330D}. Jupiter ejects its nearby asteroids
along this conveyor belt (indicated in figure\,\ref{fig:OCformation}) on
a timescale of a few million years.

Once the eccentricity of an asteroid is $\apgt 0.998$ and its semi-major
axis is $\apgt 20\,000$\,au, the Galactic tidal field starts to dominate
the orbital evolution of the  asteroid at apocenter~\citep[][see
Fig\,\ref{fig:OCformation}]{2007AJ....134.1693H}.  As soon as the
pericenter distance exceeds the semi-major axis of the  giant planet, the
eccentricity of the asteroids' orbit continues to be reduced.  By this
time, the orbital period is $\apgt 3$\,Myr, and the Galaxy starts
damping the asteroid's eccentricity and randomizing the inclination
\citep{2007AJ....134.1693H} on a timescale of
$\sim 100$\,Myr~\citep{1987AJ.....94.1330D,2020arXiv200609657D}.

Eventually this process leads to the Oort cloud. The distance travelled by an
asteroid into the Oort cloud depends on the last interaction
with the planets before its orbit detaches (see
figure\,\ref{fig:ae_randomwalk}). Once near apocenter, the Galactic
tidal field reduces the eccentricity of the orbit, causing the pericenter
distance to increase.

Our calculations did not last a sufficient length of time to circularize the
entire population, but the reduction in eccentricity converged in
$\sim 1$\,Gyr to $f(e) \propto (1-e^2)^{-2/9}$. We therefore argue
that the Oort cloud still hosts a considerable fraction of asteroids
in relatively high-eccentricity orbits.  This distribution in
eccentricity is consistent with the one found in
\cite{2015AJ....150...26H} at an age of $500$\,Myr to $1$\,Gyr. These latter authors
further argue that the eccentricity distribution in the Oort cloud
thermalizes in $\sim 5$\,Gyr. We do not observe such a thermalization
of the eccentricity distribution.
The structure of the Oort cloud can
then also inform us about what happened in the early Solar System
\citep[see also][]{2018A&A...620A..45F}.

The orbital distribution of the giant planets is not very important in
this evolutionary sequence, except that more massive planets, upon
their last interaction, can inject asteroids in wider orbits with a
higher eccentricity.

The large mass and short orbital period of Jupiter clears the local
asteroid field in as little as a few million years
\citep{1987AJ.....94.1330D,2014Icar..231...99F}. Therefore, the Oort cloud
cannot have formed from asteroids ejected by Jupiter or Saturn,
because their ejection timescale is shorter ($\aplt 10$\,Myr) than the
expected cluster lifetime of
$\apgt 100$\,Myr~\citep{2010ARA&A..48..431P}.  In
Sect.\,\ref{Sect:SunEscapersCluster} we conclude that the Sun escaped
the parent cluster even before that time, that is, in $20-50$\,Myr.
In the first million years after their formation, the giant planets
launch $\apgt 80.3$\,\% of the asteroids between 3 and 15\,au onto
the conveyor belt, and $57.0$\,\% of these escape the Solar System
within the next million years. Within $\sim 10$\,Myr, all the
asteroids born near Jupiter and Saturn are either deposited in
resonances or escape the Solar System.  Jupiter and Saturn therefore
cannot have contributed much to the formation of the Oort cloud
\citep[this was also concluded by][]{Torres2020,Torres2020b}.  However, the
details depend on the relative timing of the various events
and on the mass distribution within the disc.  Constraining those more
precisely would require more extended and sophisticated
self-consistent simulations.

In our simulations, we ignore the star, planet, and asteroid-formation
processes: all are born instantaneously. We have not explored
parameter-space exhaustively, but argue that the details regarding the
precise moment and the orbits in which the giant planets form are
relevant only to second order because they affect the timescale for
the formation of the Oort cloud but not the fundamental process.

The time scale over which asteroids are ejected by the planets is
sufficiently long that an uncertainty of few million years in the
planet-formation process \citep{2017PNAS..114.6712K} does not affect
the efficiency of the ejection process. So long as the Solar System is
a member of a star cluster, a sequence of distinct processes drive the
formation of the Oort cloud.

\section{Results and Discussion}\label{Sect:Discussion}

\subsection{The birth environment of the  Solar System}\label{Sect:Discussion}

The birth environment of the Solar System plays a major role in the formation of the Oort cloud: It
prevents the gas-giant planets from forming the Oort cloud, but at the
same time stimulates its formation by scattering the outer regions of
the circumstellar disk and by introducing new asteroids by capturing
them from other stars, or from the interstellar free-floating
population \citep[see also][]{1990CeMDA..49..265Z}.

The Oort cloud has, according to this view, only a minor contribution
from the scattered asteroids originating from the gas-giant planets, Jupiter and
Saturn. Apart from those asteroids parked in resonant orbits, these
planets cause asteroids to escape the Solar System and become
interstellar free-floating objects \citep[see
also][]{Torres_PhD2020,Torres2020}.

In contrast to Jupiter and Saturn, the ice giants, Uranus and Neptune,
are more favorable in producing the Oort cloud \citep[see
also][]{2008MNRAS.391.1350L,2010A&A...509A..48P}. They carry out the
same process of ejecting asteroids, but do this, because their lower
mass and wider orbits, on a longer timescale and with smaller
impulsive changes in eccentricity and semi-major axis of the asteroids
\cite[see
also][]{1987AJ.....94.1330D,2004Icar..172..372F,2019MNRAS.490.2495C}.

The longer timescale on which the ice-giants planets eject asteroids
is important for the formation of the Oort cloud because so long as
the Solar System is a member of the parent cluster, asteroids in wide
orbits are easily lost from the Solar System
\citep{2017A&A...603A.112N}. This process of asteroid stripping due to
encounters with nearby stars is particularly effective if the
asteroid's orbit is wide and highly eccentric. After a few tens of millions of years
to $\sim 500$\,Myr, once the Solar System has escaped the parent cluster,
such asteroids can remain bound.

The wide and highly eccentric orbits of new arrivals in the Oort cloud
can subsequently be perturbed at apocenter by the Galactic tidal
field, leading to a reduction in their eccentricity through von
Zeipel-Lidov-Kozai resonance. As a consequence, the pericenter
distance of these asteroids increases, which brings them outside the
influence of the planets and detaches them from the conveyor
belt. While the orbital eccentricity of the asteroids continues to
decay due to the Galactic tidal field, they slowly, over a timescale
of $\sim 100$\,Myr, form a distribution of very wide but relatively
low-eccentricity orbits The distribution has to shape of the tidal
lobe of the Solar System in the Galactic potential.  The size of the
Oort cloud is confined on the inner side to the distance where the
Galaxy starts to influence the orbits of the asteroids (at a relative
velocity change of $\delta v \apgt 10^{-3}$), and at the side furthest
from the Sun in the Galactic potential by the surface of the Hill
sphere (illustrated in Figs.\,\ref{fig:schematic_overview}
and\,\ref{fig:illustration_Torrest_Fig4.1}).

\begin{figure}
\includegraphics[clip, trim=0.0cm 0.0cm 0.0cm 1.3cm,width=\columnwidth]{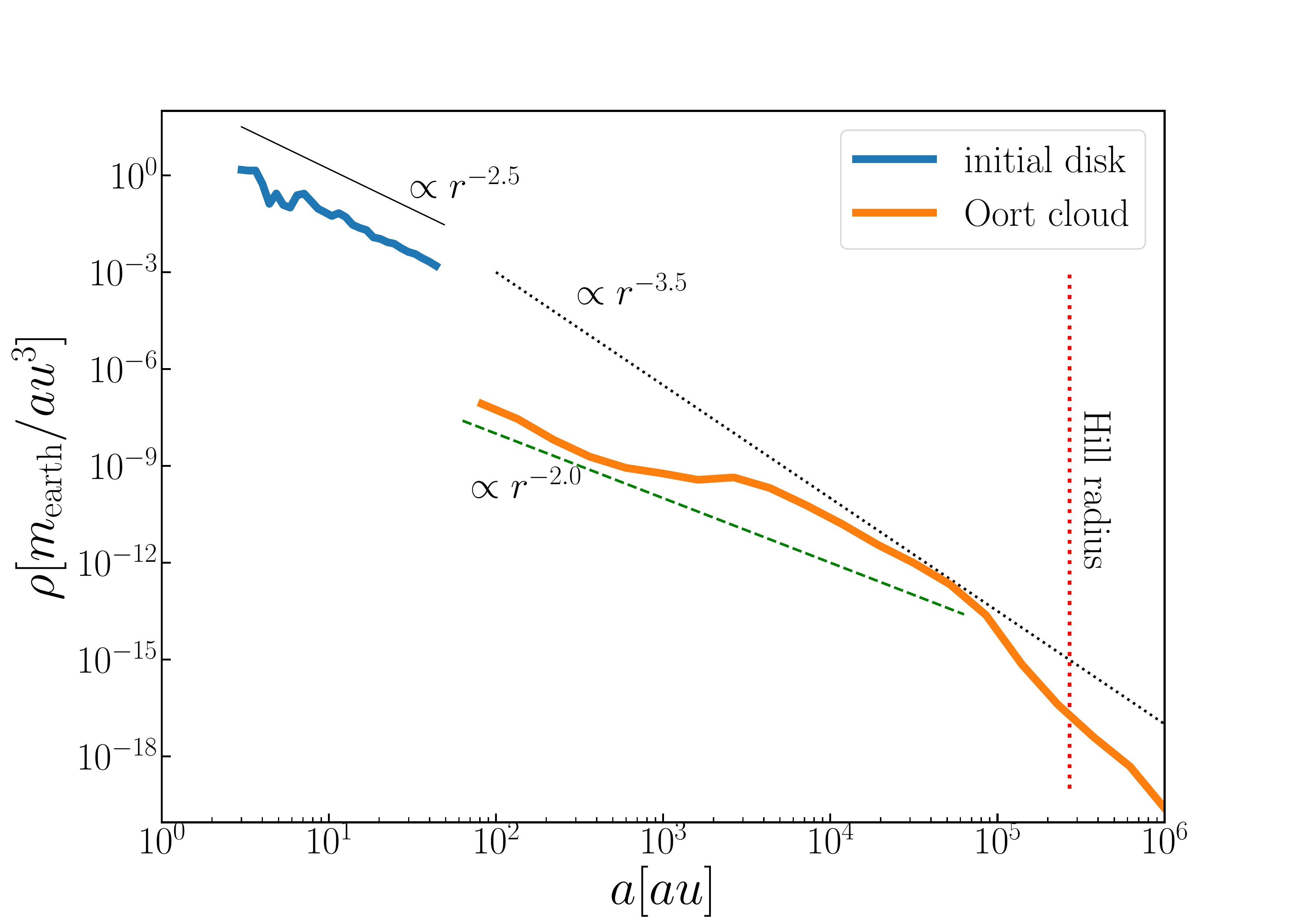}
\caption{Density profile of the initial circumstellar debris disk
  (blue, to the top left) and the eventual Oort cloud, $\sim 1$\,Gyr
  after the Solar System left the parent cluster (orange). The red
  dotted line to the right indicates the formal Hill radius of the Sun
  in its orbit around the Galactic center. The thin-solid black-dotted
  curve shows the initial density profile of the circumstellar disk of
  $\propto r^{-2.5}$ (from simulations of
  Table\,\ref{Tab:Simulations_E}). The thin dotted curve shows the
  density profile $\propto r^{-3.5}$ which, according to
  \cite{2008MNRAS.391.1350L}, resembles the slope of the density
  profile of the Oort cloud.  In our calculations, the distribution is
  somewhat shallower, but not as shallow as found by
  \cite{2015AJ....150...26H} who argue in favor of $\propto r^{-2.0}$,
  which is presented here with green dashes.  The small
  tail of asteroids beyond the Solar System Hill radius in the
  Galactic tidal field has contributions of a minority of asteroids
  that remain bound to the Sun, and those that slowly move away.
\label{fig:OC_density_profile}
}
\end{figure}

According to \cite{2008MNRAS.391.1350L}, the number density in the Oort
cloud is proportional to $n \propto r^{-3.53}$, whereas we find a
considerably more complex structure.  In the Hills cloud and Oort
cloud, we find $n \propto r^{-2.25}$, with an overall flatter slope of
$-2.57$ for the entire range from 100\,au to the Hills radius (see
figure\,\ref{fig:OC_density_profile}).  The slope of the density
profile of $\sim -2.25$ between $2\times 10^{4}$\,au and $10^{5}$\,au
is somewhat shallower than the slope of 
$\sim -4.0$ by \cite{2015AJ....150...26H}, that of $\sim -3.35$ found by
\cite{2019AJ....157..181V}, or that of $\sim -3.0$ by
\cite{2008MNRAS.391.1350L}.  \cite{2015AJ....150...26H} argue that
this slope is independent of the mechanism that brings asteroids to
the Oort cloud, but that does not explain the differences in the
density profile between these different studies.  Part of this
structure may be the result of orbital evolution of the Solar System
and its possible migration in the Galaxy
\citep{2008CeMDA.102..111R,2011DDA....42.0904K,2011Icar..215..491K},
but another part may be the result of planetary migration
\citep{2018A&A...620A..45F}.

\subsection{Results}
\subsubsection{The Sun in the Galaxy}
\label{Sect:SSinGalaxy}

A small fraction ($\sim 0.59\pm0.1$ \%)\ of the Oort cloud population
on the far side of the Hill surface remains bound to the Sun (see
figure\,\ref{fig:OC_density_profile}). At an age of 1\,Gyr in our
simulations, this population has an eccentricity of
$\langle e \rangle = 0.80 \pm 0.23$ with an inclination of
$\langle i \rangle = 96^\circ \pm 42$.  These orbital parameters are
similar to those of the Oort cloud asteroids within the Sun's Hill
surface, and they are consistent with the population presented in
figure\,7 of \cite{2019AJ....157..181V}.  The objects at the far side
of the Hill surface (beyond $\sim 200000$\,au), are not expected to
stay there long as they are ionized from the Solar System by close
stellar passages in the Galactic field \citep{2019MNRAS.490.2495C}.

Asteroids ejected from the Solar System become free-floating
interstellar objects.  This latter population forms a trail of
asteroidal objects in the Galaxy along the orbit of the parent star
\citep{2019MNRAS.490.2495C,Torres2020,
  2020ARep...64..936T,2021A&A...647A.136P}. As the asteroids have a
low velocity relative to the orbital speed of the Solar System, they
continue to follow the same orbit as the Sun (see
figure\,\ref{fig:Galactic_orbit}). Long after they have become
unbound, these asteroids still orbit in a cloud around the Solar
System distributed along extended trailing and leading arms \citep[see
also][]{2021ARep...65..305T}. Eventually, these trails may be
perturbed and scattered by passing molecular clouds or other stars,
scattering their orbits and widening their phase-space distribution
\citep[see][]{2016MNRAS.457.1062M,2020ApJ...903..114P,Torres2020}.

\subsubsection{Consequences of planets in resonant orbits\label{Sect:resonances}}

We performed several additional simulations to study the consequences
of planet migration in the early Solar System. In
Tables\,\ref{Tab:Simulations_F} and \,\ref{Tab:Simulations_G} we
present an overview of these calculations. In addition to the
currently observed distribution of orbits for the giant planets, we
also simulated a more compact configuration and a resonant
configuration.  The main problem in exploring these initial conditions
lies in the enormity of the initial parameter space to be covered; our
parameter-space coverage is far from complete. We therefore selected a
few cases which have been of some interest in the past \citep[see
e.g.,][]{2003ApJ...598.1321W,2005Natur.435..466G,1998AAS...193.9602M,
  2007epsc.conf..866C,2019MNRAS.483.5042B}.

In one of these cases, we initiate a Nice-like planetary orbital
distribution and a disk between 3\,au and 50\,au. In that
configuration, Jupiter was placed on a circular orbit at 5\,au in the
ecliptic plane, and the subsequent planets in 2:1 mean-motion
resonances with the one closer to the star. We also tried other
configurations (see Table\,\ref{Tab:Simulations_E}). The main
conclusion, namely that the gas-giant planets (Jupiter and Saturn) eject
asteroids on a timescale shorter ($\aplt 10$\,Myr) than the expected
survival timescale in the parent cluster ($\apgt 100$\,Myr), also holds
 for these initial conditions.

Changes in the orbital distribution of the giant planets affect the
location and efficiency of the conveyor belt, and the timescale on
which asteroids migrate along the conveyor belt. In our initial
resonant configuration, the entire disk, up to the orbit of the
outermost planet, at around 15\,au, is cleared within 1\,Myr. The area
slightly outside the outermost planet, up to a distance of 20\,au from
the star, is cleared within a few million years. By the time the Solar System leaves the parent cluster, there are too few asteroids left to
populate the Oort cloud.

\cite{2019A&A...623A.169P} also simulated the early Solar System
including the migration of the giant planets. Although these latter authors focused on
the inner Solar System and the populations of resonant and Trojan
orbits, their conclusions are consistent with the results of the
calculations presented here.  Planets in an initial resonant
configuration are excluded based on the consequence that the majority
of asteroids would be lost well before the Sun escapes the parent
cluster. This also accounts for Uranus and Neptune; if they are born
in (near) resonance, the chaotic reorganization of the Solar System
would eject most of the asteroids in the disk on a timescale shorter
than the expected lifetime of the Sun in the parent star cluster.
Asteroids outside the outermost planet would be affected by resonant
reorganization of the planetary orbits, but these asteroids are not
likely to migrate into the conveyor belt, and therefore are not
expected to arrive in the Oort cloud.  Fully resonant initial
conditions can therefore be omitted based on the existence of the Oort
cloud, except for the part of the Oort cloud formed from material that was accreted
from other stars or from free-floating debris.

Even if a resonant configuration in the ice-giant planets could lead
to a dramatic change in the distribution of orbital parameters after
$\sim 100$\,Myr, as is advertised in the original Nice model
\citep{2005Natur.435..466G,2005Natur.435..462M,2005Natur.435..459T},
the number of asteroids has already reduced below the number needed to
explain the Oort cloud. We therefore argue that a system in which
the planets are born close to resonance, or in which planets migrate
in the first few million years since formation \citep[such as advertised in the
grand-tack model;][]{2011Natur.475..206W}, is unlikely to lead to the
formation of an Oort cloud. Planetary migration after the Oort cloud
formed has no further consequence for the Oort cloud, and we cannot
exclude such evolution \citep{1982ApJ...255..307V}. Nevertheless, again,
resonant initial conditions for the planets cannot be excluded if the
majority of the Oort cloud material was accreted.

\begin{table*}
\begin{tabular}{p{2.0cm}p{14.0cm}}
\multicolumn{2}{l}{\bf Planets born in orbital resonance (see sect.\,\ref{Sect:resonances})}\\
Designation & Simulation D. \\
Simulations & One simulation with planets (in various configurations),
asteroids, and Galactic tidal field.
\\ \hline
Star & Single 1\,\MSun\, with four giant planets and 10000 asteroids
in the smooth potential of the Milky Way galaxy.
\\
Planets & The four planets have almost circular orbits ($e<0.006$)
in the plane ($i<0.1^\circ$) with Jupiter, Saturn, Neptune, and Uranus
in orbits with semi-major axes of
$5.1$, $9.5$, $19.2$, and $  30.1$\,au respectively.
The ecliptic was inclined by $60^\circ$ to the Galactic plane.
We performed two more simulations
with initial semi-major axes of $3.75$, $5.953$, $9,449$, and $15.0$\,au, and
with semi-major axes of $5.5$, $8.1$, $11.5$, and $14.2$\,au for the four
giant planets.
\\ 
Asteroids & Ten-thousand asteroids in circular orbits between 3\,au and
50\,au in a thick disk \citep[Toomre-Q parameter of $25$][]{1964ApJ...139.1217T}
in the plane of the planets. 
\\ 
Numerics & {\tt Huayno} \citep{2012NewA...17..711P} coupled to the Galactic
model; see Table\,\ref{Tab:galaxy_paraeters} via
bridge \citep{ZWART2020105240} in AMUSE \citep{2018araa.book.....P}.
\\ 
Computer & Run on LGM-II with GPU
\\ 
Duration & Simulation performed for a duration of 1\,Gyr.
\\ \hline
\end{tabular}
\caption{Simulations of a hypothetical young Solar System with planets in various orbital configurations.}
\label{Tab:Simulations_F}
\end{table*} 

\begin{table*}
\begin{tabular}{p{2.0cm}p{14.0cm}}
  \multicolumn{2}{l}{\bf Planets born in orbit around an isolated star
    (see Sects.\,\,\ref{Sect:Isolated_disk} and \ref{Sect:resonances})}\\
Designation & Simulations E. \\
Simulations & Twelve simulations to study the internal planetary dynamics and their response to the local asteroids.
\\ \hline
Star & Single 1\,\MSun\, with 4 planets, 2000 asteroids.
\\
Planets & Four planets in almost circular orbits ($e<0.006$)
in the plane ($i<0.1^\circ$) with Jupiter, Saturn, Neptune, and Uranus
in orbits with semi-major axes of
$5.1$, $9.5$, $19.2$, and $30.1$\,au. We performed two more simulations
with initial semi-major axes of $3.75$, $5.953$, $9,449$, and $15.0$\,au.
\\ 
Asteroids & Two-thousand asteroids in circular orbits in a thick $Q=25$ disk
in the ecliptic plane of the planets between 3\,au and 50\,au.
\\ 
Numerics & {\tt Huayno} \citep{2012NewA...17..711P}
\\ 
Computer & Run on a 192 core Intel-Xeon workstation.
\\ 
Duration & Simulation performed for a duration of 50\,Myr.
\\ \hline
\end{tabular}
\caption{Simulations of the young Solar System.}
\label{Tab:Simulations_G}
\end{table*} 

\subsubsection{Retrograde and re-entry orbits}
\label{Sect:retrogade_orbits}

We measure the fraction of asteroids in retrograde orbits. At the age
of $300$\,Myr, well after the Sun escaped the cluster, $8.1\pm0.3$\% of
the scattered asteroids ($a \apgt 100$\,au) have retrograde orbits,
and for the compact resonant initial conditions this fraction is
$7.1\pm0.5$\%.  In comparison, $71.9 \pm 0.3$\% of the captured
asteroids ($a \apgt 100$\,au) have retrograde orbits. However, this latter
fraction depends notable on the encounter parameters and
the number may vary considerably for other simulations.  Once well
along the conveyor belt (at $a(1-e) \apgt 50$\,au and eccentricity
$\apgt 0.998$), about $57\pm 1$\,\% of the asteroids have retrograde
orbits. This large fraction of retrograde orbits in the conveyor belt
indicates the moment that the Oort cloud becomes spherical and
isotropic. By this time, the orbits have also reached their terminal
distribution in eccentricity (which is not thermal; see
figure\,\ref{fig:cumulative_eccentricity_distribution}).

Because of interaction with the tidal field, a small fraction of asteroids
return to the inner Solar System after having spent some time in the
Oort cloud. We find $0.6\pm0.1$\% of the asteroids that were launched
from the conveyor belt into the Oort cloud return into the inner
10\,au of the Solar System on a timescale of 1\,Gyr. This results in a
re-entry rate of $2$ to $6 \times 10^{-12}$\,yr$^{-1}$. The re-entry
rate for approaching the Sun to within 1\,au is an order of magnitude
smaller. The observed rate of observable comets in the Solar System,
measured between 110\,BC and 1970, was remarkably constant at a rate
of $0.86 \pm 0.067$\, per year \citep{1999Icar..137..355L}.  A larger
rate of $6.4 \times 10^{-12}$ per year was derived by
\cite{2004ASPC..323..371D}, but they accounted for perturbations in
the Oort cloud due to passing stars, which we neglected in our
calculations.  With our derived re-entry rate of comets within 1 au of the Sun of $\sim 6 \times 10^{-13}$\,yr$^{-1}$,
and a total number of $\sim 10^{12}$ cometary bodies in the Oort
cloud, we derive a rate of about two new comets every 3 yr.  Even
when taking the tidal field and passing stars into account, our
derived rate is somewhat on the low side compared to
observations. The presence of a planet in the Oort cloud would pose an
interesting possibility in increasing the cometary influx rate
\citep{2014Icar..231..110F,2020DPS....5230407I,2021ApJ...910L..20B}.

\subsection{Interstellar comets}

According to the simulations presented here, the Solar System is a
copious polluter of interstellar space
\citep{2019A&A...629A.139T,2021arXiv210406845P}; interstellar comets
produced in this way could have characteristics similar to 'Oumuamua
or Borisov \citep{2020MNRAS.492..268H}.  This pollution predominantly
happens in four rather distinct phases in the evolution of the Solar System: (A) while being a cluster member, when encounters between stars
  cause debris from the circumstellar disk to escape;
(B) while the giant planets (Jupiter and Saturn) eject most
  of the asteroids within their gravitational influence;
(C) after the Sun ascapes the cluster, and the ice giant planets
  (Uranus and Neptune) start kicking out asteroids; and
(D) eventually due to the Sun's copious mass loss while
  ascending the asymptotic giant branch and evolving into a white
  dwarf, causing asteroids to become unbound
  \citep{2011MNRAS.417.2104V,2012MNRAS.422.1648V,2014MNRAS.437.1127V,2020MNRAS.493.5062V}.

According to \cite{2019MNRAS.489.4311C} $20$--$80$\% (with an average
of $50$\,\%) of the circumstellar material survives the first
$100$\,Myr of its evolution in the parent cluster.  The majority of
this mass is lost through encounters with other stars.  The amount of
material lost from the Solar System in the simulation presented here
falls within this range, meaning that the circumstellar disk has lost about
half of its mass due to interactions with other stars in the parent
cluster, or about $100$\,M$_{\oplus}$ to $3000$\,M$_{\oplus}$.  Each
of the other processes results in a mass loss of roughly $20$\%.  A
small fraction of the ejected asteroids acquire bound orbits in the
Oort cloud.

With an estimated mass of the Oort cloud of $\sim 10^{13}$\,kg/comet,
and a Hill radius of $\sim 0.65$\,pc, we arrive at a density of $\sim
10^{14}$\,kg/pc$^{3}$. If the Solar System had ejected $\sim 90$\,\%
of its asteroids into interstellar space rather than forming an Oort
cloud, interstellar space would have an average density of a few times
$\sim 10^{14}$\,kg/pc$^{3}$. This estimate is consistent with earlier
estimates for the interstellar density of interstellar objects
\citep{2018MNRAS.479L..17P,2018ApJ...855L..10D,2019NatAs...3..594O}

Regretfully, tracking a single interstellar interloper back to its
origin is hindered by the uncertainties in the orbital parameters and
the positions of stars in the past. Such tracking is limited to
at most a few tens of millions of years \citep{2018ApJ...852L..13Z}.
However, if multiple objects were found to have the same origin,
tracking the orbit back in time could lead to a more precise determination of
the original launching point \citep{2021A&A...647A.136P}. However, according to
our calculations,  most objects will have been ejected at an
early stage while planets are still forming and migrating \citep[see
also][]{2019A&A...623A.169P}. Possibly the star is even still a
cluster member, in which case internal scattering with other stars
will make it hard to trace the interstellar object  back to a single
star.

\subsection{Discussion}

\subsubsection{Interpretations and caveats in the numerical approach}\label{Sect:Caveates}

Numerical simulations are always affected by the choice of initial
conditions and by numerical errors; either from the discretization of
the underlying differential equations or by the exponential growth of
round-off in the least significant digit. There is also the possibility of  glitches or "bugs" in the simulations.

These uncertainties and errors become a particular potential cause of
concern in chaotic systems. The outer parts of the Solar System, for
example, have an estimated Lyapunov timescale ranging from
$\aplt 20$\,Myr
\citep{1986AJ.....92..176A,1988Sci...241..433S,1990Icar...88..266L,1992Sci...257...56S,1999Sci...283.1877M,2013CeMDA.116..141F}
to as much as 50\,Myr \citep{2011A&A...532A..89L} or even exceeding
the age of the Solar System
\citep{1999Icar..140..341G,2003ApJ...592..620V}.

Choices made in the initial conditions have considerable consequences
for the numerical results and their interpretation. At the same time,
numerical errors will also have repercussions for the conclusions in
Section\,\ref{Sect:Conclusions}. Similar caveats in numerical work
were acknowledged by Charles Babbage (1791-1871), the inventor of the
difference and the analytical engines.  Aware of these and other
limitations, we describe the caveats in our approach in this section
and discuss how to mitigate them.

\subsubsection{Integration, symplecticity, and numerical errors}\label{Sect:chaos}

Newton's equations of motion are intrinsically chaotic. Planetary
systems are therefore also chaotic, as in the Solar System
\citep{2001ARA&A..39..581L,2004ASPC..316...45V}. Chaos in the outer
parts of the Solar System manifests itself as phase-bound or stable
chaos \citep{1992Natur.357..569M,2019A&A...629A..95S} and although it
affects the stability of the Solar System \citep{2020SciA....6.1313T},
it will not lead to a global unstable configuration
\citep{2007PASJ...59..989T}. Chaos in the Solar System is driven by
resonance overlap \citep{1999Sci...283.1877M,2001Natur.410..773M}.
Even the Oort cloud is chaotic \citep{2019A&A...629A..95S}.

Such chaos hinders the integration of the equations of motion for the
Solar System; it renders these simulations notoriously unreliable for
individual orbits. Possibly statistically they are trustworthy,
meaning that the integrated bodies preserve the same phase space, but
this has never been tested \citep[for a discussion
see][]{2018CNSNS..61..160P}.

In this paper and many others, we rely on the statistical ensemble,
averaging by numerical integration. It is widely suggested that, in
these simulations, the conservation of energy is sufficient to
preserve the final orbital parameter space. Symplectic integration
methods are developed with precisely this objective, to preserve
energy over secular timescales. Therefore, these methods are popular
for integrating planetary systems \citep{2002MNRAS.336..483I}, but not
for integrating star clusters or galaxies.

Although details in our calculations will depend on precise initial
realization of the giant planets, the general results remain
unaffected.  We tested the effect of rather drastic variations of the
initial conditions and the effect of subtle changes (such as the 
epoch of the Solar System), but we did not explore these systematically.  The
various simulation codes to perform this study are made public via the
AMUSE framework, and we encourage others to continue exploring the
parameter space.

\section{Summary and conclusions}\label{Sect:Conclusions}

In our simulations, shortly after their birth, Jupiter and Saturn
start to eject local asteroids from the circumstellar disk along the
``conveyor belt''. Asteroids in the conveyor belt cross the orbit of
the giant planets near their pericenter. Each time this happens, they
receive a small kick causing their orbits to drift to higher
eccentricity and larger semi-major axis while preserving pericenter
distance. This process has been studied for asteroids whose pericenter
is near Jupiter \citep{1987AJ.....94.1330D,1997Icar..129..106F}. We
confirm that Jupiter, with its relatively high mass and short orbital
period, clears the local asteroid field in a few million years
\citep{1997Icar..129..106F,2019A&A...623A.169P}. In
figure\,\ref{fig:asteroid_ejection}, we illustrate this process by
presenting one calculation where a single asteroid is ejected along
the conveyor belt to reach the Oort-cloud region in a few million
years. However, for the ice giants with their smaller mass and wider
orbits, the process takes up to $\sim 100$\,Myr
\citep{1981A&A....96...26F,2013Icar..222...20F,2020arXiv200609657D},
and if the asteroid has an inclined orbit, the process may last a
gigayear.

While a member of a star cluster, an asteroid in a wide $\apgt
10^3$\,au orbit is easily lost from the Solar System. A small
perturbation in relative velocity of $\delta v \equiv dv/v \geq {\cal
  O}(10^{-4})$ is sufficient to unbind an asteroid.  It turns out
that asteroids on an eccentric orbit $\geq 0.98$ and with a semi-major
axis of $\apgt 2400$\,au are vulnerable to being stripped from the Solar System. The period of such a wide orbit ($\apgt 0.1$\,Myr) is
comparable to the mean encounter time between the Sun and another star
in the parent cluster (assuming that the encountering star induces a
similar perturbation in relative velocity). In the first million years
after their formation, the giant planets launch $\sim 80$\,\% of the
asteroids between 3 and 15\,au onto the conveyor belt (see
Sect.\,\ref{Sect:Methods_Summary}). Once there, the eccentricity and
orbital separation of the asteroids rapidly increases.

The majority of the asteroids ($\sim 57$\,\%) are moving along trajectories that
become so wide that they are stripped within a few orbits, and within $\sim 10$\,Myr most
of the asteroids born near Jupiter and Saturn are either parked in
resonant orbits (such as the Hilda family of asteroids) or their
orbits have become so wide that they are easily stripped by passing
stars in the parent cluster, and escape in about $20$ to $50$\,Myr.
The Oort cloud therefore cannot have formed in the first 10\,Myr by
Jupiter and Saturn ejecting asteroids because their ejection timescale
is smaller than the timescale on which the Solar System is expected to
be ejected from the parent cluster ($\aplt 100$\,Myr).  This statement
is unaffected by the few-million-years formation timescale for the giant
planets.

Before escaping the cluster, stars in our simulations experience
multiple encounters. The dynamic signature of multiple encounters on
the outer regions of a planetary system is distinctly different from
that of a single strong encounter \citep{2019MNRAS.490...21H}. The
outer parts of the Solar System have signatures of both: A single
strong encounter with another star can explain the Kuiper-cliff
\citep{2014MNRAS.444.2808P} and the orbit of Sedna
\citep{2011epsc.conf..633S,2015MNRAS.453.3157J}, but the complex
distribution of orbital parameters in the scattered Kuiper-belt beyond
$\sim 45$\,au \citep{2005Icar..177..246K} seems best explained with a
series of relatively weak encounters \cite[see][and
Sect.\,\ref{Sect:Scattered_asteroids}]{2012Icar..217....1B,
  2017A&A...599A..91B, 2019MNRAS.490...21H,2020ApJ...901...92M}.  We
therefore argue that the Solar System was affected by multiple
encounters in its parent cluster, until a strong encounter caused it
to escape.

In our simulations, multiple interactions lead to $\sim 5$\,\% of the
asteroids in the outer disk beyond 45\,au being scattered into the
conveyor belt region (see
Sect.\,\ref{Sect:Scattered_asteroids}), but the Sun also captures
material from the disks of the stars it encounters. The efficiencies
at which material is transported from one star to another is rather
symmetric \citep{2016A&A...594A..53B}. The scattered and captured
populations are comparable in number, but populate different regions
in orbital parameter-space around the Sun.

In a reconstruction of the encounter that brought Sedna into the Solar System as a captured asteroid from the disk of another star, $\sim
28$\% of the captured asteroids are injected directly into the
conveyor belt region \citep[see figure\,1 of][and
  Sect.\,\ref{Sect:Methods_Summary}]{2015MNRAS.453.3157J}. The high
inclination of the scattered and captured populations compared to the
native disk causes them to interact less efficiently with the giant
planets. These asteroids reach the Oort cloud on a timescale of $\sim
1$\,Gyr. Because of this long timescale, these objects remain in the parking
zone of the Solar System where they are protected against ejection caused by passing stars.  Once the Sun is isolated, asteroids in the
conveyor belt escape the Solar System only if their apocenter
distance exceeds the Hill radius of the Solar System in the Galactic
potential, at $\sim 0.65$\,pc, or $\delta v \apgt 0.1$.

Before an asteroid reaches such a wide orbit, it is subject to
eccentricity damping by the Galactic tidal field via von
Zeipel-Lidov-Kozai oscillations
\citep{1910AN....183..345V,1962PSS..9..719L,1962AJ.....67..591K}.  The
relatively high inclination of the ecliptic to the Galactic plane (of
$\sim 60^\circ$) helps to circularize their orbits and randomize their
inclinations~\citep{2007AJ....134.1693H}.  The von Zeipel-Lidov-Kozai
process causes the pericenter of an asteroid's orbit to detach from
the planetary region \citep{2007AJ....134.1693H}, preventing it from
being kicked out by a planet. Both processes, namely the ejection by the
giant planets and the eccentricity damping by the tidal field,
are essential ingredients that operate on comparable timescales. Once
the Galactic tidal field reduces the eccentricity below $0.998$ and
the semi-major axis $\apgt 20\,000$\,au ($a(1-e)\apgt 50$\,au, to the
right of the dashed curve in figure\,\ref{fig:OCformation}), the
driving force that pumps an asteroid's orbit further into the
conveyor belt switches off.  By this time, the asteroid is detached
from the inner Solar System.

The ejection of the native scattered, and captured asteroids along the
conveyor belt, and their subsequent eccentricity damping by the
Galactic tidal field lead to the formation of the Oort cloud some
100\,Myr after the Sun escaped the parent cluster
\citep{2008Icar..197..221K}.  The effects of passing stars in the
parent cluster and planet--asteroid interactions together with the
Galactic tidal field are essential ingredients that contribute to
defining the orbital structure of the outer Kuiper belt, the Hill
surface, and the Oort cloud.  In figure\,\ref{fig:OCformation}, we
illustrate the orbital parameter space at this instance (see also
Sect.\,\ref{Sect:Methods_Summary}).

The conveyor belt beyond the ice-giant planets is further depleted in
the next few $100$\,Myr, resulting in the clearance of the asteroids
beyond Saturn's orbit and the provision of material to the Oort cloud. This
could explain the dearth of Centaurs in the Solar System without
requiring a chaotic reorganization of the ice-giant planets
\citep{2014Icar..231...99F}.

Based on our calculations, $\sim 5$\,\% of the asteroids in the Oort
cloud originate from the outer disk beyond $\sim 45$\,au.  About
one-third ($\sim 30$\%) are captured from another star, and the rest
originate from the circumstellar disk between $\sim 15$\,au and
$45$\,au. The majority of scattered and captured asteroids never make
it to the Oort cloud, but linger around in the parking zone, between
the two solid curves indicated with $\delta v \sim 10^{-8}$ and
$\delta v \sim 10^{-5}$ in Figs.\,\ref{fig:OCformation} and
\,\ref{fig:ae_simulations_scattered_and_sednitos}, or scape the Solar
System. These captured objects should still be there and can be
identified based on their unorthodox orbits and composition.  The
difference in the color of Kuiper belt and Oort cloud objects
indicates that they are not related \citep{2002AJ....123.1039J}.  Once
in the Oort cloud, the origin of an asteroid can be established by
studying its kinematics, because the phase-mixing driven by the
chaotic process that injected it into the Oort cloud manifests itself
on a timescale longer than the Solar System's lifetime. The outer
Solar System origin of these asteroids can possibly be established by
spectral analysis, much in the same way as C/2019 Q4 (Borisov) was
analyzed using the OSIRIS \citep{2019RNAAS...3..131D} and MUSE
\citep{2020arXiv200111605B} instruments. One remarkable finding is the
depletion of C$_{2}$ \citep{2019A&A...631L...8O} and NH$_{2}$ in this
object \citep{2020arXiv200111605B}.

The scenario for the formation of the Oort cloud presented here is
relatively insensitive to the details of the orbits of the outer
planets.  Asteroids can be launched into the conveyor belt to reach
the Oort cloud from the current orbital distribution of the giant
planets, or from a chaotic migration of the ice-giant
planets. However, according to our simulations, if any chaotic
reorganization or migration of the giant planets happens before the
Sun escapes the parent cluster, the Oort cloud contains considerably
fewer objects because most asteroids escape; in addition, the orbital
distribution of objects in the Oort cloud would differ
\citep{2019MNRAS.485.5511S}.  We continued our calculations for
1\,Gyr, but already after a few hundred million years the Oort cloud
is fully formed, and the eccentricity distribution approaches a
probability distribution of $f(e) \propto (1-e^2)^{-2/9}$ (see
sect.\,\ref{Sect:Captured_asteroids}) with a density distribution
$\propto r^{-3}$ (see sect.\,\ref{Sect:Discussion}).

By this time, the Solar System is surrounded by a cloud of unbound
asteroids which co-move along the Sun's orbit in the Galaxy.  This
co-moving group contains $\apgt 10^{11}$ comet-mass asteroids, most of
which originated from the inner Solar System. This unbound population
is illustrated in figure\,\ref{fig:Galactic_orbit}. If other stars
have Oort clouds of their own, the Solar System moves through a sea of
Oort-cloud objects that originally belonged to other stars \citep[see
also][]{Torres2020,2021A&A...647A.136P}.

The Oort cloud gradually evaporates with a half-life of about 1\,Gyr
due to the tidal field's kinematic heating and passing stars. Lost
objects become free-floating asteroids in interstellar space
\citep{2019A&A...629A.139T}, much like 'Oumuamua and Borisov.

Interaction with the Galactic tides also causes asteroids to be
launched into the inner Solar System, where they can be observed as
comets. In our simulations, the rate at which Oort cloud objects
re-enter the inner Solar System (to within 1\,au) is $0.2$---$0.6$ per
year (see Sect.\,\ref{Sect:retrogade_orbits}), which is somewhat
lower than the empirical estimate of $0.86\pm0.07$ per year
\citep{1999Icar..137..355L}.  Nevertheless, here, we ignored the possible
ionizing effect of passing stars in the Galactic potential.

\section*{Public data}
The source code, input files, and simulation data for this manuscript
are available at {\tt 10.6084/m9.figshare.13214471}.  A tutorial on
how to run the various codes in AMUSE is available at
\url{https://github.com/spzwart/AMUSE-Tutorial}.

\section*{Acknowledgments}

It is a pleasure to thank Diptajyoti Mukherjee, Fransisca
Concha-Ram{\'\i}rez, $\ln(a)$ Sellentin, Julia Wasala, Eiichiro
Kokubo, Ramon Brasser and Tim de Zeeuw for discussions.  We also thank
the anonymous referee(s) for valuable comments on on the manuscript.
We are grateful for the support of the Mexican National Council for
Science and Technology (CONACYT) grant \#291004-410780.  This work was
performed using resources provided by the Academic Leiden
Interdisciplinary Cluster Environment (ALICE), LGM-II (NWO grant \#
621.016.701) and the Dutch national e-infrastructure with the use of
the Dutch national supercomputer Cartesius, and the support of SURF
Cooperative.

\subsection*{Energy consumption of this calculation}

Being concerned about the polluting influence of our science
\citep{2020NatAs...4..823B,2020NatAs...4..819P} we would like to raise
awareness of the environmental impact of our calculations. We run
AMUSE for about 40000 single-core CPU hours.  This results in about
2MWh of electricity \url{http://green-algorithms.org/}) being consumed
by the Dutch National supercomputer.  With our estimate of the
proportion of green electricity used, this process produces
$\sim 530$\,kg CO2, which is comparable to driving a car from Leiden
to Kars (about 4000km).  However, the same data were used in
additional research, and we argue that less than half the CO2 produced
in these calculations should be attributed to this paper.

\subsection*{Software used for this study}

This work would have been impossible without the following public open
source packages and libraries: Python \citep{vanRossum:1995:EEP},
matplotlib \citep{2007CSE.....9...90H}, numpy
\citep{Oliphant2006ANumPy}, MPI \citep{Gropp:1996:HPI,Gropp2002}, ABIE
\citep[][see
\url{https://github.com/MovingPlanetsAround/ABIE}]{ABIE2018}, {\tt
  NBODY6++GPU} \citep{2015MNRAS.450.4070W} and {\tt REBOUND}
\citep{2012A&A...537A.128R}, {\tt SeBa}
\citep{1996A&A...309..179P,2016ComAC...3....6T}, and AMUSE
\citep[][available for download at
\url{https://amusecode.org}]{portegies_zwart_simon_2018_1443252}.


\begin{thebibliography}{}
\expandafter\ifx\csname natexlab\endcsname\relax\def\natexlab#1{#1}\fi

\bibitem[{{Adams}(2010)}]{2010ARA&A..48...47A}
{Adams}, F.~C. 2010, \araa, 48, 47

\bibitem[{{Adams} {et~al.}(2004){Adams}, {Hollenbach}, {Laughlin}, \&
  {Gorti}}]{2004ApJ...611..360A}
{Adams}, F.~C., {Hollenbach}, D., {Laughlin}, G., \& {Gorti}, U. 2004, \apj,
  611, 360

\bibitem[{{Adams} {et~al.}(2006){Adams}, {Proszkow}, {Fatuzzo}, \&
  {Myers}}]{2006ApJ...641..504A}
{Adams}, F.~C., {Proszkow}, E.~M., {Fatuzzo}, M., \& {Myers}, P.~C. 2006, \apj,
  641, 504

\bibitem[{{Alexandersen} {et~al.}(2019){Alexandersen}, {Benecchi}, {Chen},
  {Eduardo}, {Thirouin}, {Schwamb}, {Lehner}, {Wang}, {Bannister}, {Gladman},
  {Gwyn}, {Kavelaars}, {Petit}, \& {Volk}}]{2019ApJS..244...19A}
{Alexandersen}, M., {Benecchi}, S.~D., {Chen}, Y.-T., {et~al.} 2019, \apjs,
  244, 19

\bibitem[{{Allen}(1973)}]{1973asqu.book.....A}
{Allen}, C.~W. 1973, {Astrophysical quantities}

\bibitem[{{Alvarez} \& {Muller}(1984)}]{1984Natur.308..718A}
{Alvarez}, W. \& {Muller}, R.~A. 1984, \nat, 308, 718

\bibitem[{{Applegate} {et~al.}(1986){Applegate}, {Douglas}, {Gursel},
  {Sussman}, \& {Wisdom}}]{1986AJ.....92..176A}
{Applegate}, J.~H., {Douglas}, M.~R., {Gursel}, Y., {Sussman}, G.~J., \&
  {Wisdom}, J. 1986, \aj, 92, 176

\bibitem[{{Astakhov} {et~al.}(2005){Astakhov}, {Lee}, \&
  {Farrelly}}]{2005MNRAS.360..401A}
{Astakhov}, S.~A., {Lee}, E.~A., \& {Farrelly}, D. 2005, \mnras, 360, 401

\bibitem[{{Bailer-Jones}(2009)}]{2009IJAsB...8..213B}
{Bailer-Jones}, C.~A.~L. 2009, International Journal of Astrobiology, 8, 213

\bibitem[{{Bannister} {et~al.}(2020){Bannister}, {Opitom}, {Fitzsimmons},
  {Moulane}, {Jehin}, {Seligman}, {Rousselot}, {Knight}, {Marsset}, {Schwamb},
  {Guilbert-Lepoutre}, {Jorda}, {Vernazza}, \&
  {Benkhaldoun}}]{2020arXiv200111605B}
{Bannister}, M.~T., {Opitom}, C., {Fitzsimmons}, A., {et~al.} 2020, arXiv
  e-prints, arXiv:2001.11605

\bibitem[{{Batygin} {et~al.}(2019){Batygin}, {Adams}, {Brown}, \&
  {Becker}}]{2019PhR...805....1B}
{Batygin}, K., {Adams}, F.~C., {Brown}, M.~E., \& {Becker}, J.~C. 2019,
  \physrep, 805, 1

\bibitem[{{Batygin} \& {Brown}(2016)}]{2016AJ....151...22B}
{Batygin}, K. \& {Brown}, M.~E. 2016, \aj, 151, 22

\bibitem[{{Batygin} \& {Brown}(2021)}]{2021ApJ...910L..20B}
{Batygin}, K. \& {Brown}, M.~E. 2021, \apjl, 910, L20

\bibitem[{{Baumgardt} \& {Kroupa}(2007)}]{2007MNRAS.380.1589B}
{Baumgardt}, H. \& {Kroupa}, P. 2007, \mnras, 380, 1589

\bibitem[{{Beckwith} {et~al.}(1990){Beckwith}, {Sargent}, {Chini}, \&
  {Guesten}}]{1990AJ.....99..924B}
{Beckwith}, S. V.~W., {Sargent}, A.~I., {Chini}, R.~S., \& {Guesten}, R. 1990,
  \aj, 99, 924

\bibitem[{{Bernardinelli} {et~al.}(2020){Bernardinelli}, {Bernstein}, {Sako},
  {Hamilton}, {Gerdes}, {Adams}, {Saunders}, {Aguena}, {Allam}, {Avila},
  {Brooks}, {Diehl}, {Doel}, {Everett}, {Garc{\'\i}a-Bellido}, {Gaztanaga},
  {Gruendl}, {Honscheid}, {Ogando}, {Palmese}, {Tucker}, {Walker}, {Wester}, \&
  {(The DES Collaboration)}}]{2020PSJ.....1...28B}
{Bernardinelli}, P.~H., {Bernstein}, G.~M., {Sako}, M., {et~al.} 2020, The
  Planetary Science Journal, 1, 28

\bibitem[{{Bhandare} {et~al.}(2016){Bhandare}, {Breslau}, \&
  {Pfalzner}}]{2016A&A...594A..53B}
{Bhandare}, A., {Breslau}, A., \& {Pfalzner}, S. 2016, \aap, 594, A53

\bibitem[{Boehnke \& Harrison(2016)}]{Boehnke10802}
Boehnke, P. \& Harrison, T.~M. 2016, Proceedings of the National Academy of
  Sciences, 113, 10802

\bibitem[{{Boehnke} \& {Harrison}(2018)}]{2018LPICo2107.2033B}
{Boehnke}, P. \& {Harrison}, T.~M. 2018, LPI Contributions, 2107, 2033

\bibitem[{{Bottke} {et~al.}(2010){Bottke}, {Nesvorn{\'y}}, {Vokrouhlick{\'y}},
  \& {Morbidelli}}]{2010AJ....139..994B}
{Bottke}, W.~F., {Nesvorn{\'y}}, D., {Vokrouhlick{\'y}}, D., \& {Morbidelli},
  A. 2010, \aj, 139, 994

\bibitem[{{Bottke} \& {Norman}(2017)}]{2017AREPS..45..619B}
{Bottke}, W.~F. \& {Norman}, M.~D. 2017, Annual Review of Earth and Planetary
  Sciences, 45, 619

\bibitem[{{Brasser}(2008)}]{2008A&A...492..251B}
{Brasser}, R. 2008, \aap, 492, 251

\bibitem[{Brasser {et~al.}(2007)Brasser, Duncan, \& Levison}]{BRASSER2007413}
Brasser, R., Duncan, M., \& Levison, H. 2007, Icarus, 191, 413

\bibitem[{{Brasser} {et~al.}(2006){Brasser}, {Duncan}, \&
  {Levison}}]{2006Icar..184...59B}
{Brasser}, R., {Duncan}, M.~J., \& {Levison}, H.~F. 2006, \icarus, 184, 59

\bibitem[{{Brasser} {et~al.}(2012){Brasser}, {Duncan}, {Levison}, {Schwamb}, \&
  {Brown}}]{2012Icar..217....1B}
{Brasser}, R., {Duncan}, M.~J., {Levison}, H.~F., {Schwamb}, M.~E., \& {Brown},
  M.~E. 2012, \icarus, 217, 1

\bibitem[{{Brasser} {et~al.}(2010){Brasser}, {Higuchi}, \&
  {Kaib}}]{2010A&A...516A..72B}
{Brasser}, R., {Higuchi}, A., \& {Kaib}, N. 2010, \aap, 516, A72

\bibitem[{{Brasser} \& {Schwamb}(2015)}]{2015MNRAS.446.3788B}
{Brasser}, R. \& {Schwamb}, M.~E. 2015, \mnras, 446, 3788

\bibitem[{{Breslau} {et~al.}(2014){Breslau}, {Steinhausen}, {Vincke}, \&
  {Pfalzner}}]{2014A&A...565A.130B}
{Breslau}, A., {Steinhausen}, M., {Vincke}, K., \& {Pfalzner}, S. 2014, \aap,
  565, A130

\bibitem[{{Breslau} {et~al.}(2017){Breslau}, {Vincke}, \&
  {Pfalzner}}]{2017A&A...599A..91B}
{Breslau}, A., {Vincke}, K., \& {Pfalzner}, S. 2017, \aap, 599, A91

\bibitem[{{Brown}(2001)}]{2001AJ....121.2804B}
{Brown}, M.~E. 2001, \aj, 121, 2804

\bibitem[{{Brown} {et~al.}(2004){Brown}, {Trujillo}, \&
  {Rabinowitz}}]{2004ApJ...617..645B}
{Brown}, M.~E., {Trujillo}, C., \& {Rabinowitz}, D. 2004, \apj, 617, 645

\bibitem[{{Brucker} {et~al.}(2009){Brucker}, {Grundy}, {Stansberry}, {Spencer},
  {Sheppard}, {Chiang}, \& {Buie}}]{2009Icar..201..284B}
{Brucker}, M.~J., {Grundy}, W.~M., {Stansberry}, J.~A., {et~al.} 2009, \icarus,
  201, 284

\bibitem[{{Brunini}(2019)}]{2019MNRAS.483.5042B}
{Brunini}, A. 2019, \mnras, 483, 5042

\bibitem[{{Burtscher} {et~al.}(2020){Burtscher}, {Barret}, {Borkar},
  {Grinberg}, {Jahnke}, {Kendrew}, {Maffey}, \&
  {McCaughrean}}]{2020NatAs...4..823B}
{Burtscher}, L., {Barret}, D., {Borkar}, A.~P., {et~al.} 2020, Nature
  Astronomy, 4, 823

\bibitem[{{Byl}(1986)}]{1986EM&P...36..263B}
{Byl}, J. 1986, Earth Moon and Planets, 36, 263

\bibitem[{Cai(2018)}]{ABIE2018}
Cai, M. 2018, Moving Planets Around (MPA) Project

\bibitem[{{Cai} {et~al.}(2017){Cai}, {Kouwenhoven}, {Portegies Zwart}, \&
  {Spurzem}}]{2017MNRAS.470.4337C}
{Cai}, M.~X., {Kouwenhoven}, M.~B.~N., {Portegies Zwart}, S.~F., \& {Spurzem},
  R. 2017, \mnras, 470, 4337

\bibitem[{{Cai} {et~al.}(2019){Cai}, {Portegies Zwart}, {Kouwenhoven}, \&
  {Spurzem}}]{2019MNRAS.489.4311C}
{Cai}, M.~X., {Portegies Zwart}, S., {Kouwenhoven}, M.~B.~N., \& {Spurzem}, R.
  2019, \mnras, 489, 4311

\bibitem[{{Cai} {et~al.}(2018){Cai}, {Portegies Zwart}, \& {van
  Elteren}}]{2018MNRAS.474.5114C}
{Cai}, M.~X., {Portegies Zwart}, S., \& {van Elteren}, A. 2018, \mnras, 474,
  5114

\bibitem[{{Calura} {et~al.}(2020){Calura}, {Bellazzini}, \&
  {D'Ercole}}]{2020MNRAS.499.5873C}
{Calura}, F., {Bellazzini}, M., \& {D'Ercole}, A. 2020, \mnras, 499, 5873

\bibitem[{{Chebotarev}(1965)}]{1965SvA.....8..787C}
{Chebotarev}, G.~A. 1965, \sovast, 8, 787

\bibitem[{{Chiang} {et~al.}(2003){Chiang}, {Jordan}, {Millis}, {Buie},
  {Wasserman}, {Elliot}, {Kern}, {Trilling}, {Meech}, \&
  {Wagner}}]{2003AJ....126..430C}
{Chiang}, E.~I., {Jordan}, A.~B., {Millis}, R.~L., {et~al.} 2003, \aj, 126, 430

\bibitem[{{Clarke}(2007)}]{2007MNRAS.376.1350C}
{Clarke}, C.~J. 2007, \mnras, 376, 1350

\bibitem[{{Clarke} \& {Pringle}(1993)}]{1993MNRAS.261..190C}
{Clarke}, C.~J. \& {Pringle}, J.~E. 1993, \mnras, 261, 190

\bibitem[{{Concha-Ram{\'\i}rez} {et~al.}(2021){Concha-Ram{\'\i}rez}, {Wilhelm},
  {Portegies Zwart}, {van Terwisga}, \& {Hacar}}]{2021MNRAS.501.1782C}
{Concha-Ram{\'\i}rez}, F., {Wilhelm}, M. J.~C., {Portegies Zwart}, S., {van
  Terwisga}, S.~E., \& {Hacar}, A. 2021, \mnras, 501, 1782

\bibitem[{{Correa-Otto} \& {Calandra}(2019)}]{2019MNRAS.490.2495C}
{Correa-Otto}, J.~A. \& {Calandra}, M.~F. 2019, \mnras, 490, 2495

\bibitem[{{Crida}(2009)}]{2009ApJ...698..606C}
{Crida}, A. 2009, \apj, 698, 606

\bibitem[{{Crida} {et~al.}(2007){Crida}, {Morbidelli}, \&
  {Tsiganis}}]{2007epsc.conf..866C}
{Crida}, A., {Morbidelli}, A., \& {Tsiganis}, K. 2007, in European Planetary
  Science Congress 2007, 866

\bibitem[{{Cuello} {et~al.}(2019){Cuello}, {Dipierro}, {Mentiplay}, {Price},
  {Pinte}, {Cuadra}, {Laibe}, {M{\'e}nard}, {Poblete}, \&
  {Montesinos}}]{2019MNRAS.483.4114C}
{Cuello}, N., {Dipierro}, G., {Mentiplay}, D., {et~al.} 2019, \mnras, 483, 4114

\bibitem[{{D'Angelo} {et~al.}(2021){D'Angelo}, {Weidenschilling}, {Lissauer},
  \& {Bodenheimer}}]{2021Icar..35514087D}
{D'Angelo}, G., {Weidenschilling}, S.~J., {Lissauer}, J.~J., \& {Bodenheimer},
  P. 2021, \icarus, 355, 114087

\bibitem[{{Davis} {et~al.}(1984){Davis}, {Hut}, \&
  {Muller}}]{1984Natur.308..715D}
{Davis}, M., {Hut}, P., \& {Muller}, R.~A. 1984, \nat, 308, 715

\bibitem[{{de Le{\'o}n} {et~al.}(2019){de Le{\'o}n}, {Licandro},
  {Serra-Ricart}, {Cabrera-Lavers}, {Font Serra}, {Scarpa}, {de la Fuente
  Marcos}, \& {de la Fuente Marcos}}]{2019RNAAS...3..131D}
{de Le{\'o}n}, J., {Licandro}, J., {Serra-Ricart}, M., {et~al.} 2019, Research
  Notes of the American Astronomical Society, 3, 131

\bibitem[{{Delbo'} {et~al.}(2017){Delbo'}, {Walsh}, {Bolin}, {Avdellidou}, \&
  {Morbidelli}}]{2017Sci...357.1026D}
{Delbo'}, M., {Walsh}, K., {Bolin}, B., {Avdellidou}, C., \& {Morbidelli}, A.
  2017, Science, 357, 1026

\bibitem[{{Di Sisto} \& {Rossignoli}(2020)}]{2020arXiv200609657D}
{Di Sisto}, R.~P. \& {Rossignoli}, N.~L. 2020, arXiv e-prints, arXiv:2006.09657

\bibitem[{{Do} {et~al.}(2018){Do}, {Tucker}, \& {Tonry}}]{2018ApJ...855L..10D}
{Do}, A., {Tucker}, M.~A., \& {Tonry}, J. 2018, \apjl, 855, L10

\bibitem[{{Dones} {et~al.}(2015){Dones}, {Brasser}, {Kaib}, \&
  {Rickman}}]{2015SSRv..197..191D}
{Dones}, L., {Brasser}, R., {Kaib}, N., \& {Rickman}, H. 2015, \ssr, 197, 191

\bibitem[{{Dones} {et~al.}(2000){Dones}, {Levison}, {Duncan}, \&
  {Weissman}}]{2000DPS....32.3602D}
{Dones}, L., {Levison}, H., {Duncan}, M., \& {Weissman}, P. 2000, in
  AAS/Division for Planetary Sciences Meeting Abstracts, Vol.~32, AAS/Division
  for Planetary Sciences Meeting Abstracts \#32, 36.02

\bibitem[{{Dones} {et~al.}(2004{\natexlab{a}}){Dones}, {Weissman}, {Levison},
  \& {Duncan}}]{2004ASPC..323..371D}
{Dones}, L., {Weissman}, P.~R., {Levison}, H.~F., \& {Duncan}, M.~J.
  2004{\natexlab{a}}, in Astronomical Society of the Pacific Conference Series,
  Vol. 323, Star Formation in the Interstellar Medium: In Honor of David
  Hollenbach, ed. D.~{Johnstone}, F.~C. {Adams}, D.~N.~C. {Lin}, D.~A.
  {Neufeeld}, \& E.~C. {Ostriker}, 371

\bibitem[{{Dones} {et~al.}(2004{\natexlab{b}}){Dones}, {Weissman}, {Levison},
  \& {Duncan}}]{2004come.book..153D}
{Dones}, L., {Weissman}, P.~R., {Levison}, H.~F., \& {Duncan}, M.~J.
  2004{\natexlab{b}}, {Oort cloud formation and dynamics}, ed. M.~C. {Festou},
  H.~U. {Keller}, \& H.~A. {Weaver}, 153

\bibitem[{{Drimmel}(2000)}]{2000A&A...358L..13D}
{Drimmel}, R. 2000, \aap, 358, L13

\bibitem[{{Duncan} {et~al.}(1987){Duncan}, {Quinn}, \&
  {Tremaine}}]{1987AJ.....94.1330D}
{Duncan}, M., {Quinn}, T., \& {Tremaine}, S. 1987, \aj, 94, 1330

\bibitem[{{Duncan} {et~al.}(2011){Duncan}, {Babcock}, {Kaib}, \&
  {Levison}}]{2011DDA....42.0903D}
{Duncan}, M.~J., {Babcock}, C., {Kaib}, N., \& {Levison}, H. 2011, in
  AAS/Division of Dynamical Astronomy Meeting \#42, AAS/Division of Dynamical
  Astronomy Meeting, 9.03

\bibitem[{{Duncan} {et~al.}(1995){Duncan}, {Levison}, \&
  {Budd}}]{1995AJ....110.3073D}
{Duncan}, M.~J., {Levison}, H.~F., \& {Budd}, S.~M. 1995, \aj, 110, 3073

\bibitem[{{Edgeworth}(1943)}]{1943JBAA...53..181E}
{Edgeworth}, K.~E. 1943, Journal of the British Astronomical Association, 53,
  181

\bibitem[{{Elliot} {et~al.}(2005){Elliot}, {Kern}, {Clancy}, {Gulbis},
  {Millis}, {Buie}, {Wasserman}, {Chiang}, {Jordan}, {Trilling}, \&
  {Meech}}]{2005AJ....129.1117E}
{Elliot}, J.~L., {Kern}, S.~D., {Clancy}, K.~B., {et~al.} 2005, \aj, 129, 1117

\bibitem[{{Emel'yanenko}(2020)}]{2020A&A...642L..20E}
{Emel'yanenko}, V.~V. 2020, \aap, 642, L20

\bibitem[{{Emery} {et~al.}(2015){Emery}, {Marzari}, {Morbidelli}, {French}, \&
  {Grav}}]{2015aste.book..203E}
{Emery}, J.~P., {Marzari}, F., {Morbidelli}, A., {French}, L.~M., \& {Grav}, T.
  2015, {The Complex History of Trojan Asteroids}, 203--220

\bibitem[{{Emsenhuber} {et~al.}(2020{\natexlab{a}}){Emsenhuber}, {Mordasini},
  {Burn}, {Alibert}, {Benz}, \& {Asphaug}}]{2020arXiv200705561E}
{Emsenhuber}, A., {Mordasini}, C., {Burn}, R., {et~al.} 2020{\natexlab{a}},
  arXiv e-prints, arXiv:2007.05561

\bibitem[{{Emsenhuber} {et~al.}(2020{\natexlab{b}}){Emsenhuber}, {Mordasini},
  {Burn}, {Alibert}, {Benz}, \& {Asphaug}}]{2020arXiv200705562E}
{Emsenhuber}, A., {Mordasini}, C., {Burn}, R., {et~al.} 2020{\natexlab{b}},
  arXiv e-prints, arXiv:2007.05562

\bibitem[{{Engelhardt} {et~al.}(2017){Engelhardt}, {Jedicke}, {Vere{\v s}},
  {Fitzsimmons}, {Denneau}, {Beshore}, \& {Meinke}}]{2017AJ....153..133E}
{Engelhardt}, T., {Jedicke}, R., {Vere{\v s}}, P., {et~al.} 2017, \aj, 153, 133

\bibitem[{{Farr{\'e}s} {et~al.}(2013){Farr{\'e}s}, {Laskar}, {Blanes}, {Casas},
  {Makazaga}, \& {Murua}}]{2013CeMDA.116..141F}
{Farr{\'e}s}, A., {Laskar}, J., {Blanes}, S., {et~al.} 2013, Celestial
  Mechanics and Dynamical Astronomy, 116, 141

\bibitem[{{Fernandez}(1981)}]{1981A&A....96...26F}
{Fernandez}, J.~A. 1981, \aap, 96, 26

\bibitem[{{Fern{\'a}ndez}(1997)}]{1997Icar..129..106F}
{Fern{\'a}ndez}, J.~A. 1997, \icarus, 129, 106

\bibitem[{{Fern{\'a}ndez} \& {Brunini}(2000)}]{2000Icar..145..580F}
{Fern{\'a}ndez}, J.~A. \& {Brunini}, A. 2000, \icarus, 145, 580

\bibitem[{{Fern{\'a}ndez} {et~al.}(2004){Fern{\'a}ndez}, {Gallardo}, \&
  {Brunini}}]{2004Icar..172..372F}
{Fern{\'a}ndez}, J.~A., {Gallardo}, T., \& {Brunini}, A. 2004, \icarus, 172,
  372

\bibitem[{{Fernandez} \& {Ip}(1981)}]{1981Icar...47..470F}
{Fernandez}, J.~A. \& {Ip}, W.~H. 1981, \icarus, 47, 470

\bibitem[{{Fouchard} {et~al.}(2020){Fouchard}, {Emel'yanenko}, \&
  {Higuchi}}]{2020CeMDA.132...43F}
{Fouchard}, M., {Emel'yanenko}, V., \& {Higuchi}, A. 2020, Celestial Mechanics
  and Dynamical Astronomy, 132, 43

\bibitem[{{Fouchard} {et~al.}(2006{\natexlab{a}}){Fouchard}, {Froeschl{\'e}},
  {Matese}, \& {Valsecchi}}]{2006CeMDA..96..341F}
{Fouchard}, M., {Froeschl{\'e}}, C., {Matese}, J.~J., \& {Valsecchi}, G.~B.
  2006{\natexlab{a}}, Celestial Mechanics and Dynamical Astronomy, 96, 341

\bibitem[{{Fouchard} {et~al.}(2006{\natexlab{b}}){Fouchard}, {Froeschl{\'e}},
  {Valsecchi}, \& {Rickman}}]{2006CeMDA..95..299F}
{Fouchard}, M., {Froeschl{\'e}}, C., {Valsecchi}, G., \& {Rickman}, H.
  2006{\natexlab{b}}, Celestial Mechanics and Dynamical Astronomy, 95, 299

\bibitem[{{Fouchard} {et~al.}(2018){Fouchard}, {Higuchi}, {Ito}, \&
  {Maquet}}]{2018A&A...620A..45F}
{Fouchard}, M., {Higuchi}, A., {Ito}, T., \& {Maquet}, L. 2018, \aap, 620, A45

\bibitem[{{Fouchard} {et~al.}(2013){Fouchard}, {Rickman}, {Froeschl{\'e}}, \&
  {Valsecchi}}]{2013Icar..222...20F}
{Fouchard}, M., {Rickman}, H., {Froeschl{\'e}}, C., \& {Valsecchi}, G.~B. 2013,
  \icarus, 222, 20

\bibitem[{{Fouchard} {et~al.}(2014{\natexlab{a}}){Fouchard}, {Rickman},
  {Froeschl{\'e}}, \& {Valsecchi}}]{2014Icar..231..110F}
{Fouchard}, M., {Rickman}, H., {Froeschl{\'e}}, C., \& {Valsecchi}, G.~B.
  2014{\natexlab{a}}, \icarus, 231, 110

\bibitem[{{Fouchard} {et~al.}(2014{\natexlab{b}}){Fouchard}, {Rickman},
  {Froeschl{\'e}}, \& {Valsecchi}}]{2014Icar..231...99F}
{Fouchard}, M., {Rickman}, H., {Froeschl{\'e}}, C., \& {Valsecchi}, G.~B.
  2014{\natexlab{b}}, \icarus, 231, 99

\bibitem[{{Fritz} {et~al.}(2014){Fritz}, {Bitsch}, {K{\"u}hrt}, {Morbidelli},
  {Tornow}, {W{\"u}nnemann}, {Fernandes}, {Grenfell}, {Rauer}, {Wagner}, \&
  {Werner}}]{2014P&SS...98..254F}
{Fritz}, J., {Bitsch}, B., {K{\"u}hrt}, E., {et~al.} 2014, \planss, 98, 254

\bibitem[{{Fujii} {et~al.}(2007){Fujii}, {Iwasawa}, {Funato}, \&
  {Makino}}]{2007PASJ...59.1095F}
{Fujii}, M., {Iwasawa}, M., {Funato}, Y., \& {Makino}, J. 2007, \pasj, 59, 1095

\bibitem[{{Fukushima} \& {Yajima}(2021)}]{2021arXiv210410892F}
{Fukushima}, H. \& {Yajima}, H. 2021, arXiv e-prints, arXiv:2104.10892

\bibitem[{{Gardner} {et~al.}(2011){Gardner}, {Nurmi}, {Flynn}, \&
  {Mikkola}}]{2011MNRAS.411..947G}
{Gardner}, E., {Nurmi}, P., {Flynn}, C., \& {Mikkola}, S. 2011, \mnras, 411,
  947

\bibitem[{{Gavagnin} {et~al.}(2017){Gavagnin}, {Bleuler}, {Rosdahl}, \&
  {Teyssier}}]{2017MNRAS.472.4155G}
{Gavagnin}, E., {Bleuler}, A., {Rosdahl}, J., \& {Teyssier}, R. 2017, \mnras,
  472, 4155

\bibitem[{{Gerhard}(2011)}]{2011MSAIS..18..185G}
{Gerhard}, O. 2011, Memorie della Societa Astronomica Italiana Supplementi, 18,
  185

\bibitem[{{Gladman} {et~al.}(2001){Gladman}, {Kavelaars}, {Petit},
  {Morbidelli}, {Holman}, \& {Loredo}}]{2001AJ....122.1051G}
{Gladman}, B., {Kavelaars}, J.~J., {Petit}, J.-M., {et~al.} 2001, \aj, 122,
  1051

\bibitem[{{Gladman} {et~al.}(2008){Gladman}, {Marsden}, \&
  {Vanlaerhoven}}]{2008ssbn.book...43G}
{Gladman}, B., {Marsden}, B.~G., \& {Vanlaerhoven}, C. 2008, {Nomenclature in
  the Outer Solar System}, ed. M.~A. {Barucci}, H.~{Boehnhardt}, D.~P.
  {Cruikshank}, A.~{Morbidelli}, \& R.~{Dotson}, 43

\bibitem[{{Glaser} {et~al.}(2020){Glaser}, {McMillan}, {Geller}, {Thornton}, \&
  {Giovinazzi}}]{2020AJ....160..126G}
{Glaser}, J.~P., {McMillan}, S. L.~W., {Geller}, A.~M., {Thornton}, J.~D., \&
  {Giovinazzi}, M.~R. 2020, \aj, 160, 126

\bibitem[{{Gomes} {et~al.}(2005){Gomes}, {Levison}, {Tsiganis}, \&
  {Morbidelli}}]{2005Natur.435..466G}
{Gomes}, R., {Levison}, H.~F., {Tsiganis}, K., \& {Morbidelli}, A. 2005, \nat,
  435, 466

\bibitem[{{Gomes} {et~al.}(2004){Gomes}, {Morbidelli}, \&
  {Levison}}]{2004Icar..170..492G}
{Gomes}, R.~S., {Morbidelli}, A., \& {Levison}, H.~F. 2004, \icarus, 170, 492

\bibitem[{{Goodman} {et~al.}(1993){Goodman}, {Heggie}, \&
  {Hut}}]{1993ApJ...415..715G}
{Goodman}, J., {Heggie}, D.~C., \& {Hut}, P. 1993, \apj, 415, 715

\bibitem[{{Grazier} {et~al.}(1999){Grazier}, {Newman}, {Kaula}, \&
  {Hyman}}]{1999Icar..140..341G}
{Grazier}, K.~R., {Newman}, W.~I., {Kaula}, W.~M., \& {Hyman}, J.~M. 1999,
  \icarus, 140, 341

\bibitem[{Gropp(2002)}]{Gropp2002}
Gropp, W. 2002, MPICH2: A New Start for MPI Implementations, ed.
  D.~Kranzlm{\"u}ller, J.~Volkert, P.~Kacsuk, \& J.~Dongarra (Berlin,
  Heidelberg: Springer Berlin Heidelberg), 7--7

\bibitem[{Gropp {et~al.}(1996)Gropp, Lusk, Doss, \& Skjellum}]{Gropp:1996:HPI}
Gropp, W., Lusk, E., Doss, N., \& Skjellum, A. 1996, Parallel Computing, 22,
  789

\bibitem[{{Hands} \& {Dehnen}(2020)}]{2020MNRAS.493L..59H}
{Hands}, T.~O. \& {Dehnen}, W. 2020, \mnras, 493, L59

\bibitem[{{Hands} {et~al.}(2019){Hands}, {Dehnen}, {Gration}, {Stadel}, \&
  {Moore}}]{2019MNRAS.490...21H}
{Hands}, T.~O., {Dehnen}, W., {Gration}, A., {Stadel}, J., \& {Moore}, B. 2019,
  \mnras, 490, 21

\bibitem[{{Hanse} {et~al.}(2018){Hanse}, {J{\'\i}lkov{\'a}}, {Portegies Zwart},
  \& {Pelupessy}}]{2018MNRAS.473.5432H}
{Hanse}, J., {J{\'\i}lkov{\'a}}, L., {Portegies Zwart}, S.~F., \& {Pelupessy},
  F.~I. 2018, \mnras, 473, 5432

\bibitem[{{Hartmann}(1965)}]{1965Icar....4..157H}
{Hartmann}, W.~K. 1965, \icarus, 4, 157

\bibitem[{{Hartmann}(1966)}]{1966Icar....5..406H}
{Hartmann}, W.~K. 1966, \icarus, 5, 406

\bibitem[{{Hayashi} {et~al.}(1985){Hayashi}, {Nakazawa}, \&
  {Nakagawa}}]{1985prpl.conf.1100H}
{Hayashi}, C., {Nakazawa}, K., \& {Nakagawa}, Y. 1985, in Protostars and
  Planets II, ed. D.~C. {Black} \& M.~S. {Matthews}, 1100--1153

\bibitem[{{Heisler} \& {Tremaine}(1986)}]{1986Icar...65...13H}
{Heisler}, J. \& {Tremaine}, S. 1986, \icarus, 65, 13

\bibitem[{{Higuchi} \& {Kokubo}(2015)}]{2015AJ....150...26H}
{Higuchi}, A. \& {Kokubo}, E. 2015, \aj, 150, 26

\bibitem[{{Higuchi} \& {Kokubo}(2020)}]{2020MNRAS.492..268H}
{Higuchi}, A. \& {Kokubo}, E. 2020, \mnras, 492, 268

\bibitem[{{Higuchi} {et~al.}(2007){Higuchi}, {Kokubo}, {Kinoshita}, \&
  {Mukai}}]{2007AJ....134.1693H}
{Higuchi}, A., {Kokubo}, E., {Kinoshita}, H., \& {Mukai}, T. 2007, \aj, 134,
  1693

\bibitem[{{Hill}(1913)}]{1913AJ.....27..171H}
{Hill}, G.~W. 1913, \aj, 27, 171

\bibitem[{{Hills}(1981)}]{1981AJ.....86.1730H}
{Hills}, J.~G. 1981, \aj, 86, 1730

\bibitem[{{Hills}(1984)}]{1984Natur.311..636H}
{Hills}, J.~G. 1984, \nat, 311, 636

\bibitem[{{Hunter}(2007)}]{2007CSE.....9...90H}
{Hunter}, J.~D. 2007, Computing in Science and Engineering, 9, 90

\bibitem[{{Hut}(1984)}]{1984Natur.311..638H}
{Hut}, P. 1984, \nat, 311, 638

\bibitem[{{Ida} {et~al.}(2000){Ida}, {Larwood}, \&
  {Burkert}}]{2000ApJ...528..351I}
{Ida}, S., {Larwood}, J., \& {Burkert}, A. 2000, \apj, 528, 351

\bibitem[{{Ito} \& {Higuchi}(2020)}]{2020DPS....5230407I}
{Ito}, T. \& {Higuchi}, A. 2020, in AAS/Division for Planetary Sciences Meeting
  Abstracts, Vol.~52, AAS/Division for Planetary Sciences Meeting Abstracts,
  304.07

\bibitem[{{Ito} \& {Ohtsuka}(2019)}]{2019MEEP....7....1I}
{Ito}, T. \& {Ohtsuka}, K. 2019, Monographs on Environment, Earth and Planets,
  7, 1

\bibitem[{{Ito} \& {Tanikawa}(2002)}]{2002MNRAS.336..483I}
{Ito}, T. \& {Tanikawa}, K. 2002, \mnras, 336, 483

\bibitem[{{Jetsu} \& {Pelt}(2000)}]{2000A&A...353..409J}
{Jetsu}, L. \& {Pelt}, J. 2000, \aap, 353, 409

\bibitem[{{Jewitt}(2002)}]{2002AJ....123.1039J}
{Jewitt}, D.~C. 2002, \aj, 123, 1039

\bibitem[{{J{\'{\i}}lkov{\'a}} {et~al.}(2012){J{\'{\i}}lkov{\'a}}, {Carraro},
  {Jungwiert}, \& {Minchev}}]{2012A&A...541A..64J}
{J{\'{\i}}lkov{\'a}}, L., {Carraro}, G., {Jungwiert}, B., \& {Minchev}, I.
  2012, \aap, 541, A64

\bibitem[{{J{\'{\i}}lkov{\'a}} {et~al.}(2016){J{\'{\i}}lkov{\'a}}, {Hamers},
  {Hammer}, \& {Portegies Zwart}}]{2016MNRAS.457.4218J}
{J{\'{\i}}lkov{\'a}}, L., {Hamers}, A., {Hammer}, M., \& {Portegies Zwart}, S.
  2016, \mnras, 457, 4218

\bibitem[{{J{\'{\i}}lkov{\'a}} {et~al.}(2015){J{\'{\i}}lkov{\'a}}, {Portegies
  Zwart}, {Pijloo}, \& {Hammer}}]{2015MNRAS.453.3157J}
{J{\'{\i}}lkov{\'a}}, L., {Portegies Zwart}, S., {Pijloo}, T., \& {Hammer}, M.
  2015, \mnras, 453, 3157

\bibitem[{{Jim{\'e}nez-Torres}(2020)}]{2020AcA....70...53J}
{Jim{\'e}nez-Torres}, J.~J. 2020, \actaa, 70, 53

\bibitem[{{Johansen} \& {Lambrechts}(2017)}]{2017AREPS..45..359J}
{Johansen}, A. \& {Lambrechts}, M. 2017, Annual Review of Earth and Planetary
  Sciences, 45, 359

\bibitem[{{Johansen} {et~al.}(2007){Johansen}, {Oishi}, {Mac Low}, {Klahr},
  {Henning}, \& {Youdin}}]{2007Natur.448.1022J}
{Johansen}, A., {Oishi}, J.~S., {Mac Low}, M.-M., {et~al.} 2007, \nat, 448,
  1022

\bibitem[{{Johansen} {et~al.}(2021){Johansen}, {Ronnet}, {Bizzarro},
  {Schiller}, {Lambrechts}, {Nordlund}, \& {Lammer}}]{2021arXiv210208611J}
{Johansen}, A., {Ronnet}, T., {Bizzarro}, M., {et~al.} 2021, arXiv e-prints,
  arXiv:2102.08611

\bibitem[{{Johnstone} {et~al.}(1998){Johnstone}, {Hollenbach}, \&
  {Bally}}]{1998ApJ...499..758J}
{Johnstone}, D., {Hollenbach}, D., \& {Bally}, J. 1998, \apj, 499, 758

\bibitem[{{Juri{\'c}} {et~al.}(2008){Juri{\'c}}, {Ivezi{\'c}}, {Brooks},
  {Lupton}, {Schlegel}, {Finkbeiner}, {Padmanabhan}, {Bond}, {Sesar},
  {Rockosi}, {Knapp}, {Gunn}, {Sumi}, {Schneider}, {Barentine}, {Brewington},
  {Brinkmann}, {Fukugita}, {Harvanek}, {Kleinman}, {Krzesinski}, {Long},
  {Neilsen}, {Nitta}, {Snedden}, \& {York}}]{2008ApJ...673..864J}
{Juri{\'c}}, M., {Ivezi{\'c}}, {\v Z}., {Brooks}, A., {et~al.} 2008, \apj, 673,
  864

\bibitem[{{Kaib} \& {Quinn}(2008)}]{2008Icar..197..221K}
{Kaib}, N.~A. \& {Quinn}, T. 2008, \icarus, 197, 221

\bibitem[{{Kaib} {et~al.}(2013){Kaib}, {Raymond}, \&
  {Duncan}}]{2013Natur.493..381K}
{Kaib}, N.~A., {Raymond}, S.~N., \& {Duncan}, M. 2013, \nat, 493, 381

\bibitem[{{Kaib} {et~al.}(2011{\natexlab{a}}){Kaib}, {Roskar}, \&
  {Quinn}}]{2011DDA....42.0904K}
{Kaib}, N.~A., {Roskar}, R., \& {Quinn}, T. 2011{\natexlab{a}}, in AAS/Division
  of Dynamical Astronomy Meeting \#42, AAS/Division of Dynamical Astronomy
  Meeting, 9.04

\bibitem[{{Kaib} {et~al.}(2011{\natexlab{b}}){Kaib}, {Ro{\v{s}}kar}, \&
  {Quinn}}]{2011Icar..215..491K}
{Kaib}, N.~A., {Ro{\v{s}}kar}, R., \& {Quinn}, T. 2011{\natexlab{b}}, \icarus,
  215, 491

\bibitem[{{Kavelaars} {et~al.}(2020){Kavelaars}, {Lawler}, {Bannister}, \&
  {Shankman}}]{2020tnss.book...61K}
{Kavelaars}, J.~J., {Lawler}, S.~M., {Bannister}, M.~T., \& {Shankman}, C.
  2020, {Perspectives on the distribution of orbits of distant Trans-Neptunian
  objects}, ed. D.~{Prialnik}, M.~A. {Barucci}, \& L.~{Young}, 61--77

\bibitem[{{Kenyon} \& {Bromley}(2006)}]{2006AJ....131.1837K}
{Kenyon}, S.~J. \& {Bromley}, B.~C. 2006, \aj, 131, 1837

\bibitem[{{Kobayashi} {et~al.}(2005){Kobayashi}, {Ida}, \&
  {Tanaka}}]{2005Icar..177..246K}
{Kobayashi}, H., {Ida}, S., \& {Tanaka}, H. 2005, \icarus, 177, 246

\bibitem[{{Kokubo} \& {Ida}(1998)}]{1998Icar..131..171K}
{Kokubo}, E. \& {Ida}, S. 1998, Icarus, 131, 171

\bibitem[{Kokubo \& Ida(2002)}]{0004-637X-581-1-666}
Kokubo, E. \& Ida, S. 2002, The Astrophysical Journal, 581, 666

\bibitem[{{Korycansky} \& {Papaloizou}(1995)}]{1995MNRAS.274...85K}
{Korycansky}, D.~G. \& {Papaloizou}, J.~C.~B. 1995, \mnras, 274, 85

\bibitem[{{Kozai}(1962)}]{1962AJ.....67..591K}
{Kozai}, Y. 1962, \aj, 67, 591

\bibitem[{{Kroupa}(2001)}]{2001MNRAS.322..231K}
{Kroupa}, P. 2001, \mnras, 322, 231

\bibitem[{{Kruijer} {et~al.}(2017){Kruijer}, {Burkhardt}, {Budde}, \&
  {Kleine}}]{2017PNAS..114.6712K}
{Kruijer}, T.~S., {Burkhardt}, C., {Budde}, G., \& {Kleine}, T. 2017,
  Proceedings of the National Academy of Science, 114, 6712

\bibitem[{{Kuiper}(1951)}]{1951astr.conf..357K}
{Kuiper}, G.~P. 1951, in 50th Anniversary of the Yerkes Observatory and Half a
  Century of Progress in Astrophysics, ed. J.~A. {Hynek}, 357

\bibitem[{{Lada} \& {Lada}(2003)}]{2003ARA&A..41...57L}
{Lada}, C.~J. \& {Lada}, E.~A. 2003, \araa, 41, 57

\bibitem[{{Laskar}(1990)}]{1990Icar...88..266L}
{Laskar}, J. 1990, \icarus, 88, 266

\bibitem[{{Laskar} {et~al.}(2011){Laskar}, {Fienga}, {Gastineau}, \&
  {Manche}}]{2011A&A...532A..89L}
{Laskar}, J., {Fienga}, A., {Gastineau}, M., \& {Manche}, H. 2011, \aap, 532,
  A89

\bibitem[{{Laughlin} \& {Adams}(1998)}]{1998ApJ...508L.171L}
{Laughlin}, G. \& {Adams}, F.~C. 1998, \apjl, 508, L171

\bibitem[{{Lecar} {et~al.}(2001){Lecar}, {Franklin}, {Holman}, \&
  {Murray}}]{2001ARA&A..39..581L}
{Lecar}, M., {Franklin}, F.~A., {Holman}, M.~J., \& {Murray}, N.~J. 2001,
  \araa, 39, 581

\bibitem[{{Leto} {et~al.}(2008){Leto}, {Jakub{\'\i}k}, {Paulech},
  {Neslu{\v{s}}an}, \& {Dybczy{\'n}ski}}]{2008MNRAS.391.1350L}
{Leto}, G., {Jakub{\'\i}k}, M., {Paulech}, T., {Neslu{\v{s}}an}, L., \&
  {Dybczy{\'n}ski}, P.~A. 2008, \mnras, 391, 1350

\bibitem[{{Leto} {et~al.}(2009){Leto}, {Jakub{\'\i}k}, {Paulech},
  {Neslu{\v{s}}an}, \& {Dybczy{\'n}ski}}]{2009EM&P..105..263L}
{Leto}, G., {Jakub{\'\i}k}, M., {Paulech}, T., {Neslu{\v{s}}an}, L., \&
  {Dybczy{\'n}ski}, P.~A. 2009, Earth Moon and Planets, 105, 263

\bibitem[{{Levison} {et~al.}(2001){Levison}, {Dones}, \&
  {Duncan}}]{2001AJ....121.2253L}
{Levison}, H.~F., {Dones}, L., \& {Duncan}, M.~J. 2001, \aj, 121, 2253

\bibitem[{{Levison} {et~al.}(2010{\natexlab{a}}){Levison}, {Duncan}, {Brasser},
  \& {Kaufmann}}]{2010Sci...329..187L}
{Levison}, H.~F., {Duncan}, M.~J., {Brasser}, R., \& {Kaufmann}, D.~E.
  2010{\natexlab{a}}, Science, 329, 187

\bibitem[{{Levison} {et~al.}(2008){Levison}, {Morbidelli}, {Van Laerhoven},
  {Gomes}, \& {Tsiganis}}]{2008Icar..196..258L}
{Levison}, H.~F., {Morbidelli}, A., {Van Laerhoven}, C., {Gomes}, R., \&
  {Tsiganis}, K. 2008, \icarus, 196, 258

\bibitem[{{Levison} {et~al.}(2010{\natexlab{b}}){Levison}, {Thommes}, \&
  {Duncan}}]{2010AJ....139.1297L}
{Levison}, H.~F., {Thommes}, E., \& {Duncan}, M.~J. 2010{\natexlab{b}}, \aj,
  139, 1297

\bibitem[{{Li} {et~al.}(2006){Li}, {Zhou}, \& {Sun}}]{2006ChJAA...6..588L}
{Li}, J., {Zhou}, L.-Y., \& {Sun}, Y.-S. 2006, \cjaa, 6, 588

\bibitem[{{Licht}(1999)}]{1999Icar..137..355L}
{Licht}, A.~L. 1999, \icarus, 137, 355

\bibitem[{{Lidov}(1962)}]{1962PSS..9..719L}
{Lidov}, M. 1962, Planet. Space Sci., 9, 719

\bibitem[{{Lineweaver}(2010)}]{2010LPICo1538.5226L}
{Lineweaver}, C.~H. 2010, in Astrobiology Science Conference 2010: Evolution
  and Life: Surviving Catastrophes and Extremes on Earth and Beyond, Vol. 1538,
  5226

\bibitem[{{L{\"u}ghausen} {et~al.}(2012){L{\"u}ghausen}, {Parmentier},
  {Pflamm-Altenburg}, \& {Kroupa}}]{2012MNRAS.423.1985L}
{L{\"u}ghausen}, F., {Parmentier}, G., {Pflamm-Altenburg}, J., \& {Kroupa}, P.
  2012, \mnras, 423, 1985

\bibitem[{{Luhman}(2014)}]{2014ApJ...781....4L}
{Luhman}, K.~L. 2014, \apj, 781, 4

\bibitem[{{Luu} \& {Jewitt}(2002)}]{2002ARA&A..40...63L}
{Luu}, J.~X. \& {Jewitt}, D.~C. 2002, \araa, 40, 63

\bibitem[{{Lykawka} \& {Mukai}(2008)}]{2008AJ....135.1161L}
{Lykawka}, P.~S. \& {Mukai}, T. 2008, \aj, 135, 1161

\bibitem[{{Madigan} \& {McCourt}(2016)}]{2016MNRAS.457L..89M}
{Madigan}, A.-M. \& {McCourt}, M. 2016, \mnras, 457, L89

\bibitem[{{Malhotra}(1993)}]{1993Natur.365..819M}
{Malhotra}, R. 1993, \nat, 365, 819

\bibitem[{{Malhotra}(1998)}]{1998AAS...193.9602M}
{Malhotra}, R. 1998, in American Astronomical Society Meeting Abstracts, Vol.
  193, American Astronomical Society Meeting Abstracts, 96.02

\bibitem[{{Malhotra}(2019)}]{2019GSL.....6...12M}
{Malhotra}, R. 2019, Geoscience Letters, 6, 12

\bibitem[{{Mart{\'{\i}}nez-Barbosa} {et~al.}(2016){Mart{\'{\i}}nez-Barbosa},
  {Brown}, {Boekholt}, {Portegies Zwart}, {Antiche}, \&
  {Antoja}}]{2016MNRAS.457.1062M}
{Mart{\'{\i}}nez-Barbosa}, C.~A., {Brown}, A.~G.~A., {Boekholt}, T., {et~al.}
  2016, \mnras, 457, 1062

\bibitem[{{Mart{\'{\i}}nez-Barbosa} {et~al.}(2015){Mart{\'{\i}}nez-Barbosa},
  {Brown}, \& {Portegies Zwart}}]{2015MNRAS.446..823M}
{Mart{\'{\i}}nez-Barbosa}, C.~A., {Brown}, A.~G.~A., \& {Portegies Zwart}, S.
  2015, \mnras, 446, 823

\bibitem[{{Mart{\'{\i}}nez-Barbosa} {et~al.}(2017){Mart{\'{\i}}nez-Barbosa},
  {J{\'{\i}}lkov{\'a}}, {Portegies Zwart}, \& {Brown}}]{2017MNRAS.464.2290M}
{Mart{\'{\i}}nez-Barbosa}, C.~A., {J{\'{\i}}lkov{\'a}}, L., {Portegies Zwart},
  S., \& {Brown}, A.~G.~A. 2017, \mnras, 464, 2290

\bibitem[{{McKee} \& {Ostriker}(2007)}]{2007ARA&A..45..565M}
{McKee}, C.~F. \& {Ostriker}, E.~C. 2007, \araa, 45, 565

\bibitem[{{Melott} \& {Bambach}(2010)}]{2010MNRAS.407L..99M}
{Melott}, A.~L. \& {Bambach}, R.~K. 2010, \mnras, 407, L99

\bibitem[{{Milani} \& {Nobili}(1992)}]{1992Natur.357..569M}
{Milani}, A. \& {Nobili}, A.~M. 1992, \nat, 357, 569

\bibitem[{{Miller}(1964)}]{1964ApJ...140..250M}
{Miller}, R.~H. 1964, \apj, 140, 250

\bibitem[{{Monari} {et~al.}(2014){Monari}, {Helmi}, {Antoja}, \&
  {Steinmetz}}]{2014A&A...569A..69M}
{Monari}, G., {Helmi}, A., {Antoja}, T., \& {Steinmetz}, M. 2014, \aap, 569,
  A69

\bibitem[{{Moore} {et~al.}(2020){Moore}, {Li}, \&
  {Adams}}]{2020ApJ...901...92M}
{Moore}, N. W.~H., {Li}, G., \& {Adams}, F.~C. 2020, \apj, 901, 92

\bibitem[{{Morbidelli} {et~al.}(2009){Morbidelli}, {Brasser}, {Tsiganis},
  {Gomes}, \& {Levison}}]{2009A&A...507.1041M}
{Morbidelli}, A., {Brasser}, R., {Tsiganis}, K., {Gomes}, R., \& {Levison},
  H.~F. 2009, \aap, 507, 1041

\bibitem[{{Morbidelli} \& {Levison}(2004)}]{2004AJ....128.2564M}
{Morbidelli}, A. \& {Levison}, H.~F. 2004, \aj, 128, 2564

\bibitem[{{Morbidelli} {et~al.}(2005){Morbidelli}, {Levison}, {Tsiganis}, \&
  {Gomes}}]{2005Natur.435..462M}
{Morbidelli}, A., {Levison}, H.~F., {Tsiganis}, K., \& {Gomes}, R. 2005, \nat,
  435, 462

\bibitem[{{Morbidelli} \& {Nesvorn{\'y}}(2020)}]{2020tnss.book...25M}
{Morbidelli}, A. \& {Nesvorn{\'y}}, D. 2020, {Kuiper belt: formation and
  evolution}, ed. D.~{Prialnik}, M.~A. {Barucci}, \& L.~{Young}, 25--59

\bibitem[{{Morbidelli} {et~al.}(2007){Morbidelli}, {Tsiganis}, {Crida},
  {Levison}, \& {Gomes}}]{2007AJ....134.1790M}
{Morbidelli}, A., {Tsiganis}, K., {Crida}, A., {Levison}, H.~F., \& {Gomes}, R.
  2007, \aj, 134, 1790

\bibitem[{{Movshovitz} {et~al.}(2010){Movshovitz}, {Bodenheimer}, {Podolak}, \&
  {Lissauer}}]{2010Icar..209..616M}
{Movshovitz}, N., {Bodenheimer}, P., {Podolak}, M., \& {Lissauer}, J.~J. 2010,
  \icarus, 209, 616

\bibitem[{{Murray} \& {Holman}(1999)}]{1999Sci...283.1877M}
{Murray}, N. \& {Holman}, M. 1999, Science, 283, 1877

\bibitem[{{Murray} \& {Holman}(2001)}]{2001Natur.410..773M}
{Murray}, N. \& {Holman}, M. 2001, \nat, 410, 773

\bibitem[{{Mustill} {et~al.}(2016){Mustill}, {Raymond}, \&
  {Davies}}]{2016MNRAS.460L.109M}
{Mustill}, A.~J., {Raymond}, S.~N., \& {Davies}, M.~B. 2016, \mnras, 460, L109

\bibitem[{{Napier} {et~al.}(2021{\natexlab{a}}){Napier}, {Adams}, \&
  {Batygin}}]{2021PSJ.....2...53N}
{Napier}, K.~J., {Adams}, F.~C., \& {Batygin}, K. 2021{\natexlab{a}}, The
  Planetary Science Journal, 2, 53

\bibitem[{{Napier} {et~al.}(2021{\natexlab{b}}){Napier}, {Gerdes}, {Lin},
  {Hamilton}, {Bernstein}, {Bernardinelli}, {Abbott}, {Aguena}, {Annis},
  {Avila}, {Bacon}, {Bertin}, {Brooks}, {Burke}, {Carnero Rosell}, {Carrasco
  Kind}, {Carretero}, {Costanzi}, {da Costa}, {De Vicente}, {Diehl}, {Doel},
  {Everett}, {Ferrero}, {Fosalba}, {Garc{\'\i}a Bellido}, {Gruen}, {Gruendl},
  {Gutierrez}, {Hollowood}, {Honscheid}, {Hoyle}, {James}, {Kent}, {Kuehn},
  {Kuropatkin}, {Maia}, {Menanteau}, {Miquel}, {Morgan}, {Palmese},
  {Paz-Chinch{\'o}n}, {Plazas}, {Sanchez}, {Scarpine}, {Serrano},
  {Sevilla-Noarbe}, {Smith}, {Suchyta}, {Swanson}, {To}, {Walker}, \&
  {Wilkinson}}]{2021arXiv210205601N}
{Napier}, K.~J., {Gerdes}, D.~W., {Lin}, H.~W., {et~al.} 2021{\natexlab{b}},
  arXiv e-prints, arXiv:2102.05601

\bibitem[{{Neslu{\v{s}}an}(2000)}]{2000A&A...361..369N}
{Neslu{\v{s}}an}, L. 2000, \aap, 361, 369

\bibitem[{{Neslu{\v{s}}an} {et~al.}(2009){Neslu{\v{s}}an}, {Dybczy{\'n}ski},
  {Leto}, {Jakub{\'\i}k}, \& {Paulech}}]{2009EM&P..105..257N}
{Neslu{\v{s}}an}, L., {Dybczy{\'n}ski}, P.~A., {Leto}, G., {Jakub{\'\i}k}, M.,
  \& {Paulech}, T. 2009, Earth Moon and Planets, 105, 257

\bibitem[{{Nesvorn{\'y}}(2020)}]{2020RNAAS...4..212N}
{Nesvorn{\'y}}, D. 2020, Research Notes of the American Astronomical Society,
  4, 212

\bibitem[{{Neukum} {et~al.}(2001){Neukum}, {Ivanov}, \&
  {Hartmann}}]{2001SSRv...96...55N}
{Neukum}, G., {Ivanov}, B.~A., \& {Hartmann}, W.~K. 2001, \ssr, 96, 55

\bibitem[{Newton(1687)}]{Newton:1687}
Newton, I. 1687, Philosophiae Naturalis Principia Mathematica, Vol.~1

\bibitem[{{Nordlander} {et~al.}(2017){Nordlander}, {Rickman}, \&
  {Gustafsson}}]{2017A&A...603A.112N}
{Nordlander}, T., {Rickman}, H., \& {Gustafsson}, B. 2017, \aap, 603, A112

\bibitem[{{Nurmi} {et~al.}(2002){Nurmi}, {Valtonen}, {Zheng}, \&
  {Rickman}}]{2002MNRAS.333..835N}
{Nurmi}, P., {Valtonen}, M.~J., {Zheng}, J.~Q., \& {Rickman}, H. 2002, \mnras,
  333, 835

\bibitem[{{Offner} \& {Arce}(2015)}]{2015ApJ...811..146O}
{Offner}, S. S.~R. \& {Arce}, H.~G. 2015, \apj, 811, 146

\bibitem[{Oliphant(2006)}]{Oliphant2006ANumPy}
Oliphant, T.~E. 2006, {A guide to NumPy}, Vol.~1 (Trelgol Publishing USA)

\bibitem[{{Oort}(1950)}]{1950BAN....11...91O}
{Oort}, J.~H. 1950, \bain, 11, 91

\bibitem[{{{\"O}pik}(1932)}]{1932PAAAS..67..169O}
{{\"O}pik}, E. 1932, Proceedings of the American Academy of Arts and Sciences,
  67, 169

\bibitem[{{Opitom} {et~al.}(2019){Opitom}, {Fitzsimmons}, {Jehin}, {Moulane},
  {Hainaut}, {Meech}, {Yang}, {Snodgrass}, {Micheli}, {Keane}, {Benkhaldoun},
  \& {Kleyna}}]{2019A&A...631L...8O}
{Opitom}, C., {Fitzsimmons}, A., {Jehin}, E., {et~al.} 2019, \aap, 631, L8

\bibitem[{{'Oumuamua ISSI Team} {et~al.}(2019){'Oumuamua ISSI Team},
  {Bannister}, {Bhandare}, {Dybczy{\'n}ski}, {Fitzsimmons},
  {Guilbert-Lepoutre}, {Jedicke}, {Knight}, {McNeill}, {Pfalzner}, {Raymond},
  {Snodgrass}, {Trilling}, \& {Ye}}]{2019NatAs...3..594O}
{'Oumuamua ISSI Team}, {Bannister}, M.~T., {Bhandare}, A., {et~al.} 2019,
  Nature Astronomy, 3, 594

\bibitem[{{Parker}(2020)}]{2020RSOS....701271P}
{Parker}, R.~J. 2020, Royal Society Open Science, 7, 201271

\bibitem[{{Patterson} \& {Smith}(1987)}]{1987Natur.330..248P}
{Patterson}, C. \& {Smith}, A.~B. 1987, \nat, 330, 248

\bibitem[{{Paulech} {et~al.}(2010){Paulech}, {Jakub{\'\i}k}, {Neslu{\v{s}}an},
  {Dybczy{\'n}ski}, \& {Leto}}]{2010A&A...509A..48P}
{Paulech}, T., {Jakub{\'\i}k}, M., {Neslu{\v{s}}an}, L., {Dybczy{\'n}ski},
  P.~A., \& {Leto}, G. 2010, \aap, 509, A48

\bibitem[{{Pelupessy} {et~al.}(2012){Pelupessy}, {J{\"a}nes}, \& {Portegies
  Zwart}}]{2012NewA...17..711P}
{Pelupessy}, F.~I., {J{\"a}nes}, J., \& {Portegies Zwart}, S. 2012, \na, 17,
  711

\bibitem[{{Perets} \& {Kouwenhoven}(2012)}]{2012ApJ...750...83P}
{Perets}, H.~B. \& {Kouwenhoven}, M.~B.~N. 2012, \apj, 750, 83

\bibitem[{{Pfalzner} {et~al.}(2021){Pfalzner}, {Aizpuru Vargas}, {Bhandare}, \&
  {Veras}}]{2021arXiv210406845P}
{Pfalzner}, S., {Aizpuru Vargas}, L., {Bhandare}, A., \& {Veras}, D. 2021,
  arXiv e-prints, arXiv:2104.06845

\bibitem[{{Pfalzner} {et~al.}(2020){Pfalzner}, {Davies}, {Kokaia}, \&
  {Bannister}}]{2020ApJ...903..114P}
{Pfalzner}, S., {Davies}, M.~B., {Kokaia}, G., \& {Bannister}, M.~T. 2020,
  \apj, 903, 114

\bibitem[{{Pfalzner} \& {Vincke}(2020)}]{2020ApJ...897...60P}
{Pfalzner}, S. \& {Vincke}, K. 2020, \apj, 897, 60

\bibitem[{{Pfalzner} {et~al.}(2005){Pfalzner}, {Vogel}, {Scharw{\"a}chter}, \&
  {Olczak}}]{2005A&A...437..967P}
{Pfalzner}, S., {Vogel}, P., {Scharw{\"a}chter}, J., \& {Olczak}, C. 2005,
  \aap, 437, 967

\bibitem[{{Pirani} {et~al.}(2018){Pirani}, {Johansen}, {Bitsch}, {Mustill}, \&
  {Turrini}}]{2018DPS....5020001P}
{Pirani}, S., {Johansen}, A., {Bitsch}, B., {Mustill}, A.~J., \& {Turrini}, D.
  2018, in AAS/Division for Planetary Sciences Meeting Abstracts, Vol.~50,
  AAS/Division for Planetary Sciences Meeting Abstracts \#50, 200.01D

\bibitem[{{Pirani} {et~al.}(2019){Pirani}, {Johansen}, {Bitsch}, {Mustill}, \&
  {Turrini}}]{2019A&A...623A.169P}
{Pirani}, S., {Johansen}, A., {Bitsch}, B., {Mustill}, A.~J., \& {Turrini}, D.
  2019, \aap, 623, A169

\bibitem[{{Plummer}(1911)}]{1911MNRAS..71..460P}
{Plummer}, H.~C. 1911, \mnras, 71, 460

\bibitem[{{Popovas} {et~al.}(2018){Popovas}, {Nordlund}, {Ramsey}, \&
  {Ormel}}]{2018MNRAS.479.5136P}
{Popovas}, A., {Nordlund}, {\r{A}}., {Ramsey}, J.~P., \& {Ormel}, C.~W. 2018,
  \mnras, 479, 5136

\bibitem[{{Portegies Zwart}(2019)}]{2019A&A...622A..69P}
{Portegies Zwart}, S. 2019, \aap, 622, A69

\bibitem[{{Portegies Zwart}(2020)}]{2020NatAs...4..819P}
{Portegies Zwart}, S. 2020, Nature Astronomy, 4, 819

\bibitem[{{Portegies Zwart}(2021)}]{2021A&A...647A.136P}
{Portegies Zwart}, S. 2021, \aap, 647, A136

\bibitem[{{Portegies Zwart} \& {McMillan}(2018)}]{2018araa.book.....P}
{Portegies Zwart}, S. \& {McMillan}, S. 2018, {Astrophysical Recipes; The art
  of AMUSE}

\bibitem[{{Portegies Zwart} {et~al.}(2009){Portegies Zwart}, {McMillan},
  {Harfst}, {Groen}, {Fujii}, {Nuall{\'a}in}, {Glebbeek}, {Heggie}, {Lombardi},
  {Hut}, {Angelou}, {Banerjee}, {Belkus}, {Fragos}, {Fregeau}, {Gaburov},
  {Izzard}, {Juri{\'c}}, {Justham}, {Sottoriva}, {Teuben}, {van Bever},
  {Yaron}, \& {Zemp}}]{2009NewA...14..369P}
{Portegies Zwart}, S., {McMillan}, S., {Harfst}, S., {et~al.} 2009, New
  Astronomy, 14, 369

\bibitem[{{Portegies Zwart} {et~al.}(2020){Portegies Zwart}, {Pelupessy},
  {Martínez-Barbosa}, \& McMillan}]{ZWART2020105240}
{Portegies Zwart}, S., {Pelupessy}, I., {Martínez-Barbosa}, C.~{van Elteren},
  A., \& McMillan, S. 2020, Communications in Nonlinear Science and Numerical
  Simulation, 105240

\bibitem[{{Portegies Zwart} {et~al.}(2018){Portegies Zwart}, {Torres},
  {Pelupessy}, {B{\'e}dorf}, \& {Cai}}]{2018MNRAS.479L..17P}
{Portegies Zwart}, S., {Torres}, S., {Pelupessy}, I., {B{\'e}dorf}, J., \&
  {Cai}, M.~X. 2018, \mnras, 479, L17

\bibitem[{Portegies~Zwart {et~al.}(2018)Portegies~Zwart, van Elteren,
  Pelupessy, McMillan, Rieder, de~Vries, Marosvolgyi, Whitehead, Wall, Drost,
  Jilkova, Martinez~Barbosa, van~der Helm, Beedorf, Bos, Boekholt, van
  Werkhoven, Wijnen, Hamers, Caputo, Ferrari, Toonen, Gaburov, Paardekooper,
  Janes, Punzo, Kruip, \& Altay}]{portegies_zwart_simon_2018_1443252}
Portegies~Zwart, S., van Elteren, A., Pelupessy, I., {et~al.} 2018, {AMUSE: the
  Astrophysical Multipurpose Software Environment}

\bibitem[{{Portegies Zwart}(2009)}]{2009ApJ...696L..13P}
{Portegies Zwart}, S.~F. 2009, \apjl, 696, L13

\bibitem[{{Portegies Zwart} \& {Boekholt}(2018)}]{2018CNSNS..61..160P}
{Portegies Zwart}, S.~F. \& {Boekholt}, T. C.~N. 2018, Communications in
  Nonlinear Science and Numerical Simulations, 61, 160

\bibitem[{{Portegies Zwart} \&
  {J{\'{\i}}lkov{\'a}}(2015)}]{2015MNRAS.451.4663P}
{Portegies Zwart}, S.~F. \& {J{\'{\i}}lkov{\'a}}, L. 2015, \mnras, 451, 4663

\bibitem[{{Portegies Zwart} {et~al.}(2010){Portegies Zwart}, {McMillan}, \&
  {Gieles}}]{2010ARA&A..48..431P}
{Portegies Zwart}, S.~F., {McMillan}, S.~L.~W., \& {Gieles}, M. 2010, \araa,
  48, 431

\bibitem[{{Portegies Zwart} \& {Verbunt}(1996)}]{1996A&A...309..179P}
{Portegies Zwart}, S.~F. \& {Verbunt}, F. 1996, \aap, 309, 179

\bibitem[{{Punzo} {et~al.}(2014){Punzo}, {Capuzzo-Dolcetta}, \& {Portegies
  Zwart}}]{2014MNRAS.444.2808P}
{Punzo}, D., {Capuzzo-Dolcetta}, R., \& {Portegies Zwart}, S. 2014, \mnras,
  444, 2808

\bibitem[{{Rabinowitz} {et~al.}(2006){Rabinowitz}, {Barkume}, {Brown}, {Roe},
  {Schwartz}, {Tourtellotte}, \& {Trujillo}}]{2006ApJ...639.1238R}
{Rabinowitz}, D.~L., {Barkume}, K., {Brown}, M.~E., {et~al.} 2006, \apj, 639,
  1238

\bibitem[{{Rampino} \& {Prokoph}(2020)}]{2020AsBio..20.1097R}
{Rampino}, M.~R. \& {Prokoph}, A. 2020, Astrobiology, 20, 1097

\bibitem[{{Raup} \& {Sepkoski}(1984)}]{1984PNAS...81..801R}
{Raup}, D.~M. \& {Sepkoski}, J.~J. 1984, Proceedings of the National Academy of
  Science, 81, 801

\bibitem[{{Rawiraswattana} {et~al.}(2016){Rawiraswattana}, {Hubber}, \&
  {Goodwin}}]{2016MNRAS.460.3505R}
{Rawiraswattana}, K., {Hubber}, D.~A., \& {Goodwin}, S.~P. 2016, \mnras, 460,
  3505

\bibitem[{{Rein} \& {Liu}(2012)}]{2012A&A...537A.128R}
{Rein}, H. \& {Liu}, S.-F. 2012, \aap, 537, A128

\bibitem[{{Rein} \& {Spiegel}(2015)}]{2015MNRAS.446.1424R}
{Rein}, H. \& {Spiegel}, D.~S. 2015, \mnras, 446, 1424

\bibitem[{{Rickman}(2014)}]{2014M&PS...49....8R}
{Rickman}, H. 2014, Meteoritics and Planetary Science, 49, 8

\bibitem[{{Rickman} {et~al.}(2008){Rickman}, {Fouchard}, {Froeschl{\'e}}, \&
  {Valsecchi}}]{2008CeMDA.102..111R}
{Rickman}, H., {Fouchard}, M., {Froeschl{\'e}}, C., \& {Valsecchi}, G.~B. 2008,
  Celestial Mechanics and Dynamical Astronomy, 102, 111

\bibitem[{{Rickman} {et~al.}(1987){Rickman}, {Kamel}, {Festou}, \&
  {Froeschle}}]{1987ESASP.278..471R}
{Rickman}, H., {Kamel}, L., {Festou}, M.~C., \& {Froeschle}, C. 1987, in ESA
  Special Publication, Vol. 278, Diversity and Similarity of Comets, ed. E.~J.
  {Rolfe}, B.~{Battrick}, M.~{Ackerman}, M.~{Scherer}, \& R.~{Reinhard},
  471--481

\bibitem[{{Romero-G{\'o}mez} {et~al.}(2011){Romero-G{\'o}mez}, {Athanassoula},
  {Antoja}, \& {Figueras}}]{2011MNRAS.418.1176R}
{Romero-G{\'o}mez}, M., {Athanassoula}, E., {Antoja}, T., \& {Figueras}, F.
  2011, \mnras, 418, 1176

\bibitem[{{Ryder}(2002)}]{2002JGRE..107.5022R}
{Ryder}, G. 2002, Journal of Geophysical Research (Planets), 107, 5022

\bibitem[{{Saillenfest}(2020)}]{2020CeMDA.132...12S}
{Saillenfest}, M. 2020, Celestial Mechanics and Dynamical Astronomy, 132, 12

\bibitem[{{Saillenfest} {et~al.}(2019){Saillenfest}, {Fouchard}, {Ito}, \&
  {Higuchi}}]{2019A&A...629A..95S}
{Saillenfest}, M., {Fouchard}, M., {Ito}, T., \& {Higuchi}, A. 2019, \aap, 629,
  A95

\bibitem[{{Schlecker} {et~al.}(2020){Schlecker}, {Mordasini}, {Emsenhuber},
  {Klahr}, {Henning}, {Burn}, {Alibert}, \& {Benz}}]{2020arXiv200705563S}
{Schlecker}, M., {Mordasini}, C., {Emsenhuber}, A., {et~al.} 2020, arXiv
  e-prints, arXiv:2007.05563

\bibitem[{{Scott}(1992)}]{GaussianKDE1992S}
{Scott}, D. 1992, Multivariate Density Estimation: Theory, Practice, and
  Visualization (Wiley Series in Probability and Statistics, John Wiley \&
  Sons, Inc. 1992)

\bibitem[{{Shankman} {et~al.}(2011){Shankman}, {Gladman}, \&
  {Kaib}}]{2011epsc.conf..633S}
{Shankman}, C., {Gladman}, B., \& {Kaib}, N. 2011, in EPSC-DPS Joint Meeting
  2011, Vol. 2011, 633

\bibitem[{{Shannon} {et~al.}(2019){Shannon}, {Jackson}, \&
  {Wyatt}}]{2019MNRAS.485.5511S}
{Shannon}, A., {Jackson}, A.~P., \& {Wyatt}, M.~C. 2019, \mnras, 485, 5511

\bibitem[{{Sheppard} {et~al.}(2014){Sheppard}, {Trujillo}, \&
  {Williams}}]{2014MPEC....F...40S}
{Sheppard}, S.~S., {Trujillo}, C., \& {Williams}, G.~V. 2014, Minor Planet
  Electronic Circulars, 2014-F40

\bibitem[{{Sheppard} {et~al.}(2019){Sheppard}, {Trujillo}, {Tholen}, \&
  {Kaib}}]{2019AJ....157..139S}
{Sheppard}, S.~S., {Trujillo}, C.~A., {Tholen}, D.~J., \& {Kaib}, N. 2019, \aj,
  157, 139

\bibitem[{{Siraj} \& {Loeb}(2020)}]{2020ApJ...899L..24S}
{Siraj}, A. \& {Loeb}, A. 2020, \apjl, 899, L24

\bibitem[{{Sosa} \& {Fern{\'a}ndez}(2009)}]{2009MNRAS.393..192S}
{Sosa}, A. \& {Fern{\'a}ndez}, J.~A. 2009, \mnras, 393, 192

\bibitem[{{Sosa} \& {Fern{\'a}ndez}(2011)}]{2011MNRAS.416..767S}
{Sosa}, A. \& {Fern{\'a}ndez}, J.~A. 2011, \mnras, 416, 767

\bibitem[{{Stevenson} \& {Lunine}(1988)}]{1988Icar...75..146S}
{Stevenson}, D.~J. \& {Lunine}, J.~I. 1988, \icarus, 75, 146

\bibitem[{{Stock} {et~al.}(2020){Stock}, {Cai}, {Spurzem}, {Kouwenhoven}, \&
  {Portegies Zwart}}]{2020MNRAS.497.1807S}
{Stock}, K., {Cai}, M.~X., {Spurzem}, R., {Kouwenhoven}, M.~B.~N., \&
  {Portegies Zwart}, S. 2020, \mnras, 497, 1807

\bibitem[{{St{\"o}ffler} \& {Ryder}(2001)}]{2001SSRv...96....9S}
{St{\"o}ffler}, D. \& {Ryder}, G. 2001, \ssr, 96, 9

\bibitem[{{Sussman} \& {Wisdom}(1988)}]{1988Sci...241..433S}
{Sussman}, G.~J. \& {Wisdom}, J. 1988, Science, 241, 433

\bibitem[{{Sussman} \& {Wisdom}(1992)}]{1992Sci...257...56S}
{Sussman}, G.~J. \& {Wisdom}, J. 1992, Science, 257, 56

\bibitem[{{Tanikawa} \& {Ito}(2007)}]{2007PASJ...59..989T}
{Tanikawa}, K. \& {Ito}, T. 2007, \pasj, 59, 989

\bibitem[{{Tegler} \& {Romanishin}(2000)}]{2000Natur.407..979T}
{Tegler}, S.~C. \& {Romanishin}, W. 2000, \nat, 407, 979

\bibitem[{{Thommes} {et~al.}(2008){Thommes}, {Nagasawa}, \&
  {Lin}}]{2008ApJ...676..728T}
{Thommes}, E., {Nagasawa}, M., \& {Lin}, D.~N.~C. 2008, \apj, 676, 728

\bibitem[{{Todorovi{\'c}} {et~al.}(2020){Todorovi{\'c}}, {Wu}, \&
  {Rosengren}}]{2020SciA....6.1313T}
{Todorovi{\'c}}, N., {Wu}, D., \& {Rosengren}, A.~J. 2020, Science Advances, 6,
  eabd1313

\bibitem[{{Toomre}(1964)}]{1964ApJ...139.1217T}
{Toomre}, A. 1964, \apj, 139, 1217

\bibitem[{{Toonen} {et~al.}(2016){Toonen}, {Hamers}, \& {Portegies
  Zwart}}]{2016ComAC...3....6T}
{Toonen}, S., {Hamers}, A., \& {Portegies Zwart}, S. 2016, Computational
  Astrophysics and Cosmology, 3, 6

\bibitem[{{Torres}(2020)}]{Torres_PhD2020}
{Torres}, S. 2020, PhD thesis: Dynamics of the Oort Cloud and the Formation of
  Interstellar Comets (Leiden Observatory)

\bibitem[{{Torres} {et~al.}(2019){Torres}, {Cai}, {Brown}, \& {Portegies
  Zwart}}]{2019A&A...629A.139T}
{Torres}, S., {Cai}, M.~X., {Brown}, A.~G.~A., \& {Portegies Zwart}, S. 2019,
  \aap, 629, A139

\bibitem[{Torres {et~al.}(2020{\natexlab{a}})Torres, Cai, Mukherjee, {Portegies
  Zwart}, \& Brown}]{Torres2020}
Torres, S., Cai, M.~X., Mukherjee, D., {Portegies Zwart}, S., \& Brown, A.
  G.~A. 2020{\natexlab{a}}, Astron. Astrophys., submitted, 17

\bibitem[{Torres {et~al.}(2020{\natexlab{b}})Torres, {Portegies Zwart}, \&
  Brown}]{Torres2020b}
Torres, S., {Portegies Zwart}, S., \& Brown, A. G.~A. 2020{\natexlab{b}},
  Astron. Astrophys., submitted, 10

\bibitem[{{Trujillo} {et~al.}(2000){Trujillo}, {Jewitt}, \&
  {Luu}}]{2000ApJ...529L.103T}
{Trujillo}, C.~A., {Jewitt}, D.~C., \& {Luu}, J.~X. 2000, \apjl, 529, L103

\bibitem[{{Trujillo} \& {Sheppard}(2014)}]{2014Natur.507..471T}
{Trujillo}, C.~A. \& {Sheppard}, S.~S. 2014, \nat, 507, 471

\bibitem[{{Tsiganis} {et~al.}(2005){Tsiganis}, {Gomes}, {Morbidelli}, \&
  {Levison}}]{2005Natur.435..459T}
{Tsiganis}, K., {Gomes}, R., {Morbidelli}, A., \& {Levison}, H.~F. 2005, \nat,
  435, 459

\bibitem[\protect\citeauthoryear{Tutukov, Dremova, \& Dremov}{2020}]{2020ARep...64..936T}
  Tutukov A.~V., Dremova G.~N., Dremov V.~V., 2020, ARep, 64, 936

\bibitem[\protect\citeauthoryear{Tutukov, Sizova, \&
    Vereshchagin}{2021}]{2021ARep...65..305T} Tutukov A.~V., Sizova
  M.~D., Vereshchagin S.~V., 2021, ARep, 65, 305
  
\bibitem[{{Valtonen} {et~al.}(2004){Valtonen}, {Myll{\"a}ri}, {Orlov}, \&
  {Rubinov}}]{2004ASPC..316...45V}
{Valtonen}, M., {Myll{\"a}ri}, A., {Orlov}, V., \& {Rubinov}, A. 2004, in
  Astronomical Society of the Pacific Conference Series, Vol. 316, Order and
  Chaos in Stellar and Planetary Systems, ed. G.~G. {Byrd}, K.~V.
  {Kholshevnikov}, A.~A. {Myllri}, I.~I. {Nikiforov}, \& V.~V. {Orlov}, 45

\bibitem[{{Valtonen} \& {Innanen}(1982)}]{1982ApJ...255..307V}
{Valtonen}, M.~J. \& {Innanen}, K.~A. 1982, \apj, 255, 307

\bibitem[{{Valtonen} {et~al.}(1992){Valtonen}, {Zheng}, \&
  {Mikkola}}]{1992CeMDA..54...37V}
{Valtonen}, M.~J., {Zheng}, J.-Q., \& {Mikkola}, S. 1992, Celestial Mechanics
  and Dynamical Astronomy, 54, 37

\bibitem[{{van Rossum}(1995)}]{vanRossum:1995:EEP}
{van Rossum}, G. 1995, Extending and embedding the {Python} interpreter, Report
  CS-R9527

\bibitem[{{Varadi} {et~al.}(2003){Varadi}, {Runnegar}, \&
  {Ghil}}]{2003ApJ...592..620V}
{Varadi}, F., {Runnegar}, B., \& {Ghil}, M. 2003, \apj, 592, 620

\bibitem[{{Vargya} \& {Sanderson}(2020)}]{2020DDA....5120005V}
{Vargya}, D. \& {Sanderson}, R. 2020, in AAS/Division of Dynamical Astronomy
  Meeting, Vol.~52, AAS/Division of Dynamical Astronomy Meeting, 200.05

\bibitem[{{Veras} {et~al.}(2014){Veras}, {Evans}, {Wyatt}, \&
  {Tout}}]{2014MNRAS.437.1127V}
{Veras}, D., {Evans}, N.~W., {Wyatt}, M.~C., \& {Tout}, C.~A. 2014, \mnras,
  437, 1127

\bibitem[{{Veras} {et~al.}(2020){Veras}, {Reichert}, {Flammini Dotti}, {Cai},
  {Mustill}, {Shannon}, {McDonald}, {Portegies Zwart}, {Kouwenhoven}, \&
  {Spurzem}}]{2020MNRAS.493.5062V}
{Veras}, D., {Reichert}, K., {Flammini Dotti}, F., {et~al.} 2020, \mnras, 493,
  5062

\bibitem[{{Veras} \& {Tout}(2012)}]{2012MNRAS.422.1648V}
{Veras}, D. \& {Tout}, C.~A. 2012, \mnras, 422, 1648

\bibitem[{{Veras} {et~al.}(2011){Veras}, {Wyatt}, {Mustill}, {Bonsor}, \&
  {Eldridge}}]{2011MNRAS.417.2104V}
{Veras}, D., {Wyatt}, M.~C., {Mustill}, A.~J., {Bonsor}, A., \& {Eldridge},
  J.~J. 2011, \mnras, 417, 2104

\bibitem[{{Vincke} {et~al.}(2015){Vincke}, {Breslau}, \&
  {Pfalzner}}]{2015A&A...577A.115V}
{Vincke}, K., {Breslau}, A., \& {Pfalzner}, S. 2015, \aap, 577, A115

\bibitem[{{Vincke} \& {Pfalzner}(2016)}]{2016ApJ...828...48V}
{Vincke}, K. \& {Pfalzner}, S. 2016, \apj, 828, 48

\bibitem[{{Vokrouhlick{\'y}} {et~al.}(2019){Vokrouhlick{\'y}}, {Nesvorn{\'y}},
  \& {Dones}}]{2019AJ....157..181V}
{Vokrouhlick{\'y}}, D., {Nesvorn{\'y}}, D., \& {Dones}, L. 2019, \aj, 157, 181

\bibitem[{{Volk} \& {Malhotra}(2017)}]{2017AJ....154...62V}
{Volk}, K. \& {Malhotra}, R. 2017, \aj, 154, 62

\bibitem[{{von Zeipel}(1910)}]{1910AN....183..345V}
{von Zeipel}, H. 1910, Astronomische Nachrichten, 183, 345

\bibitem[{{Walsh} {et~al.}(2011){Walsh}, {Morbidelli}, {Raymond}, {O'Brien}, \&
  {Mandell}}]{2011Natur.475..206W}
{Walsh}, K.~J., {Morbidelli}, A., {Raymond}, S.~N., {O'Brien}, D.~P., \&
  {Mandell}, A.~M. 2011, \nat, 475, 206

\bibitem[{{Wang} {et~al.}(2015){Wang}, {Spurzem}, {Aarseth}, {Nitadori},
  {Berczik}, {Kouwenhoven}, \& {Naab}}]{2015MNRAS.450.4070W}
{Wang}, L., {Spurzem}, R., {Aarseth}, S., {et~al.} 2015, \mnras, 450, 4070

\bibitem[{{Weissman}(1983)}]{1983A&A...118...90W}
{Weissman}, P.~R. 1983, \aap, 118, 90

\bibitem[{{Weissman}(1996)}]{1996ASPC..107..265W}
{Weissman}, P.~R. 1996, in Astronomical Society of the Pacific Conference
  Series, Vol. 107, Completing the Inventory of the Solar System, ed.
  T.~{Rettig} \& J.~M. {Hahn}, 265--288

\bibitem[{{Williams} \& {Cieza}(2011)}]{2011ARA&A..49...67W}
{Williams}, J.~P. \& {Cieza}, L.~A. 2011, \araa, 49, 67

\bibitem[{{Winter} {et~al.}(2018){Winter}, {Clarke}, {Rosotti}, {Ih},
  {Facchini}, \& {Haworth}}]{2018MNRAS.478.2700W}
{Winter}, A.~J., {Clarke}, C.~J., {Rosotti}, G., {et~al.} 2018, \mnras, 478,
  2700

\bibitem[{{Wong} {et~al.}(2019){Wong}, {Brasser}, \&
  {Werner}}]{2019E&PSL.506..407W}
{Wong}, E.~W., {Brasser}, R., \& {Werner}, S.~C. 2019, Earth and Planetary
  Science Letters, 506, 407

\bibitem[{{Wyatt}(2003)}]{2003ApJ...598.1321W}
{Wyatt}, M.~C. 2003, \apj, 598, 1321

\bibitem[{{Wyatt}(2008)}]{2008ARA&A..46..339W}
{Wyatt}, M.~C. 2008, \araa, 46, 339

\bibitem[{{Xiang-Gruess}(2016)}]{2016MNRAS.455.3086X}
{Xiang-Gruess}, M. 2016, \mnras, 455, 3086

\bibitem[{{Zderic} \& {Madigan}(2020)}]{2020AJ....160...50Z}
{Zderic}, A. \& {Madigan}, A.-M. 2020, \aj, 160, 50

\bibitem[{{Zellner}(2017)}]{2017OLEB...47..261Z}
{Zellner}, N. E.~B. 2017, Origins of Life and Evolution of the Biosphere, 47,
  261

\bibitem[{{Zhang}(2018)}]{2018ApJ...852L..13Z}
{Zhang}, Q. 2018, \apjl, 852, L13

\bibitem[{{Zheng} {et~al.}(1990){Zheng}, {Valtonen}, \&
  {Valtaoja}}]{1990CeMDA..49..265Z}
{Zheng}, J.-Q., {Valtonen}, M.~J., \& {Valtaoja}, L. 1990, Celestial Mechanics
  and Dynamical Astronomy, 49, 265

\bibitem[{{Zink} {et~al.}(2020){Zink}, {Batygin}, \&
  {Adams}}]{2020AJ....160..232Z}
{Zink}, J.~K., {Batygin}, K., \& {Adams}, F.~C. 2020, \aj, 160, 232

\end{thebibliography}
\end{document}